%% file: main.tex
\documentclass[aps,prx,onecolumn,10pt,superscriptaddress,nofootinbib,english]{revtex4-2}

\makeatletter

\def\p@subsection{}

\def\p@subsubsection{}
\makeatother

\input{preamble.tex}      
\input{commands.tex}      

\begin{document}

\title{\yb Reference Data}
\author{Ronen~M.~Kroeze\orcidlink{0009-0007-9976-5124}}
\email{r.kroeze@lmu.de}
\affiliation{Fakult{\"a}t f{\"u}r Physik, Ludwig-Maximilians-Universit{\"a}t, 80799 M{\"u}nchen, Germany}
\affiliation{Munich Center for Quantum Science and Technology, 80799 M{\"u}nchen, Germany}
\author{Sofus~Laguna~Kristensen\orcidlink{0000-0002-2463-9883}}
\email{skristensen@atom-computing.com}
\altaffiliation{Present address: \href{https://ror.org/05vxfdm89}{Atom Computing, Inc.}, 2500 55th St, Boulder, Colorado 80301, USA}
\affiliation{Munich Center for Quantum Science and Technology, 80799 M{\"u}nchen, Germany}
\affiliation{Max-Planck-Institut f{\"u}r Quantenoptik, 85748 Garching, Germany}
\author{Sebastian Pucher\orcidlink{0000-0001-5559-0114}}
\email{sebastian.pucher@protonmail.com}
\affiliation{Munich Center for Quantum Science and Technology, 80799 M{\"u}nchen, Germany}
\affiliation{Max-Planck-Institut f{\"u}r Quantenoptik, 85748 Garching, Germany}
\affiliation{\href{https://ror.org/05vxfdm89}{Atom Computing, Inc.}, 2500 55th St, Boulder, Colorado 80301, USA}

\date{\today}

\begin{abstract}
\noindent Ytterbium-171 is a versatile atomic species often used in quantum optics, precision metrology, and quantum computing.
Consolidated atomic data is essential for the planning, execution, and evaluation of experiments.
In this reference, we present physical and optical properties of neutral \yb relevant to these applications.
We emphasize experimental results and supplement these with theoretical estimates.
We present equations to convert values and derive important parameters.
Tabulated results include key parameters for commonly used transitions in \yb ($\sS{0}\to\sP{1}$, $\sS{0}\to\tP{0,1,2}$, $\tP{0,2}\to\tS{1}$, and $\tP{0}\to\tD{1}$).
This dataset serves as an up-to-date reference for studies involving fermionic \yb.
\end{abstract}

\maketitle
\vspace{-3mm}

\section{Introduction}\label{sec:intro}

In this reference, we present the physical and optical properties of ytterbium-171 (\yb) relevant to various quantum optics, laser cooling, and precision measurement experiments.
This document is inspired by the widely used references on alkali atoms by Daniel A.\ Steck~\cite{Steck2024SodiumDLine, Steck2024Rb85DLine, Steck2024Rb87DLine, Steck2024CesiumDLine}, and follows the structure of a similar document on $^{88}\text{Sr}$~\cite{Pucher_Sr_2025}.
Our goal is to provide experimentalists and theorists with consolidated data and calculated parameters that are important for understanding the interaction of light with ytterbium atoms.

We consider the relevant electronic level structure of \yb, in the context of laser cooling, optical trapping, coherent manipulation, and optical clock operation.
In particular, we summarize the following optical transitions:
\begin{itemize}
    \item The broad, dipole-allowed $\sS{0}\to\sP{1}$ transition at a wavelength $\lambda$ of about \SI{399}{\nano\meter}, which is used for first-stage laser cooling and Zeeman slowing (blue MOT transition), but also commonly used for atomic imaging schemes.
    \item The narrow spin-forbidden intercombination $\sS{0}\to\tP{1}$ transition at $\lambda\approx\SI{556}{\nano\meter}$, used for second-stage cooling to microkelvin temperatures (green MOT transition) and additionally for coherent control or imaging of the nuclear spin state.
    \item The ultra-narrow $\sS{0}\to\tP{0}$ transition at $\lambda\approx\SI{578}{\nano\meter}$, used in optical lattice clocks for precision timekeeping and as an optical qubit for quantum information applications (yellow clock transition).
    \item The ultra-narrow $\sS{0}\to\tP{2}$ transition at $\lambda\approx\SI{507}{\nano\meter}$, relevant for, e.g., quantum simulations (green clock transition).
    \item The additional dipole-allowed transitions at $\lambda\approx\SI{649}{\nano\meter}$, \SI{770}{\nano\meter}, and \SI{1389}{\nano\meter} to excite atoms into the \tS{1} or \tD{1} states, primarily used to manipulate and repump population of the long-lived metastable \tP{0} and \tP{2} states back into the ground state.
\end{itemize}
Notable omissions are the $\tP{1}\to\tD{1}$ transition used for state preparation into the long-lived \tP{0,2} states, the $\tP{2}\to\tD{3}$ cycling transition that is relevant to quantum-optics applications~\cite{Masson_Dicke_2024}, and the $\sS{0}\to4f^{13}\mathrm{5d6s}^2(J=2)$ transition to a core-excited state that is relevant for studying beyond-Standard Model physics~\cite{Safronova_Two_2018, Dzuba_Testing_2018, Ishiyama_Observation_2023}.
These might be included in future updates.

Measured values are provided with their corresponding references to original experimental sources, while calculated quantities are derived using established physical models, with citations to more detailed theoretical discussions.
The data provided here are not guaranteed to be complete or error-free.
For parameters critical to your research, you should consult the primary literature.

This document is intended as a community resource, following the example of Steck's \textit{Alkali D Line Data} series, available at \url{http://steck.us/alkalidata}.
We hope that this summary will serve experimental and theoretical researchers working with ytterbium, whether for atomic physics, quantum simulation, or metrology.
Comments, corrections, and suggestions for improving this document are welcomed and encouraged.

\section{Statistical Methods}\label{sec:statistics}

This \namecref{sec:statistics} outlines the statistical methods used to combine and report the numerical values derived from multiple experimental or theoretical sources.
If more than one value is cited for quantities such as the transition frequency, the lifetime of a state, the isotope shift, or the scattering length, the quoted value is a weighted mean of the values from the references, calculated as
\begin{equation}
    \mu = \frac{\sum_{j=1}^n x_j/\sigma_j^2}{\sum_{j=1}^n 1/\sigma_j^2}~,
\end{equation}
where $x_j$ are the measurement values with uncertainties $\sigma_j$, and $(j=1,\dots,n)$ indexes the individual references.
Here, the weights are the inverse squares of the uncertainties.
The corresponding standard uncertainty in the weighted mean is
\begin{equation}
    \sigma_\mu = \sqrt{ \frac{1}{\sum_{j=1}^n 1/\sigma_j^2} }~.
\end{equation}
If the measurements are not statistically consistent (i.e., the scatter exceeds what is expected from the reported uncertainties), we scale the uncertainty in the mean by the Birge ratio~\cite{Birge_Calculation_1932}, defined as
\begin{equation}
    \birge = \sqrt{ \frac{1}{n-1} \sum_{j=1}^n \frac{(x_j - \mu)^2}{\sigma_j^2} } = \sqrt{ \frac{\chi^2}{n-1} }~,
\end{equation}
where $\chi^2$ is the chi-squared value and $n$ is the number of data points.
The final uncertainty in the mean is then
\begin{equation}
    \sigma_\mu^{\mathrm{adj}} = \birge \cdot \sigma_\mu~.
\end{equation}
If $\birge > 1$ for one dataset, we show \birge~next to the reported weighted mean to indicate that the measured values are statistically inconsistent, and that we adjusted the uncertainty with the Birge ratio.
We note that alternative techniques for assessing the uncertainty of a weighted mean are actively being investigated and may be employed in future updates~\cite{Trassinelli_minimalistic_2024}.

When an experimental reference reports both a statistical uncertainty $\sigma_{\rm stat}$ and a systematic uncertainty $\sigma_{\rm sys}$ for a single measurement, the uncertainties are assumed to be uncorrelated.
Hence, we combine the two contributions in quadrature to give the total one‐sigma uncertainty
\begin{equation}
    \sigma = \sqrt{\sigma_{\rm stat}^2 + \sigma_{\rm sys}^2}~.
\end{equation}
In cases where only asymmetric limits are provided, e.g., ${x_0}^{+\Delta x_+\!}_{-\Delta x_-\!}$, we obtain a single symmetric uncertainty by taking the larger deviation
\begin{equation}
    \sigma = \max\bigl(\Delta x_+,\,\Delta x_-\bigr)~.
\end{equation}
Reporting $x_0(\sigma) = x_0 \pm \sigma$ then guarantees coverage of both published bounds.
Although this approach overestimates the error when the true distribution is asymmetric, it provides a straightforward, conservative uncertainty for tables and figures.

If an individual experimental reference reports multiple results for the same observable, and does not further combine these into a single value, we first perform this step ourselves using the framework described above, prior to comparing this average with other references.
We note that if all results are mutually consistent, the order of this operation is irrelevant and the same $\mu$ and $\sigma_\mu$ would be obtained if all results would be averaged together once.
If there are inconsistencies, this can lead to subtle differences in $\sigma_\mu$.

\section{Ytterbium Properties}

Ytterbium (Yb) is a soft, malleable, and ductile rare-earth metal with a bright silvery luster.
Although it oxidizes slowly in air, it should be stored in sealed containers to prevent degradation from moisture and oxygen.
The metal reacts slowly with water but is readily attacked by both dilute and concentrated mineral acids~\cite{Haynes_CRC_2016}.
Due to its reactivity, contaminated tools and surfaces can be cleaned with care using water or dilute acid solutions.
In powdered form (i.e., small grains or dust), ytterbium is significantly more reactive due to its increased surface area.
Ytterbium has 70 electrons, two of which occupy the outermost shell.
Natural ytterbium consists of seven stable isotopes, alongside more than 32 known unstable isotopes~\cite{Fry_Discovery_2013, Kondev_NUBASE2020_2021}.
This review focuses on the stable isotope \yb.

\Cref{tab:fundamental_constants} lists selected fundamental physical constants from the 2022 CODATA recommended values~\cite{Mohr_CODATA_2025} that we use in this reference.
In \cref{tab:physical_properties}, key physical properties of the isotope \yb are summarized.
This isotope has a nuclear spin of $1/2$, resulting in fermionic particle statistics.
The mass is the recommended value from the 2020 Atomic Mass Evaluation (AME2020)~\cite{Wang_AME_2021}, and the relative natural abundance is from the characterization in Ref.~\cite{Wang_absolute_2015} which was used to update the IUPAC standard atomic weight of Yb~\cite{Prohaska_Standard_2022}.
Abundance, mass, and nuclear spin of all stable Yb isotopes is provided in \cref{tab:isotopes}.

\begin{figure}[ht]
    \centering
    \includegraphics[width=0.72\textwidth]{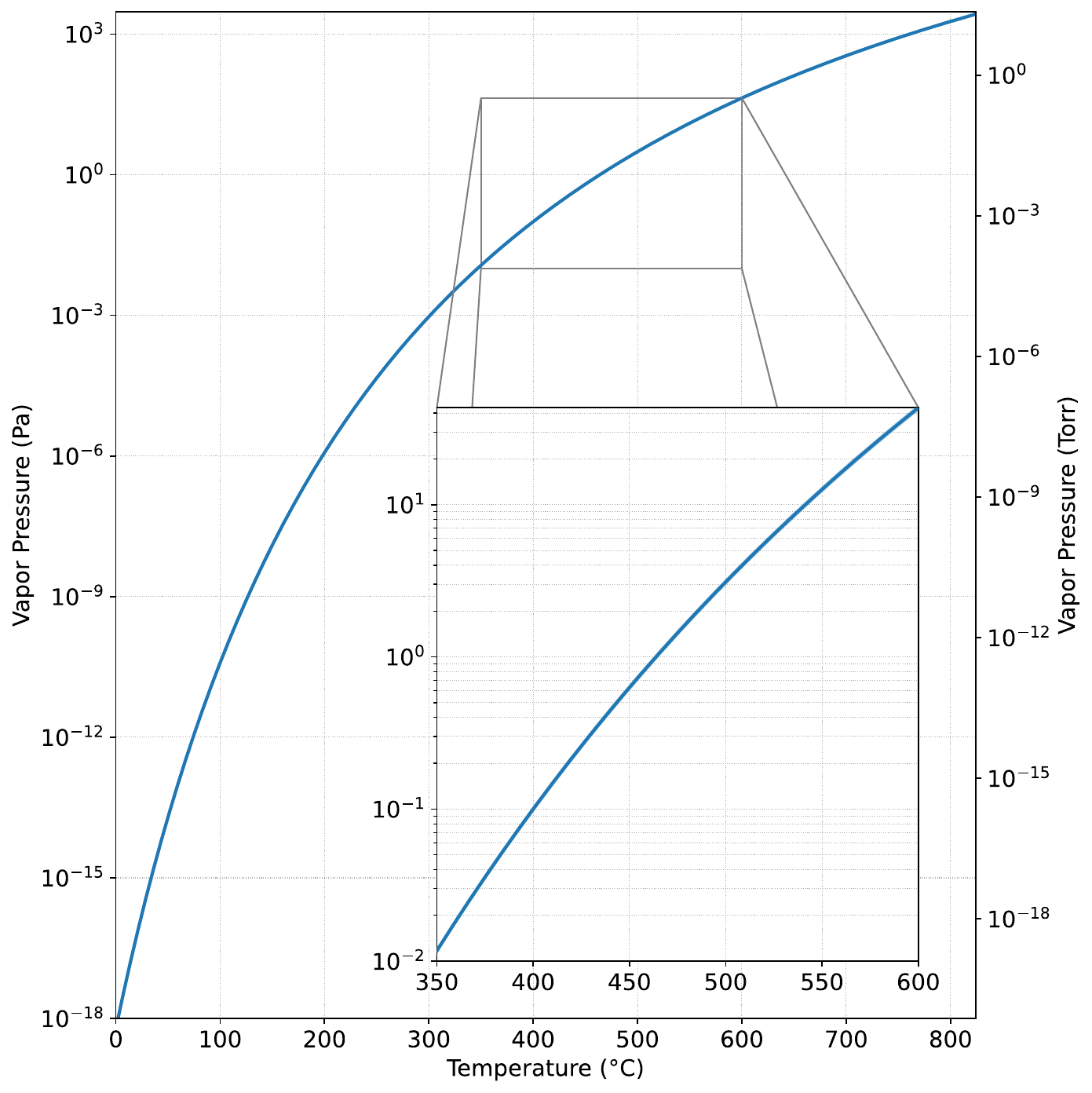}
    \caption{Vapor pressure of ytterbium as a function of temperature, calculated using \cref{eq:vapor_pressure}.}
    \label{fig:vapor_pressure}
\end{figure}

Thermodynamic properties such as density, molar volume, melting and boiling points as well as heat capacity are taken from Refs.~\cite{Haynes_CRC_2016, Arblaster_Selected_2018}, and are given for naturally occurring Yb rather than the specific isotope \yb.
The vapor pressure at a temperature of $T=\SI{25}{\celsius}$, as well as the temperature-dependent vapor pressure curve shown in \cref{fig:vapor_pressure}, are calculated using the semi-empirical equation describing solid Yb provided in Ref.~\cite{Alcock_Vapour_1984}:
\begin{equation}\label{eq:vapor_pressure}
    \log_{10}(P) = \num{9.111} + \log_{10}(\num{101325}) - \frac{\num{8111}}{T} - \num{1.0849} \log_{10}(T)~,
\end{equation}
where $P$ is the vapor pressure in pascals, and $T$ is the temperature in Kelvin.
This expression is valid from $T=\SI{298}{\kelvin}$ to $T=\SI{900}{\kelvin}$.
Over this range, it reproduces the experimental vapor pressure with an accuracy better than $\pm$\SI{5}{\percent}.
The first ionization energy (also called the ionization potential) of Yb has been measured through a variety of experimental techniques~\cite{Kaja_Characterization_2024, Lehec_Laser_2018, Chhetri_Investigation_2018, Aymar_Three-step_1984, Camus_Highly_1980, Camus_Spectre_1969, Hertel_Surface_1968, Zmbov_First_1966, Dresser_Surface_1965}.
All but one of these measurements have insufficient precision to resolve isotope effects: based on mass-scaling one expects a relative difference in ionization energy of, e.g., $m_e/m\mathopen{}\left(\yb\right) - m_e/m\mathopen{}\left(\ybb\right) \approx \num{5.5e-8}$~\cite{Ueda_Ionization_1969}.
The only high-precision measurement in Ref.~\cite{Lehec_Laser_2018}, performed for \ybb, has an uncertainty of about one order of magnitude smaller than this expected isotope effect.
We thus take the isotope dependence into account by converting to \yb prior to combining with the other measurements, resulting in the center-of-gravity ionization energy $E_I$ in \yb as listed in \cref{tab:physical_properties}.
Due to hyperfine interaction, two separate ionization energies can be found, as we further discuss in \cref{sec:energy_level_splittings}.

Low-energy elastic collisions between atoms are characterized by the $s$-wave scattering length $a$ and play a central role in ultracold gas experiments~\cite{Braaten_Universality_2006}.
For example, the magnitude of the scattering length governs the rate of thermalization during evaporative cooling, and the sign of $a$ determines stability and collective phenomena of degenerate gases~\cite{Chin_Feshbach_2010}.
In alkali atoms, $a$ can be magnetically tuned via Feshbach resonances to nearly arbitrary values $(-\infty, +\infty)$~\cite{Chin_Feshbach_2010}.
In contrast, alkaline-earth atoms such as Yb have a ground state with zero electronic angular momentum, making them largely insensitive to magnetic tuning.
For these systems, the ground-state scattering length is given by the binding energy of the highest vibrational state of the ground-state molecular potential.
(We note that Feshbach resonances, however, have been shown to exist in a cold-atom ensemble of mixtures of electronic states, as described below.)
Thanks to this simple dependence, experimentally accessible scattering lengths can be extrapolated to other isotope combinations using the Born-Oppenheimer approximation and a mass-scaling argument~\cite{Gribakin_Calculation_1993, Verhaar_Predicting_2009}.
Thus far, measurements have been limited to collisions between identical isotopes, as in Refs.~\cite{Borkowski_Beyond-Born-Oppenheimer_2017, Kitagawa_Two-color_2008, Fukuhara_Bose-Einstein_2007, Fukuhara_Degenerate_2007, Takasu_Spin-Singlet_2003, Enomoto_Determination_2007}, which are summarized on the diagonal of \Cref{tab:scattering_lengths}.
For collisions between different isotopes, corresponding to the off-diagonal elements of \cref{tab:scattering_lengths}, the results follow from mass-scaling as described in Refs.~\cite{Borkowski_Beyond-Born-Oppenheimer_2017, Kitagawa_Two-color_2008}.
This inter-isotope scattering is frequently used in cooling to degeneracy of \yb, which can be achieved via sympathetic evaporative cooling with \ybb or \ybs~\cite{Fukuhara_Quantum_2007, Fukuhara_All-optical_2009, Taie_Realization_2010}.
(Other isotopes have likewise been sympathetically cooled~\cite{Sugawa_ULTRACOLD_2013}.)
Both \ybb and \ybs themselves can be evaporatively cooled to degeneracy, the former straightforwardly~\cite{Takasu_Spin-Singlet_2003}, the latter thanks to its large nuclear-spin subspace enabling collisions~\cite{Fukuhara_Degenerate_2007}.

In addition to the electronic ground state, the metastable \tP{0} and \tP{2} states are sufficiently long-lived that it is meaningful to consider their scattering properties.
Elastic $s$-wave scattering has been quantified in, e.g., Refs.~\cite{Tomita_Dissipative_2019, Franchi_State-dependent_2017, Bouganne_Clock_2017}.
Interorbital scattering, that is, scattering between one atom in \sS{0} and another in \tP{0,2}, hosts rich physics and Feshbach resonances~\cite{Zhang_Controlling_2020, Bettermann_Clock-line_2023, Ono_Antiferromagnetic_2019, Tomita_Dissipative_2019, Takasu_Magnetoassociation_2017, Franchi_State-dependent_2017, Bouganne_Clock_2017, Scazza_Observation_2014, Kato_Control_2013, Yamaguchi_High-resolution_2010} and goes beyond the scope of this document.
Finally, we note that \textit{inelastic} scattering involving one or multiple atoms in the metastable states can induce significant loss, especially at high density.
Again, recent work has quantified some of these loss rates~\cite{Bettermann_Clock-line_2023, Tomita_Dissipative_2019, Takasu_Magnetoassociation_2017, Franchi_State-dependent_2017, Bouganne_Clock_2017, Scazza_Observation_2014, Khramov_Ultracold_2014, Uetake_Spin-dependent_2012, Ludlow_Cold-collision-shift_2011, Yamaguchi_Inelastic_2008} and we refrain from a detailed treatment.

\section{Level Structure}

\begin{figure}[ht]
  \centering
  \includegraphics[width=0.99\textwidth]{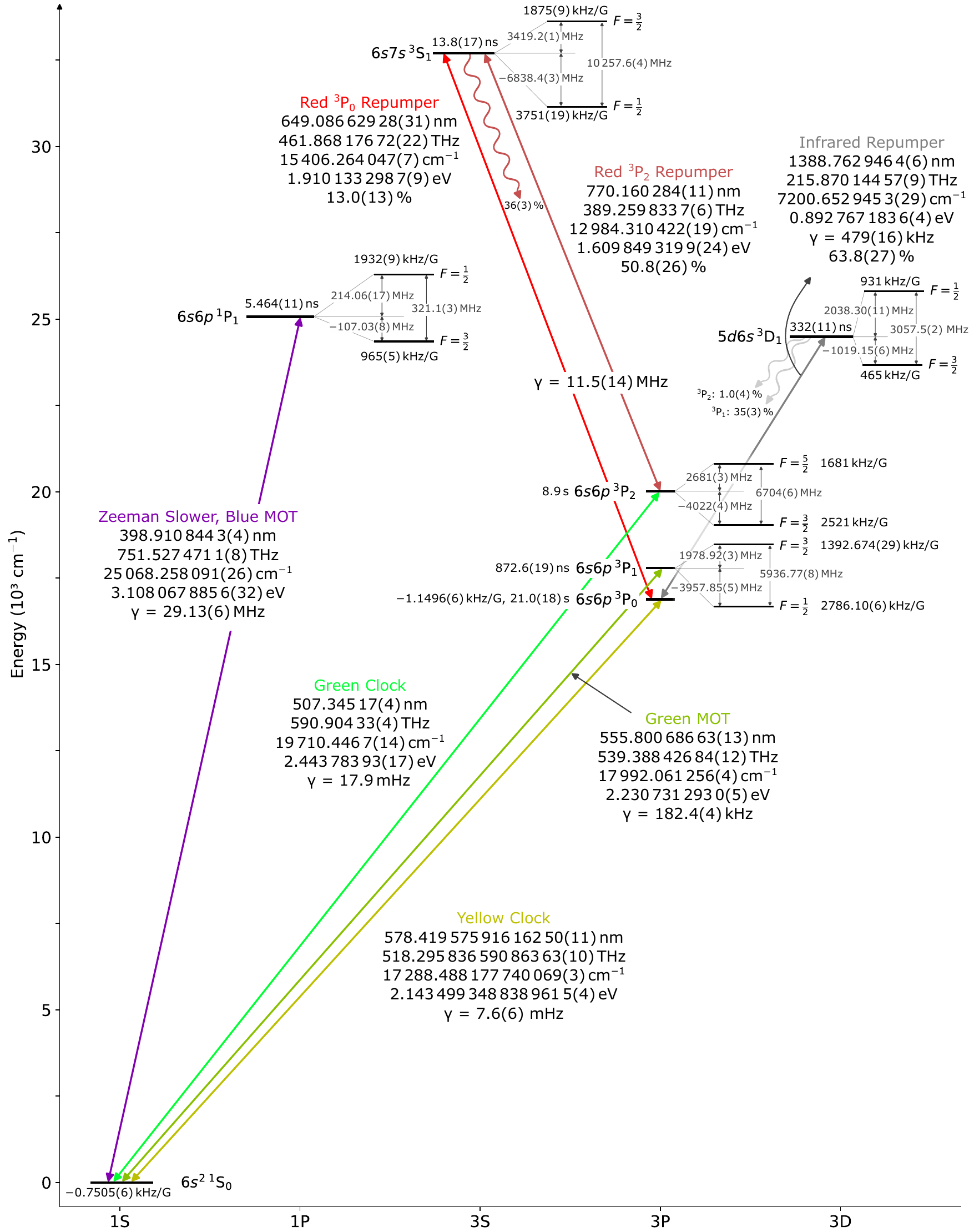}
  \caption{Data of the transitions discussed in this reference.
  \Crefrange{tab:blue_MOT}{tab:magnetic_clock_repumper} list the frequencies, wavelengths, energies, linewidths, and branching ratios of the transitions as well as the lifetime of the states with the corresponding references.
  The Zeeman splittings between adjacent magnetic sublevels can be found in \cref{tab:zeeman_and_hyperfine}.}
  \label{fig:Yb_I_energy_levels_reference}
\end{figure}

\begin{figure}[ht]
  \centering
  \includegraphics[width=0.93\textwidth]{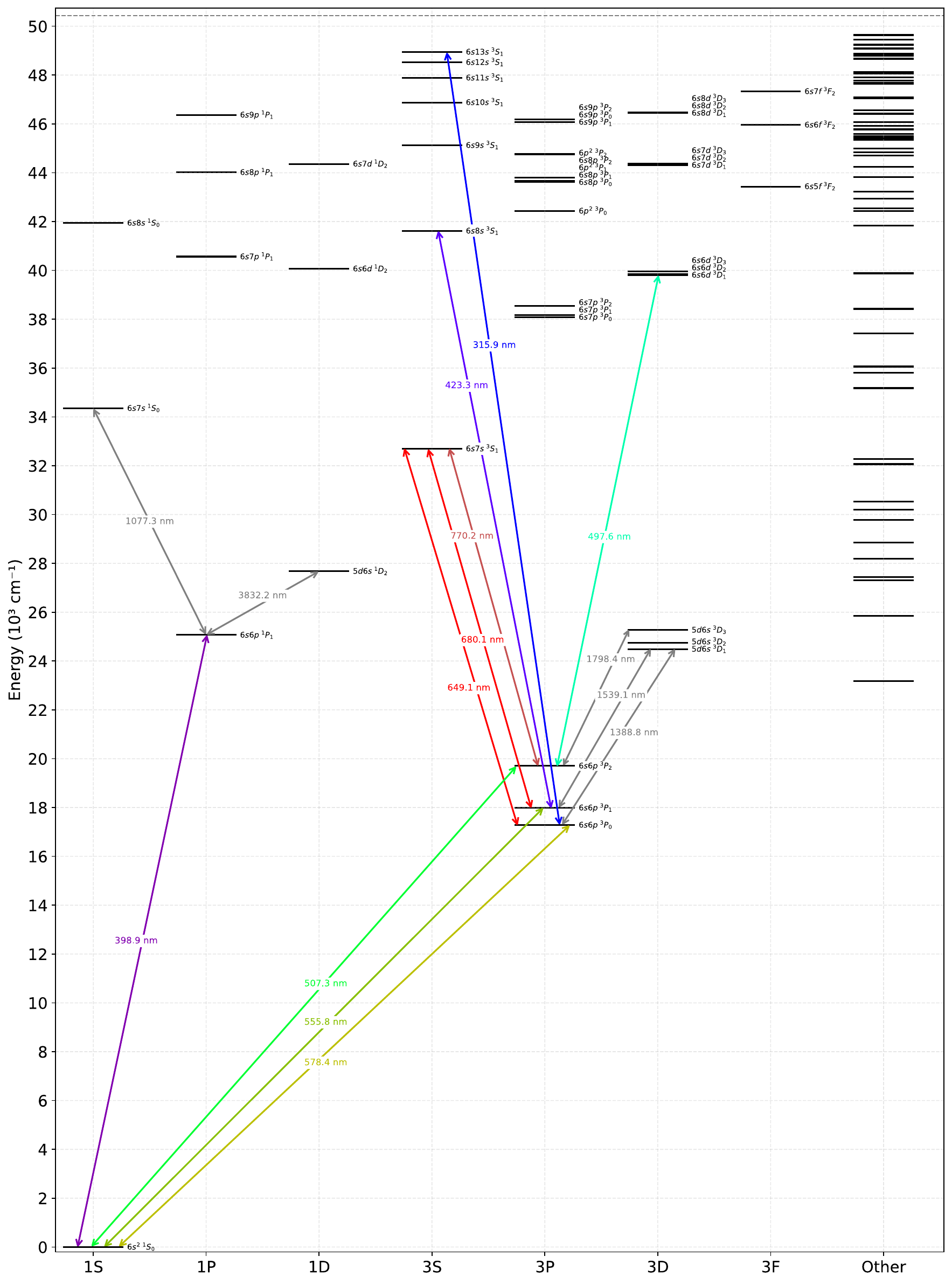}
  \caption{Neutral Yb\,I energy level diagram with selected optical transitions.
  Level energies and transition wavelengths are taken from the NIST Atomic Spectra Database~\cite{NIST_ASD}.
  The dashed gray line shows the ionization energy.
  The `Other' column summarizes all other levels that include, e.g., an inner-shell excited electron.}
  \label{fig:YbI-levels}
\end{figure}

Ytterbium's level structure is governed by its electronic configuration, which in its ground state is $\rm[Xe]4f^{14}6s^2$.
Its closed $\mathrm{f}$-shell and two valence electrons in the $\mathrm{6s}$ orbital result in an electronic structure that closely resembles that of alkaline-earth atoms.
For this reason, Yb is often referred to as an \textit{alkaline-earth-like} element.
In analogy to helium and alkaline-earth atoms such as Sr and Mg, the two-electron configuration gives rise to singlet and triplet manifolds, well described in the Russell-Saunders (L-S) coupling scheme~\cite{Russell_New_1925}.
\Cref{fig:Yb_I_energy_levels_reference} shows the energy levels and transitions discussed in this reference, while \cref{fig:YbI-levels} displays all Yb levels that can be found in the NIST Atomic Spectra Database~\cite{NIST_ASD}.
Levels without an assigned LS-coupling term are grouped in the rightmost column, labeled `Other'.
This includes, for example, states involving inner-shell excitations or states identified only by j–j coupling.
The electronic level structure features both broad electric-dipole-allowed transitions with linewidths on the order of megahertz and narrow intercombination lines between singlet and triplet manifolds, with natural linewidths spanning millihertz to kilohertz.

Atomic-level notation generally consists of two parts: the electronic configuration and the term symbol.
For the configuration, a sequence such as $n_1l_1\,n_2l_2\,\ldots$ indicates which orbitals are occupied by the electrons.
For example, $\mathrm{6s6p}$ means that one electron is in the $\mathrm{6s}$ orbital and one in the $\mathrm{6p}$ orbital.
In the term symbol ${}^{2S+1}L_J$, the superscript $2S+1$ is the spin multiplicity, where $S$ is the total (electronic) spin quantum number ($2S+1=1$ for singlet, $2S+1=3$ for triplet).
The letter $L$ encodes the total (electronic) orbital angular momentum (S for $L=0$, P for $L=1$, D for $L=2$, \textit{etc.}).
The subscript $J$ is the total (electronic) angular momentum.
The resulting label
\begin{equation}
    n_1l_1\,n_2l_2\,\ldots\,^{2S+1}L_J
\end{equation}
specifies both which orbitals the electrons occupy as well as the combined spin and orbital angular momentum quantum numbers of the entire atom (except its nucleus).
In practice, and as already was done in \cref{sec:intro}, when the electrons unambiguously occupy the lowest-possible energy configuration for a given term symbol, these get omitted from the notation.
As such, the electronic ground state $\mathrm{[Xe]4f^{14}6s^2}\:\sS{0}$ will be shortened to \sS{0} and, e.g., \tD{1} denotes $\mathrm{[Xe]4f^{14}5d6s}\:\tD{1}$.

\subsection{Energy Level Splittings}\label{sec:energy_level_splittings}

In Yb, each electronic term is split by spin-orbit (fine‐structure) interactions into levels of different total electronic angular momentum $\mathbf{J}$, which arises from coupling the orbital ($\mathbf{L}$) and spin ($\mathbf{S}$) angular momenta of the  electrons:
\begin{equation}\label{eq:jls}
    \mathbf{J} = \mathbf{L} + \mathbf{S}~.
\end{equation}
The magnitude of $\mathbf{J}$ is given by $\hbar\sqrt{J(J+1)}$, where the integer quantum number $J$ lies in the range
\begin{equation}
    \lvert L - S\rvert \le J \le L + S~.
\end{equation}
The ground state \sS{0} has $L=0$ and $S=0$, yielding $J=0$ and thus no fine-structure splitting.
The lowest electronic excited states, where one of the electrons is promoted from $\mathrm{6s}$ to $\mathrm{6p}$, has $L=1$ and $S=1$ so that $J=0,1,2$ resulting in the fine-structure levels \tP{0}, \tP{1}, and \tP{2}.

Beyond the electronic configuration, the nucleus can also contribute angular momentum.
The fermionic isotope \yb has a nuclear spin $I=1/2$, which couples with the electronic angular momentum to form the total atomic angular momentum
\begin{equation}\label{eq:fji}
    \mathbf{F} = \mathbf{J} + \mathbf{I}~,
\end{equation}
where the quantum number $F$ is restricted to
\begin{equation}
    \lvert J-I\rvert \le F \le J+I~,
\end{equation}
and again $F$ sets the magnitude of $\mathbf{F}$ via $\hbar\sqrt{F(F+1)}$.
States with $J=0$ have $F=I=1/2$.
Hence, states such as \sS{0} and \tP{0} exhibit no hyperfine splitting and therefore form an ideal nuclear-spin qubit~\cite{Jenkins_Ytterbium_2022, Ma_Universal_2022, Peper_Spectroscopy_2025}.
States with $J>0$, however, are separated into hyperfine multiplets.
To understand the shift of these levels, only the magnetic dipole interaction has to be considered, since $I=1/2$ causes higher order hyperfine interactions such as the electric quadrupole to vanish.
In that case, the hyperfine Hamiltonian is given by
\begin{equation}\label{eq:hyperfine_hamiltonian}
    H_\text{hfs} = A_\text{hfs}\mathbf{I}\cdot\mathbf{J}~,
\end{equation}
where $A_\text{hfs}$ is the magnetic dipole constant (which itself depends on $L$, $S$ and $J$).
The resulting energy shift for a hyperfine level is given by~\cite{Foot_Atomic_2004}
\begin{equation}\label{eq:E_hfs}
    \Delta E_\text{hfs} = \frac{1}{2}A_\text{hfs}K~,
\end{equation}
where
\begin{equation}
    K = F(F+1) - J(J+1) - I(I+1)~.
\end{equation}
It can be straightforwardly verified that the average of all hyperfine shifts, weighted by the number of sublevels $2F+1$ in each manifold, is zero (even when higher-order hyperfine interactions are included).

As stated, the magnetic dipole constant $A_\text{hfs}$ varies between electronic levels of \yb, and can be measured through spectroscopic investigations.
The results are summarized in \cref{tab:zeeman_and_hyperfine}.
For \sP{1}, no less than 11 papers have reported original values for the hyperfine coefficient.
These are combined according to \cref{sec:statistics}, but are dominated by the result from Ref.~\cite{Das_Absolute_2005}.
We note that the accuracy of that measurement was subsequently questioned, see for example the discussion in Ref.~\cite{Kleinert_Measurement_2016}.
If Ref.~\cite{Das_Absolute_2005} is excluded the remaining results nonetheless remain inconsistent, yielding $A_\text{hfs}(\sP{1}) / h = \SI{-212.2(4)}{\mega\hertz}$ (\birge: \num{1.8}).
The literature also yields inconsistent results for many of the other levels, though not with a clear single outlier.
Finally, we note that we were not able to verify Refs.~\cite{Chaiko_Isotope_1966, Topper_unpublished} in any original source material and instead rely on the values reported in Ref.~\cite{Kleinert_Measurement_2016} and Ref.~\cite{Bowers_Experimental_1999}, respectively.

The presence of hyperfine coupling leads to an ionization structure in \yb that is qualitatively different from that of \ybb, which has $I=0$.
In \yb, the Rydberg states exhibit strong hyperfine interactions, originating from the interplay between the hyperfine coupling in the ionic Yb$^+$ core and the exchange interaction with the Rydberg electron~\cite{Peper_Spectroscopy_2025}.
This situation is nearly the opposite of alkali-metal atoms: in alkali-metal atoms, the neutral ground state has strong hyperfine coupling while the Rydberg states are largely unaffected, whereas in \yb the ground state has no hyperfine interaction and the Rydberg states split strongly~\cite{Saffman_Quantum_2016, Peper_Spectroscopy_2025}.
As a result, the Rydberg series of \yb converge to two distinct ionization thresholds $E_{\rm I}^0$ and $E_{\rm I}^1$~\cite{Peper_Spectroscopy_2025}.
These thresholds can be associated with the hyperfine-splitting of the $^{171}$Yb$^+$ ground state $\rm 6s\,^2S_{1/2}$, which splits into $F_\text{core}=0$ and $F_\text{core}=1$ states separated by the ionic hyperfine constant $A_\text{hfs}(\rm 6s\,^2S_{1/2})$~\cite{Blatt_Precise_1983}.
By combining high-resolution spectroscopy with multichannel quantum defect theory, Ref.~\cite{Peper_Spectroscopy_2025} determined both thresholds, the values of which are listed in~\cref{tab:physical_properties}.

\subsection{Interaction with Magnetic Fields}

External magnetic fields will further affect the energy levels beyond the fine-structure and hyperfine-structure interactions considered above.
The Hamiltonian describing the interaction of atomic electrons and nuclear spins with a static magnetic field $\mathbf{B}$, which defines a quantization axis taken here to be the $z$-axis, is given by
\begin{equation}\label{eq:hamiltonian_magnetic_field}
    H_B = \frac{\muB}{\hbar}\left(g_S\mathbf{S} + g_L\mathbf{L} + g_I\mathbf{I}\right)\cdot\mathbf{B} = \frac{\muB}{\hbar}\left(g_SS_z + g_LL_z + g_II_z\right)B_z~,
\end{equation}
where $\muB$ is the Bohr magneton, $g_S$, $g_L$, and $g_I$ are the electron spin, orbital and nuclear $g$-factors respectively, and $S_z$, $L_z$, and $I_z$ denote angular momentum components along the $z$-axis.
We note explicitly that the nuclear term here is written with the prefactor $\muB$; in an alternative convention, commonly found in nuclear physics, the prefactor is $\muN$, resulting in a value of $g_I$ larger by a factor $\muB/\muN$, where $\muN$ is the nuclear magneton.

The value for $g_S$ has been measured precisely, and we list the CODATA recommended value in \cref{tab:physical_properties}.
The orbital $g$-factor is corrected for finite nuclear mass by~\cite{Bethe_Quantum_1957}
\begin{equation}
    g_L = 1 - \frac{m_{\rm e}}{m_\text{nuc}}~,
\end{equation}
where $m_{\rm e}$ is the electron mass and $m_\text{nuc}$ is the nuclear mass, accurate to leading order in $m_{\rm e}/m_\text{nuc}$.
The nuclear $g$-factor $g_I$ is experimentally determined from measurements of the nuclear magnetic moment $\mu$.
Given the nuclear spin $I$, $g_I$ can be calculated using\footnote{The minus sign here is due to our sign convention in the magnetic Hamiltonian, where we have the term $+g_I \mathbf{I}\cdot\mathbf{B}$, and is consistent with Refs.~\cite{Steck2024SodiumDLine, Steck2024Rb85DLine, Steck2024Rb87DLine, Steck2024CesiumDLine} versus Ref.~\cite{Stone_Table_2019} where $\mu$ and $g_I$ have opposite sign.}
\begin{equation}
    g_I = -\frac{\mu}{\muB I}~.
\end{equation}
For \yb, $\mu$ has been measured in Refs.~\cite{Olschewski_Messung_1972, Olschewski_Bestimmung_1967, Gossard_Ytterbium_1964} and subsequently tabulated in, e.g., Refs.~\cite{Fuller_Nuclear_1976, Raghavan_Table_1989, Stone_Table_2005}.
We report the value stated in the most recent tabulation of Ref.~\cite{Stone_Table_2019}, which corrects the earlier results for updates to fundamental constants, various nuclear magnetic effects, and aims to establish a `best' value, see the discussion therein.
It states
\begin{equation}
    \mu = \num{0.4923(4)}\muN~,
\end{equation}
so that $g_I = \num{-0.0005362(4)}$.

\begin{figure}[ht]
  \centering
  \includegraphics[width=1\textwidth]{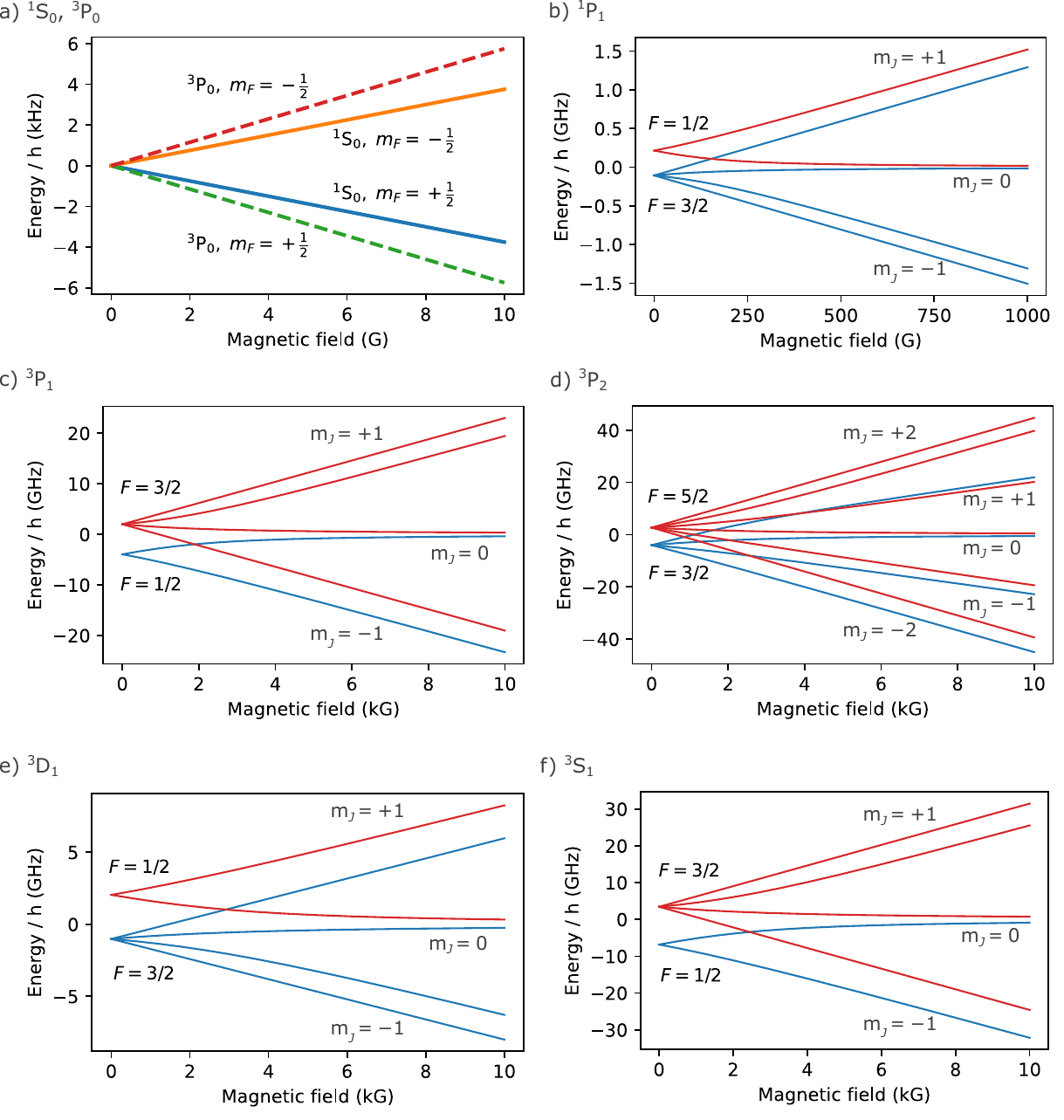}
  \caption{Eigenvalues of the combined hyperfine and Zeeman Hamiltonian, $H = H_{\mathrm{hfs}} + H_{B}$, as a function of magnetic field strength.
  In the low‐field (Zeeman) regime, each level is labeled with their respective quantum number $F$; in the high‐field (hyperfine Paschen–Back) regime, levels are labeled with the electronic magnetic quantum number $m_J$.
  For the \sS{0} and \tP{0} states, we use the experimental values from \cref{tab:zeeman_and_hyperfine}.}
  \label{fig:zeeman_shifts}
\end{figure}

In general, calculating the energy levels involves diagonalizing the combined hyperfine and magnetic Hamiltonians, $H_\text{hfs} + H_B$.
The resulting energy shifts of the hyperfine structure in an external magnetic field are shown in Fig.~\ref{fig:zeeman_shifts}.
We calculate shifts for all hyperfine ($F$) levels within a given fine-structure state, neglecting any coupling to other fine-structure states.
As a result, the calculations are valid only in the regime where the Zeeman interaction is small compared to the fine-structure splitting.
At higher magnetic fields, particularly for the \tP{0} clock state, fine-structure mixing becomes non-negligible, and a more comprehensive treatment is required to accurately capture the Zeeman effect (see, e.g., Ref.~\cite{Boyd_Nuclear_2007}).
We now turn to a discussion of low- and high-magnetic-field regimes where approximations can be found.

\subsubsection{Weak Field Limit -- Zeeman Effect}
In the limit of weak magnetic fields, where energy shifts are small compared to hyperfine splittings, $F$ is a good quantum number.
The magnetic Hamiltonian can then be written as
\begin{equation}
    H_B = \frac{\muB}{\hbar}g_FF_zB_z~,
\end{equation}
where the hyperfine Land\'e factor $g_F$ is given by
\begin{equation}\label{eq:Lande_gF}
    g_F = g_J \frac{F(F+1) - I(I+1) + J(J+1)}{2F(F+1)} + g_I \frac{F(F+1) + I(I+1) - J(J+1)}{2F(F+1)}~,
\end{equation}
and the electronic Land\'e factor is
\begin{equation}
    g_J = g_L\frac{J(J+1) - S(S+1) + L(L+1)}{2J(J+1)} + g_S\frac{J(J+1) + S(S+1) - L(L+1)}{2J(J+1)}~.
\end{equation}
The operator $F_z$ acts as
\begin{equation}
    F_z \ket{F, m_F} = \hbar m_F \ket{F, m_F}~,
\end{equation}
where $m_F$ is the magnetic quantum number for the projection of $\mathbf{F}$ on the quantization axis ($\lvert m_F\rvert\leq F$ so that there are $2F+1$ different $m_F$-levels).
The degeneracy of each hyperfine manifold is thus lifted via the first-order Zeeman shift,
\begin{equation}
    \Delta E_{F, m_F} = \hbar \Delta\omega_{F, m_F} = \muB g_F m_F B_z~.
\end{equation}

For states with $J\neq0$, the value of $g_F$ is dominated by $g_J$.
We note however, that the above framework arises from an effectively first-order perturbation theory treatment.
Further corrections to $g_J$ arise from the multi-electron structure, and additional contributions stem from, e.g., QED effects~\cite{Judd_Theory_1961, Bethe_Quantum_1957, Labzowsky_Estimates_1999}.
As such, in \cref{tab:zeeman_and_hyperfine} we summarize the $g_J$ coefficient as well as the Zeeman coefficient for each level considered in this document.
Where possible, these follow from an experimental characterization of $g_J$, with indicated references.
If no such data exists, we revert back to the above expressions, and shall indicate this with an asterisk.
For the optical clock states \sS{0} and \tP{0}, there is no $g_J$ and \cref{eq:Lande_gF} yields $g_F = g_I$ for both states.
However, these states have an easily resolvable differential Zeeman shift~\cite{Bettermann_Clock-line_2023, Zhang_Precise_2023, Ono_Antiferromagnetic_2019, McGrew_Atomic_2018}.
This arises due to the hyperfine interaction that induces state mixing (see also \cref{sec:mixing})~\cite{Boyd_Nuclear_2007}, and yields a nuclear Land\'e factor for \tP{0} that is given by $g_{I}\mathopen{}\left(\tP{0}\right) = g_I\mathopen{}\left(\sS{0}\right) + g_\text{hfs}$.
Combining the results from the above references, we find $g_\text{hfs} = \num{-0.0001425496(14)}$; the reported Zeeman coefficient in \cref{tab:zeeman_and_hyperfine} takes this effect into account.
For \tD{1}, measured Zeeman shifts are reported in Refs.~\cite{Ai_Absolute_2023, Ai_Erratum_2024}, however we note that the measurements therein are far away from the theoretical values without any given explanation.
For this level, we thus choose to report the theoretical values in \cref{tab:zeeman_and_hyperfine}.

Of further importance to the \sS{0} and \tP{0} states is the higher-order, quadratic Zeeman shift, arising due to the fine-structure interaction inducing state mixing (see again \cref{sec:mixing})~\cite{Boyd_Nuclear_2007}.
The total magnetic field shift to the clock transition is thus given by
\begin{equation}
    \Delta\omega_B(\mathbf{B}) = \muB g_I (m_{F'} - m_F)B + \muB g_\text{hfs} m_{F'}B + \eta B^2~,
\end{equation}
where $B$ is the magnitude of the magnetic field, and $\eta$ is the quadratic Zeeman shift coefficient.
Optical lattice clocks have characterized this effect in detail.
Since a negligible difference is expected between isotopes~\cite{Boyd_Nuclear_2007}, we include measurements from across all isotopes~\cite{Bettermann_Clock-line_2023, Zhang_Precise_2023, Ono_Observation_2022, Luo_Absolute_2020, Riegger_Interorbital_2019, Gao_Systematic_2018, McGrew_Atomic_2018, Lemke_Spin-1/2_2009, Poli_Frequency_2008, Barber_Direct_2006}.
This results in $\eta=\SI{-0.06084(6)}{\hertz\per\gauss\squared}$ ($\birge$: \num{1.2}).
The value reported in Ref.~\cite{McGrew_Atomic_2018}, which most strongly influences this average, has recently been corrected in Ref.~\cite{Bothwell_Lattice_2025}.
With it, the average changes to $\eta=\SI{-0.06021(12)}{\hertz\per\gauss\squared}$ ($\birge$: \num{2.1}).
Similar corrections arising from updates to, e.g., nuclear magnetic moments and diamagnetic correction factors might apply to the other references as well, however this is beyond the scope of this document.

\subsubsection{Strong Field Limit -- Paschen-Back Regime}
For strong magnetic fields, the strength of $H_B$ can exceed the hyperfine interaction.
In this Paschen-Back regime, $F$ is no longer the appropriate quantum number with which to index the sublevels.
(We note that this regime does not exist for \sS{0} and \tP{0} as they lack hyperfine structure, or in other words $L=S=0$ so that $F=I$ is always a good quantum number.)
Provided the magnetic interaction is not so strong that it also overwhelms the fine-structure coupling, $J$ and $I$ along with their projections $m_J$ and $m_I$ determine the eigenstates of the atom.
This results in energy shifts given by
\begin{equation}
    \Delta E_\text{Paschen-Back} = A_\text{hfs}m_Im_J + \muB (g_Jm_J + g_Im_I) B.
\end{equation}
The relevant quantities for this expression have been discussed above and can be found in \cref{tab:zeeman_and_hyperfine}.

\subsection{State Mixing}\label{sec:mixing}

In fermionic \yb, state mixing in the $\mathrm{6s6p}$ manifold arises from three key mechanisms: spin-orbit coupling, hyperfine coupling, and Zeeman interactions.
Each contributes to the admixture of nominally pure states $^{2S+1}L_J^{(0)}$ leading to the physical states $^{2S+1}L_J$.
This mixing enables otherwise forbidden transitions.
Here, we describe the state mixing following the methods in Refs.~\cite{Breit_Hyperfine_1933, Lurio_Second-Order_1962, Riegger_Interorbital_2019}.

\subsubsection{Spin–Orbit Coupling}
Spin–orbit coupling is the fundamental source of singlet–triplet mixing in Yb, present even in the absence of nuclear spin or external fields.
In the $\mathrm{6s6p}$ manifold, the $J=1$ subspace splits into \tP{1} and \sP{1} states, which are related to the pure spin–orbit basis by the unitary transform
\begin{equation}\label{eq:spin_orbit_mixing_equation}
    \begin{pmatrix}
        \ket{\sPso{1}} \\[6pt]
        \ket{\tPso{0}} \\[6pt]
        \ket{\tPso{1}} \\[6pt]
        \ket{\tPso{2}}
    \end{pmatrix}
    =
    \begin{pmatrix}
        \alpha & 0 & -\beta & 0 \\[4pt]
        0 & 1 & 0 & 0 \\[4pt]
        \beta & 0 & \alpha & 0 \\[4pt]
        0 & 0 & 0 & 1
    \end{pmatrix}
    \,
    \begin{pmatrix}
        \ket{\sPb{1}} \\[6pt]
        \ket{\tPb{0}} \\[6pt]
        \ket{\tPb{1}} \\[6pt]
        \ket{\tPb{2}}
    \end{pmatrix}~,
\end{equation}
with $\alpha^2+\beta^2=1$.
The singlet–triplet mixing in the $J=1$ manifold can be extracted directly from measured lifetimes and transition frequencies by assuming that the bare \tPb{1} state has zero E1 decay rate and that the entire observed decay rate of \tP{1} arises from spin-orbit coupling with the \sPb{1} state.
One can then find the relation~\cite{Lurio_Second-Order_1962, Boyd_Nuclear_2007}
\begin{equation}
    \frac{\beta^{2}}{\alpha^{2}} \;=\; \frac{\Gamma_{3\text{P}1}\;\nu^{3}\bigl(\sS{0} - \sP{1} \bigr)}{\Gamma_{1\text{P}1}\;\nu^{3}\bigl(\sS{0} - \tP{1} \bigr)}~,
\end{equation}
where $\Gamma_{i}$ are the decay rates of the states and $\nu_i$ the transition frequencies between the ground and the excited states.
Using the values of \cref{tab:blue_MOT,tab:green_MOT}, we find
\[
    \beta = \num{-0.12905(19)}\,, 
    \quad
    \text{and}
    \quad
    \alpha = \sqrt{1-\beta^{2}} \;=\; \num{0.991638(25)}\,,
\]
corresponding to a singlet admixture of approximately $\SI{1.7}{\percent}$ in the \tP{1} state.
Note that the NIST database for Yb lists leading percentages for various states, which indicate the dominant configuration components of each eigenstate and thus quantify the degree of state mixing~\cite{NIST_ASD}.
The negative sign of $\beta$ can be inferred from the derivation of the intermediate coupling coefficients in~\cite{Breit_Hyperfine_1933}.
As a result, the otherwise forbidden intercombination line acquires a finite electric‐dipole matrix element
\begin{equation}
   \braket{\sS{0} | \hat{\mathbf{d}} | \tPso{1}} = \beta\,\braket{\sS{0} | \hat{\mathbf{d}} | \sPb{1}}\neq0~.
\end{equation}

\subsubsection{Hyperfine Coupling}
The nonzero nuclear spin $I=1/2$ of \yb gives rise to the hyperfine Hamiltonian in \cref{eq:hyperfine_hamiltonian}.
This leads to a hyperfine coupling between states with different $J$ but the same total $F$ and $m_F$.
For example, to first order, the clock state is given by
\begin{equation}
    \ket{\tP{0}} = \ket{\tPso{0}} + \alpha_0\,\ket{\tPso{1}} + \beta_0\,\ket{\sPso{1}}~.
\end{equation}
Together with \cref{eq:spin_orbit_mixing_equation}, we can relate the physical states to the pure states
\begin{equation}
\renewcommand{\arraystretch}{1.4}
\begin{aligned}[t]
    \begin{pmatrix}
        \ket{\sP{1},\,F=\tfrac12}  \\
        \ket{\sP{1},\,F=\tfrac32}  \\
        \ket{\tP{0},\,F=\tfrac12}  \\
        \ket{\tP{1},\,F=\tfrac12}  \\
        \ket{\tP{1},\,F=\tfrac32}  \\
        \ket{\tP{2},\,F=\tfrac32}  \\
        \ket{\tP{2},\,F=\tfrac52}
    \end{pmatrix}
    &=
    \begin{pmatrix}
        \alpha & 0 & -\beta_0 & -\beta & 0 & 0 & 0  \\
        0 & \alpha & 0 & 0 & -\beta & -\beta_2 & 0  \\
        \beta_0\alpha+\alpha_0\beta & 0 & 1 & -\,\beta_0\beta+\alpha_0\alpha & 0 & 0 & 0  \\
        \beta & 0 & -\alpha_0 & \alpha & 0 & 0 & 0  \\
        0 & \beta & 0 & 0 & \alpha & -\alpha_2 & 0  \\
        0 & \beta_2\alpha+\alpha_2\beta & 0 & 0 & -\,\beta_2\beta+\alpha_2\alpha & 1 & 0  \\
        0 & 0 & 0 & 0 & 0 & 0 & 1
    \end{pmatrix}
    \,
    \begin{pmatrix}
        \ket{\sPb{1},\,F=\tfrac12}  \\
        \ket{\sPb{1},\,F=\tfrac32}  \\
        \ket{\tPb{0},\,F=\tfrac12}  \\
        \ket{\tPb{1},\,F=\tfrac12}  \\
        \ket{\tPb{1},\,F=\tfrac32}  \\
        \ket{\tPb{2},\,F=\tfrac32}  \\
        \ket{\tPb{2},\,F=\tfrac52}
    \end{pmatrix}~.
\end{aligned}
\end{equation}
Here, the hyperfine coupling constants can be inferred from experimentally measured hyperfine splittings in the singlet and triplet manifold using the Breit-Wills theory, see e.g., Ref.~\cite{Boyd_Nuclear_2007}.
The dominant dipole‐mediated term $\alpha_0$ is almost two orders of magnitude larger than the singlet term $\beta_0$~\cite{Riegger_Interorbital_2019}, and it produces an intrinsic electric‐dipole matrix element for the clock transition which corresponds to a zero‐field linewidth of $\sim\,$\SI{8}{\milli\hertz}.
This hyperfine‐induced E1 coupling is therefore sufficient for direct clock spectroscopy without any external mixing magnetic field, unlike in bosonic isotopes.

\subsubsection{Zeeman Coupling}
An external magnetic field $\mathbf{B}$ couples states via the Hamiltonian $H_B$ in \cref{eq:hamiltonian_magnetic_field}.
In the mixed basis, $H_B$ produces both Zeeman shifts and further state admixture.
A static magnetic field $B$ can be applied to mix the $\ket{\tP{0}}$ state with the $\ket{\tP{1}}$ state via the magnetic‐dipole operator $\hat\mu$~\cite{Taichenachev_Magnetic_2006}
\begin{equation}\label{eq:Bmix}
    \ket{\tPp{0}} = \ket{\tP{0}} + \frac{\braket{\tP{1} | \hat\mu\!\cdot\!B | \tP{0}}}{\Delta_{10}}\,\ket{\tP{1}}~,
\end{equation}
where $\Delta_{10}\equiv\omega_{3\mathrm{P}1} - \omega_{3\mathrm{P}0} \approx2\pi \times \SI{21.1}{\tera\hertz}$, see values in \cref{tab:clock,tab:green_MOT}.
Since $\ket{\tP{1}}$ itself carries the small $\beta$‐admixture of $\ket{\sPb{1}}$, the doubly‐mixed clock state becomes
\begin{equation}\label{eq:Bmix_clock_full}
    \ket{\tPp{0}} = \ket{\tP{0}} + \frac{\braket{\tP{1} | \hat\mu\!\cdot\!B | \tP{0}}}{\Delta_{10}}\bigl(\alpha\,\ket{\tPb{1}} + \;\beta\,\ket{\sPb{1}} - \,\alpha_0\,\ket{\tPb{0}}\bigr)~.
\end{equation}
This second‐order effect produces the quadratic Zeeman shift of the clock frequency, and a residual field‐dependent linewidth
\begin{equation}
    \Gamma_{\rm clock}(B) = \Gamma_{3\mathrm{P}0} \;+\; \Gamma_{3\mathrm{P}1}\,\frac{\bigl(\mu_C\,B\bigr)^{2}}{2\,\Delta_{10}^{2}}~,
    \qquad
    \mu_C = \sqrt{\tfrac{2}{3}}\;(g_L - g_S)\,\mu_B~,
\end{equation}
where the zero–field decay rate $\Gamma_{3\mathrm{P}0}\approx2\pi\times\SI{8}{\milli\hertz}$ arises from hyperfine‐induced mixing, and the field‐dependent term remains at the \unit{\milli\hertz} level even for $B\sim\SI{1}{\kilo\gauss}$.

\section{Transition Properties}\label{sec:transitions}

This \namecref{sec:transitions} examines the key atomic transitions.
Measured transition frequencies are presented and converted to wavelengths and transition energies.
Experimentally determined lifetimes of the excited states are reviewed.
Equations for obtaining additional parameters from these transition frequencies and lifetimes are introduced.

We note that we only consider direct measurements of the quantity of interest, with the caveat that this approach can lead to inconsistencies when combining different results.
For example, the energy difference between \tS{1} and \sS{0} can be obtained by using the absolute transition frequency measurements on the pathways \sS{0}~$\to$~\tP{0}~$\to$~\tS{1} versus \sS{0}~$\to$~\tP{2}~$\to$~\tS{1}, yielding a difference of $2\pi\times$\SI{-0.15(4)}{\giga\hertz}.
Alternatively, some experiments, such as in Ref.~\cite{Nenadovic_Clock_2016}, directly measure energy-separation of two states without absolute accuracy of either transition, though these then are omitted from inclusion in any of the \namecrefs{tab:blue_MOT}.
A few exceptions to this general approach are noted below.

\subsection{Transition Frequencies and Wavelengths}

The following relations allow one to convert a transition frequency $\omega_0 = 2 \pi \times \nu$ to the corresponding transition energy, vacuum wavelength, air wavelength, and wavenumber.
The photon energy $E$ is
\begin{equation}
    E = h\nu = \hbar\omega_0~.
\end{equation}
This transition energy can be converted into an energy in eV using
\begin{equation}
    E_{\mathrm{eV}} = \frac{h\nu}{e} = \frac{\hbar\omega_0}{e}~.
\end{equation}
The vacuum wavelength $\lambda$ and the angular wavenumber $k_\mathrm{L}$ are
\begin{align}
    \lambda &= \frac{c}{\nu} = \frac{2\pi c}{\omega_0}~,\\
    k_\mathrm{L} &= \frac{2\pi}{\lambda} = \frac{\omega_0}{c}~.
\end{align}
The wavelength in air is given by
\begin{equation}
    \lambda_{\rm air} = \frac{\lambda}{n_{\rm air}}~,
\end{equation}
where $n_{\rm air}$ is the refractive index of air.
To compute the refractive index of air under typical laboratory conditions, the Edl\'en’s equation is a common approximation~\cite{Edlen_Refractive_1966, Peck_Dispersion_1972}
\begin{equation}\label{eq:n_air}\begin{split}
    n_{\rm air} = 1 + \Biggl[\biggl(\num{8342.54} + \frac{\num{2406147}}{\num{130} - \kappa^2} + \frac{\num{15998}}{\num{38.9} - \kappa^2}\biggr) \frac{P}{\num{96095.43}} &\frac{1 + 10^{-8}(\num{0.601} - \num{0.00972}~T)P}{1 + \num{0.0036610}~T} \\
    &\hspace{3em}- f(\num{0.037345} - \num{0.000401}\kappa^2) \Biggr]\times10^{-8}~,
\end{split}\end{equation}
where $\kappa = k_L/(2\pi)$ is the vacuum wavenumber in~\unit{\per\micro\meter}, $P$ is the total air pressure in~\unit{\pascal}, $T$ is the temperature in~\unit{\celsius}, and $f$ is the partial pressure of water vapor in~\unit{\pascal} (obtainable from relative humidity via the Buck equation~\cite{Buck_New_1981, BuckResearch2012}).
Under standard conditions ($P=\SI{101325}{\pascal}$, $T=\SI{25}{\celsius}$, \SI{50}{\percent} relative humidity), this yields $n_{\rm air}\approx \num{1.00027}$ in the visible range with a $1\sigma$ uncertainty of \num{1e-8} which has been used throughout this reference.
In real‐world applications, additional uncertainties due to fluctuations in pressure, temperature, and humidity must be considered and are not included here, see for example Refs.~\cite{Steck2024SodiumDLine, Steck2024Rb85DLine, Steck2024Rb87DLine, Steck2024CesiumDLine}.

In precision spectroscopy one often measures transition frequencies 
$\nu_{F_g\to F_e}$ between specific hyperfine levels $F_g$ and $F_e$ of the ground and excited state, respectively.
Each of these measured transitions is (potentially) shifted due to the hyperfine shift, \cref{eq:E_hfs}.
We instead report the absolute transition frequencies between the `bare' levels, corresponding to the center-of-gravity or centroid of the hyperfine levels.
As described after \cref{eq:E_hfs}, this center-of-gravity energy level can be determined by a weighted sum over the hyperfine manifolds.
Likewise, the center-of-gravity transition frequency can be determined via a weighted sum of the individual transition frequencies $\nu_{F_g\to F_e}$.
However, to correctly perform this weighted average, the weights $W_{F_g\to F_e}$ additionally need to take into account the selection rules that forbid certain transitions.
The correct weights to achieve this are given by~\cite{Enomoto_Comparison_2016}
\begin{equation}
    W_{F_g\to F_e} = (2F_g+1)(2F_e+1)
    \begin{Bmatrix}
    J_e & F_e & I\\
    F_g & J_g & 1
    \end{Bmatrix}^2,
\end{equation}
where the term in the curly brackets is the Wigner $6$-$j$ symbol and depends also on the total electronic angular moment $J_{g,e}$ of both states.
The transition frequency is then given by
\begin{equation}\label{eq:nu_cg}
    \nu_\text{cg} = \left.\sum\limits_{F_g, F_e} W_{F_g\to F_e}\nu_{F_g\to F_e} \middle/ \sum\limits_{F_g, F_e} W_{F_g\to F_e}\right..
\end{equation}
If a reference does not explicitly state $\nu_\text{cg}$ but reports all required transition frequencies $\nu_{F_g\to F_e}$, we consider all corresponding uncertainties as independent and evaluate $\nu_\text{cg}$ using the above expression.
The propagated uncertainty of the center‐of‐gravity frequency is then given by
\begin{equation}\label{eq:sigma_cg}
    \sigma_\text{cg}
    = \left.\sqrt{ \sum_{F_g,F_e}W_{F_g\to F_e}^2\sigma_{F_g\to F_e}^2 } \middle/ \sum_{F_g,F_e}W_{F_g\to F_e}\right..
\end{equation}

The following comments apply for the absolute frequencies listed in \crefrange{tab:blue_MOT}{tab:magnetic_clock_repumper}:
\begin{itemize}
    \item For the \sS{0} $\to$ \sP{1} transition, Refs.~\cite{Laupretre_Absolute_2020, Tanabe_Frequency-stabilized_2018, Kleinert_Measurement_2016, Nizamani_Doppler-free_2010, Das_Absolute_2005} report absolute frequencies, giving an average of $2\pi\times\SI{751.52706(12)}{\tera\hertz}$ (\birge: \num{1895}).
    The extremely large \birge~is primarily caused by Ref.~\cite{Das_Absolute_2005}, which has by far the smallest uncertainty but reports an absolute frequency different from the others by many times the natural linewidth of $\sim$ \SI{30}{\mega\hertz}, suggesting some systematic discrepancies.
    As discussed in the context of hyperfine splitting, the accuracy of the measurements reported in Ref.~\cite{Das_Absolute_2005} has been questioned.
    For these reasons, in \cref{tab:blue_MOT} we exclude Ref.~\cite{Das_Absolute_2005} for figuring the absolute transition frequency.
    \item The \sS{0} $\to$ \tP{0} clock transition frequency has been subject of a large number of experimental studies~\cite{Kobayashi_Improved_2025, Goti_Absolute_2023, Kobayashi_Search_2022, Clivati_Coherent_2022, Kim_Absolute_2021, Luo_Absolute_2020, Pizzocaro_Absolute_2020, Kobayashi_Demonstration_2020, McGrew_Towards_2019, Kim_Improved_2017, Pizzocaro_Absolute_2017, Park_Absolute_2013, Yasuda_Improved_2012, Lemke_Spin-1/2_2009, Kohno_One-Dimensional_2009, Hoyt_Observation_2005}.
    Their combined absolute frequency is \SI{518295836590863.62(5)}{\hertz}.
    However, in \cref{tab:clock}, we adopt the value from the 2021 CIPM recommendation published by the BIPM~\cite{Margolis_CIPM_2024}, which is endorsed by the CCTF as a secondary representation of the SI second~\cite{CCTF2021PSFS2}.
    \item For the \sS{0}~$\to$~\tP{2} transition, we use the value from Ref.~\cite{Yamaguchi_Metastable_2008}, derived by combining the absolute frequency of the \sS{0} $\to$ \tP{0} transition from Ref.~\cite{Hoyt_Observation_2005} with absolute frequency measurements of the \tP{0}~$\to$~\tS{1} and \tP{2}~$\to$~\tS{1} transitions.
\end{itemize}

Differences in nuclear mass and charge distribution result in changes of the transition frequencies across isotopes of Yb.
These isotope shifts play important roles in beyond-Standard Model physics~\cite{Frugiuele_Constraining_2017, Berengut_Probing_2018, Figueroa_Precision_2022}.
With its many stable isotopes, Yb is well-suited for such studies.
\Cref{tab:isotope_shifts} summarizes experimental measurements of these isotope shifts for the states considered here.
All shifts are expressed relative to \ybb, as is customary in the literature.
Note that we list the isotope shift of the \textit{state} relative to the \sS{0} ground state, not the transition frequency.
These are of course equivalent for transitions out of \sS{0}, but for \tS{1} and \tD{1} we include measurements performed on transitions out of \tP{0,2}.
The latter are converted into state-shifts by adding the isotope shift of the corresponding lower state as listed elsewhere in the \namecref{tab:isotope_shifts}.
To be abundantly clear, the listed quantities are $\Delta E_\text{iso}/h$, where
\begin{equation}
    \Delta E_\text{iso} = \left[E\mathopen{}\left(^{2S+1}L_J, \ybi{A}\right) - E\mathopen{}\left(\sS{0}, \ybi{A}\right)\right] - \left[E\mathopen{}\left(^{2S+1}L_J, \ybb\right) - E\mathopen{}\left(\sS{0}, \ybb\right)\right].
\end{equation}
For the isotopes with hyperfine structure (i.e., \yb and \ybs) we use the center-of-gravity $E_\text{cg}$ as described following \cref{eq:E_hfs}.
The following comments apply:
\begin{itemize}
    \item Across the \tP{0,1,2} states, the isotope shifts are expected to be nearly identical, and the same applies for \tD{1,2,3}.
    Indeed the measured shifts for \tP{0} versus \tP{1} versus \tP{2} and \tD{1} versus \tD{2} reflect this.
    \item For \tS{1}, when combining the pathways $\sS{0}\rightarrow\tP{0}\rightarrow\tS{1}$ and $\sS{0}\rightarrow\tP{2}\rightarrow\tS{1}$ as described above, we find significant Birge ratios.
    If only the former pathway is considered, the respective measurements are mutually consistent.
    We do not speculate on the exact cause of the inconsistencies.
    \item Again, the source material for Ref.~\cite{Chaiko_Isotope_1966} could not be consulted, so here we rely on the values reported in Refs.~\cite{vanWijngaarden_Measurement_1994, Clark_Optical_1979}.
    \item For the lowest-abundance isotope \ybi{168} not all isotope shifts have been measured.
    Likewise, for the fermionic isotopes \yb and \ybs, no isotope shifts for \tP{2} have been reported.
    Those cases where no measurement was found are indicated by a question mark.
\end{itemize}

\subsection{Decay of Excited States}

An excited state $\ket{e}$ that decays to several lower levels $\{\ket{g_i}\}$ is described by Einstein $A$‐coefficients $A_{e\to g_i}$.
The total spontaneous decay rate is
\begin{equation}
    \Gamma = \sum_i A_{e\to g_i}~.
\end{equation}
The excited‐state lifetime $\tau$ is directly related to the total spontaneous decay rate
\begin{equation}
    \tau = \frac{1}{\Gamma}~.
\end{equation}
The corresponding full‐width at half‐maximum of the transition's line shape is
\begin{equation}
    \Delta\nu_{\rm FWHM} = \frac{\Gamma}{2\pi}~.
\end{equation}
Each decay channel $i$ carries a branching ratio
\begin{equation}\label{eq:branching_definition}
    \beta_i = \frac{A_{e\to g_i}}{\Gamma}~,
\end{equation}
so its partial decay rate is
\begin{equation}
    \Gamma_{\rm partial} = \beta_i\,\Gamma = A_{e\to g_i}~.
\end{equation}
Thus, combining a precise measurement of $\tau$ with known $\beta_i$ directly yields the individual $A$‐coefficients.

The following comments apply for the lifetimes and branching ratios listed in \crefrange{tab:blue_MOT}{tab:magnetic_clock_repumper}:
\begin{itemize}
    \item The metastable \tP{2} state can decay to the \sS{0} ground level via a magnetic–quadrupole (M2) transition.
    Additional decay arises from magnetic-dipole (M1) decay to \tP{1} and black-body-induced excitation to higher-lying triplet levels.
    In odd isotopes as \yb, however, hyperfine mixing with the nearby \tP{1} and \sP{1} manifolds as described in \cref{sec:mixing} opens electric–dipole (E1) decay channels that dominate the radiative lifetime.
    For the \tP{2} state, to our knowledge no experimental measurement of its radiative lifetime has been performed.
    We therefore adopt a theoretical value in \cref{tab:magnetic_clock}~\cite{Mishra_Radiative_2001}.
    Note that the lifetime in even isotopes is predicted to be significantly longer~\cite{Mishra_Radiative_2001, Porsev_Hyperfine_2004}.
    \item Likewise, we are unaware of any experimental results that completely characterize the branching ratios for any excited state.
    Ref.~\cite{Honda_Magneto-optical_1999} contains a partial characterization of the branching out of \sP{1}, finding a lower bound of \num{1.2(4)e-7}.
    This is in line with theoretical calculations~\cite{Porsev_Electric-dipole_1999}, and most notably is two orders of magnitude smaller than the comparable atom Sr~\cite{Pucher_Sr_2025}.
    As such, the \sS{0} $\to$ \sP{1} transition can be effectively considered a cycling transition, allowing laser cooling without the need for additional repumping lasers.
    All stated branching ratios follow from theoretical calculations as detailed below.
    \item We were unable to find source material for Refs.~\cite{Blagoev_1978, Burshtein_Lifetimes_1974, Lange_1970, Komarovskii_Oscillator_1969}.
    Instead, we rely on values reported in Refs.~\cite{Bowers_Experimental_1996, Blagoev_Lifetimes_1994, Rambow_Radiative_1976}.
\end{itemize}

\subsection{Additional Transition Parameters}

The spontaneous emission rate $\Gamma$ determines the relative intensity of a spectral line.
For a $J\to J'$ electric-dipole transition, $\Gamma$ is directly linked to the absorption oscillator strength $f$ via~\cite{Corney_Atomic_1977}
\begin{equation}
    \Gamma = \frac{e^2\omega_0^2}{2\pi\varepsilon_0 m_{\rm e}c^3}\frac{2J+1}{2J'+1}f~.
\end{equation}
The spontaneous decay rate for a two‐level atom is related to the dipole matrix element $d$ via~\cite{Demtroder_Laser_2008}
\begin{equation}\label{eq:gamma_matrix_element}
    \Gamma \;=\; \frac{\omega_0^3\,\lvert d\rvert^2}{3\pi\varepsilon_0\,\hbar\,c^3}  \quad\Longrightarrow\quad  \lvert d\rvert^2 \;=\; \frac{3\pi\varepsilon_0\,\hbar\,c^3}{\omega_0^3}\,\Gamma~.
\end{equation}
For a multi-level atom, the spontaneous decay rate between an excited state with total electronic angular momentum $J_e$ and one of the lower levels with $J_{g_i}$ is instead given by
\begin{equation}
    \Gamma_\text{partial} = A_{e\to g_i} = \frac{\omega_{eg_i}^3}{3\pi\varepsilon_0\hbar c^3}\frac{2J_{g_i}+1}{2J_e+1}\lvert\braket{J_{g_i} \| e \mathbf{r} \| J_e}\rvert^2  = \frac{\omega_{eg_i}^3}{3\pi\varepsilon_0\hbar c^3}\frac{1}{2J_e+1}\lvert\braket{J_{g_i} \| e \mathbf{r} \| J_e}_R\rvert^2~.
\end{equation}
Here, $\braket{J \| e \mathbf{r} \| J'}$ is the \textit{reduced} dipole matrix element using the convention of, e.g., Refs.~\cite{Brink_Angular_1968, SteckBook}, whereas the notation $\braket{J \| e \mathbf{r} \| J'}_R = \sqrt{2J+1}\braket{J \| e \mathbf{r} \| J'}$ is reserved for the Racah-Wigner convention, used for example in Ref.~\cite{UDportal}.
These reduced dipole matrix elements relate to dipole matrix elements through invocation of the Wigner-Eckart theorem, described in detail in \cref{sec:Rabi}.
The above expression can then be used together with \cref{eq:branching_definition} to calculate the branching ratios $\beta_i$ between fine-structure states as
\begin{equation}
    \beta_i = \frac{\lvert\braket{J_{g_i} \| e \mathbf{r} \| J_e}_R\rvert^2\,\omega_{eg_i}^3}{\sum_{g_i} \lvert \braket{J_{g_i} \| e\mathbf{r} \| J_e}_R\rvert^2 \, \omega_{eg_i}^3}~,
\end{equation}
where the sum in the denominator runs over all allowed final states $\{\ket{g_i}\}$.
We use this relation together with the values given in Ref.~\cite{Porsev_Electric-dipole_1999} to state the branching ratios in \crefrange{tab:blue_MOT}{tab:magnetic_clock_repumper}.
The reduced matrix elements are then found by reversing the above relation, to yield (up to a phase factor)
\begin{equation}
    \braket{J_{g_i} \| er \| J_e}_R = \sqrt{2J_e+1}\sqrt{\frac{3\pi\varepsilon_0\hbar c^3}{\omega_{eg_i}^3}\frac{\beta_i}{\tau_e}},
\end{equation}
which we evaluate using the experimental transition frequency $\omega_{eg_i}$ and excited state lifetime $\tau_e$.

Absorbing or emitting a resonant photon of momentum $\hbar k_L$ changes the velocity of an atom by the recoil velocity
\begin{equation}
    v_r = \frac{\hbar k_L}{m}~.
\end{equation}
The corresponding change in kinetic energy $\frac{1}{2} m v_r^2 = \frac{\hbar^2k_L^2}{2m}$ can be written as a recoil energy $E_{\rm rec}=\hbar\omega_r$, which defines the recoil frequency $\omega_r$
\begin{equation}\label{eq:recoil_energy}
    E_{\rm rec} = \hbar\omega_r = \frac{\hbar^2k_L^2}{2m}\quad\Rightarrow\quad\omega_r = \frac{\hbar k_L^2}{2m}~.
\end{equation}
An atom moving at velocity $v \ll c$ sees the optical frequency $\omega_\mathrm{L}$ shifted by a Doppler shift
\begin{equation}
    \Delta\omega_\mathrm{d} = \frac{v}{c}\omega_\mathrm{L}~.
\end{equation}
In particular, for $v=v_r$ one finds $\Delta\omega_d=2\omega_r$, showing that a single recoil kick corresponds to a twice as high Doppler shift change.

From these motions arise two key temperatures.
The recoil temperature
\begin{equation}
    T_r = \frac{\hbar^2\,k_L^2}{m\,k_{\rm B}}
\end{equation}
is the temperature corresponding to an ensemble with a one-dimensional root-mean-square momentum of one photon recoil~\cite{Steck2024CesiumDLine}.
Note that the single‐photon recoil energy $E_{\rm rec}$ (see \cref{eq:recoil_energy}) can be expressed as a temperature via $k_{\rm B} T = E_{\rm rec}$, giving
\begin{equation}
    T = \frac{\hbar^2\,k_L^2}{2m\,k_{\rm B}}~,
\end{equation}
which is exactly half of the rms‐defined recoil temperature $T_r$.
The Doppler temperature
\begin{equation}
    T_\mathrm{D} = \frac{\hbar\,\Gamma}{2\,k_{\rm B}}
\end{equation}
is the lowest temperature that Doppler cooling of a two‐level atom can reach, set by the balance of Doppler cooling and recoil heating~\cite{Lett_Observation_1988}.
Fermionic \yb possesses Zeeman substructure in its \sS{0} ground state.
This enables sub-Doppler cooling mechanisms (e.g., Sisyphus or polarization-gradient cooling) on the narrow $\sS{0} \to \tP{1}$ intercombination line.

\subsection{Magneto-Optical Traps}

Magneto–optical traps (MOTs) use laser light and magnetic fields to cool and confine neutral atoms.
For the fermionic isotope \yb, two transitions are typically used to cool the atoms in sequence to bring them from high thermal velocities to cold temperatures suitable for optical trapping or precision measurement.

The first cooling stage typically uses the strong \sS{0}~$\to$~\sP{1} transition at a wavelength of $\lambda \approx \SI{399}{\nano\meter}$, with a natural linewidth of $\Gamma \approx 2\pi \times \SI{29}{\mega\hertz}$.
This broad linewidth is well suited for decelerating atoms emerging from a thermal source at temperatures around \SI{350}{\celsius} or higher.
A Zeeman slower, based on a single laser beam and a spatially varying magnetic field, is commonly used to reduce the longitudinal velocity of the atoms.
Alternatively, a 2D MOT can provide transverse cooling and collimation.
In both cases, radiation pressure combined with magnetic forces slows the atoms for subsequent capture.

These slowed atoms are loaded into a blue three-dimensional (3D) MOT operating on the $\lambda \approx \SI{399}{\nano\meter}$ transition~\cite{Honda_Magneto-optical_1999}.
Although the wavelength of \SI{399}{\nano\meter} technically falls within the ultraviolet spectral range, it is sometimes referred to as a ``violet MOT'' in the literature~\cite{Park_Efficient_2003}.
However, the majority of sources refer to this trapping phase as a ``blue MOT,'' a convention we adopt in this work.
Due to the large linewidth, the blue MOT features a high capture velocity, making it effective for loading atoms from high-temperature sources.
To achieve lower temperatures, the atoms are transferred into a second-stage ``green MOT'' based on the narrow intercombination line \sS{0}~$\to$~\tP{1} at $\lambda \approx \SI{556}{\nano\meter}$, with a linewidth of $\Gamma \approx 2\pi \times \SI{182}{\kilo\hertz}$.
This allows Doppler cooling to the microkelvin regime.
Due to the difference in linewidth between the two cooling transitions, the magnetic field gradient must be adjusted between the MOT stages.
The narrower green MOT requires much smaller gradients than the blue MOT to avoid excessive Zeeman shifts that would push atoms out of resonance.
As a result, the MOT coil current is typically ramped down during the transition from the blue to the green MOT.

In ytterbium, the \sD{2} state lies higher in energy than the \sP{1} state, unlike in strontium where the \sD{2} state is lower~\cite{Pucher_Sr_2025}.
As a result, decay from the \sP{1} state predominantly returns to the ground state~\cite{Honda_Magneto-optical_1999}, enabling efficient accumulation of atoms in the MOT without the need for repumping lasers to load and maintain a large atom number.

The relatively broad linewidth of the intercombination transition compared to strontium (\SI{182}{\kilo\hertz} versus \SI{7.5}{\kilo\hertz}~\cite{Pucher_Sr_2025}) allows direct loading of the green MOT from a Zeeman-slowed atomic beam, where typically frequency broadening is used to increase the capture velocity~\cite{Kuwamoto_Magneto-optical_1999, Dorscher_Creation_2013, Guttridge_Direct_2016}.
Moreover, core-shell MOT configurations can enhance loading rates~\cite{Lee_Core-shell_2015}.
Direct loading from an oven into a blue MOT, bypassing both Zeeman slowers and 2D MOTs was also demonstrated using Yb~\cite{Letellier_Loading_2023}.

\section[Electric Fields]{Interaction with Electric Fields}

The interaction between an atom and an external electric field underpins both static energy shifts and coherent optical manipulations in modern atomic‐physics experiments.
We first derive the DC Stark effect, separating the induced energy shift into scalar, vector, and tensor contributions based on the atomic polarizability in a static field.
These shifts give rise to state‐dependent optical dipole potentials and define special operating wavelengths, the so-called “magic,” “anti-magic,” and “tune-out” wavelengths.
Finally, we address the dynamic response of the atom to near‐resonant light, deriving the Rabi frequency that governs coherent population oscillations under the rotating‐wave approximation.

\subsection{Polarizability}

In its most general form, the interaction Hamiltonian between an atom and a classical electric field is
\begin{equation}
    H_E = -\hat{\mathbf{d}}\cdot\mathbf{E}~,
\end{equation}
where $\hat{\mathbf{d}}$ is the electric dipole operator and $\mathbf{E}$ is the applied field.
We choose the quantization axis along the $z$–direction, and assume that this interaction Hamiltonian does not mix states any further than described in \cref{sec:mixing}, so that $F$ and $m_F$ remain good quantum numbers.
Far from resonance, the components of the dipole operator are related to the electric field via~\cite{SteckBook}
\begin{equation}
    d_\mu(\mathbf{r}) = \sum\limits_{\nu=x,y,z}\alpha_{\mu\nu}(\omega)E_\nu(\mathbf{r})~.
\end{equation}
Here, we explicitly denote positional dependence and
\begin{equation}
    \alpha_{\mu\nu}(\omega) = \Re\{\alpha_{\mu\nu}(\omega)\} + i\Im\{\alpha_{\mu\nu}(\omega)\}
\end{equation}
is the (complex) dynamic polarizability tensor at angular frequency $\omega$.
The real part $\Re\{\alpha_{\mu\nu}(\omega)\}$ relates to the energy shift, while the imaginary part $\Im\{\alpha_{\mu\nu}(\omega)\}$ accounts for absorption and scattering.
Hence, the resulting (time-averaged) energy shift for a state $\ket{i}=\ket{F, m_F}$ can be written as~\cite{Grimm_Optical_2000}
\begin{equation}\label{eq:lightshift}
    V_\text{ac}^i(\mathbf{r}; \omega) \equiv \braket{i | H_E | i} = - \alpha_{\rm tot}^i(\omega)\frac{I(\mathbf{r})}{2c\varepsilon_0},
\end{equation}
where $I(\mathbf{r}) = \frac{1}{2}c\varepsilon_0\lvert E(\mathbf{r})\rvert^2$ is the time-averaged local light intensity.
The dynamic polarizability $\alpha_{\rm tot}^i(\omega)$ can be related to the irreducible parts of the polarizability tensor $\alpha_{\mu\nu}$ so that it can be expressed as a sum of contributions from the scalar, vector, and tensor parts~\cite{LeKien_Dynamical_2013, Tsyganok_Scalar_2019}
\begin{equation}\label{eq:alpha_tot}
\begin{aligned}
    \alpha_{\rm tot}^i(\omega) &= 
    \underbrace{\alpha_s^i(\omega)}_{\text{scalar}}
    + \underbrace{\alpha_v^i(\omega)\sin(2\gamma)\frac{m_F}{2F}}_{\text{vector}}
    + \underbrace{\alpha_t^i(\omega)\frac{3m_F^2 - F(F+1)}{2F(2F-1)}(3\cos^2\beta -1)}_{\text{tensor}}~,
\end{aligned}
\end{equation}
where $\gamma$ is the ellipticity angle of the light polarization and $\cos(\beta)$ is the angle between $\mathbf{E}$ and the quantization axis.
In the generalized form of the polarizability, it is necessary to include the contribution of multiple transitions as well as the frequency of the electric field, $\omega$.
The scalar polarizability of the atomic state $\ket{i}$ does not depend on its quantum number $m_F$ and is
\begin{equation}
    \alpha_s^i(\omega) = \frac{2}{\hbar} \cdot \frac{1}{3(2F + 1)} \sum_{F'} \frac{\lvert\braket{F \| \mathbf{d} \| F'}_R\rvert^2 \omega_{F'F}}{\omega_{F'F}^2 - \omega^2}~,
\end{equation}
where $\braket{F \| \mathbf{d} \| F'}_R$ is the reduced electric dipole matrix element\footnote{In the Racah convention, polarizability expressions are invariant under the interchange of states since its conjugation rule is $\braket{F' \| \mathbf{d} \| F}_R = (-1)^{F'-F}\braket{F \| \mathbf{d} \| F'}_R^*$~\cite{SteckBook}.} and $\hbar\omega_{F'F} = E_{F'} - E_F$ is the energy difference between the excited state $\ket{F'}$ and the initial state $\ket{F}$ (see \cref{eq:E_hfs}).
The sum runs over all dipole-allowed transitions to states $\ket{F'}$.
The vector polarizability is given by
\begin{equation}
    \alpha_v^i(\omega) = -\frac{2}{\hbar}\,\sqrt{\frac{6F}{(F+1)(2F+1)}}\sum_{F'}(-1)^{F+F'}
    \begin{Bmatrix}
        1 & 1 & 1\\
        F & F' & F
    \end{Bmatrix}
    \frac{\lvert\braket{F \| \mathbf{d} \| F'}_R\rvert^2 \omega}{\omega_{F'F}^2-\omega^2}~,
\end{equation}
where the curly brackets are the Wigner 6-j symbols as before.
Finally, the tensor polarizability is
\begin{equation}
    \alpha_t^i(\omega) = \frac{2}{\hbar}\sqrt{ \frac{10 F (2F - 1)}{3(F + 1)(2F + 1)(2F + 3)} } \sum_{F'} (-1)^{F + F'} 
    \begin{Bmatrix}
        1 & 2 & 1 \\
        F & F' & F
    \end{Bmatrix}
    \frac{\lvert\braket{F \| \mathbf{d} \| F'}_R\rvert^2 \omega_{F'F}}{\omega_{F'F}^2 - \omega^2}~.
\end{equation}
The static polarizability can be found by setting $\omega=0$.

The reduced hyperfine matrix element $\braket{F \| \mathbf{d} \| F'}_R$ can be further simplified, leaving a reduced matrix element involving only the fine-structure quantum numbers $J$ and $J'$~\cite{SteckBook}
\begin{equation}\label{eq:reduced_dipole_matrix_element_J}
    \braket{F \| \mathbf{d} \| F'}_R = \braket{J \| \mathbf{d} \| J'}_R(-1)^{F'+J+I+1}\sqrt{(2F'+1)(2F+1)}
    \begin{Bmatrix}
        J & J' & 1 \\
        F' & F & I
    \end{Bmatrix}~.
\end{equation}
Recall that the reduced matrix elements $\braket{J \| \mathbf{d} \| J'}_R$ are listed in \crefrange{tab:blue_MOT}{tab:magnetic_clock_repumper} for the transitions considered in this reference.
A special case is provided by stretched hyperfine states, which are a single uncoupled product state of their electronic and nuclear components, $\ket{F=J+I,m_F=\pm F}=\ket{J,m_J=\pm J}\otimes\ket{I,m_I=\pm I}$, so that the light shift (and hence the magic condition) is set by the electronic component $m_J=\pm J$ rather than by hyperfine mixing.
Provided the light-shift operator is effectively nuclear-spin independent and diagonal in $\ket{J,m_J}$ for the experimental geometry (so that hyperfine-dependent corrections are negligible), the corresponding stretched-state polarizabilities match across all isotopes to within small residual effects from isotope shifts and hyperfine-resolved intermediate-state structure~\cite{MuziFalconi_Trapping_2025}.

\begin{figure}[ht]
  \centering
  \includegraphics[width=1\textwidth]{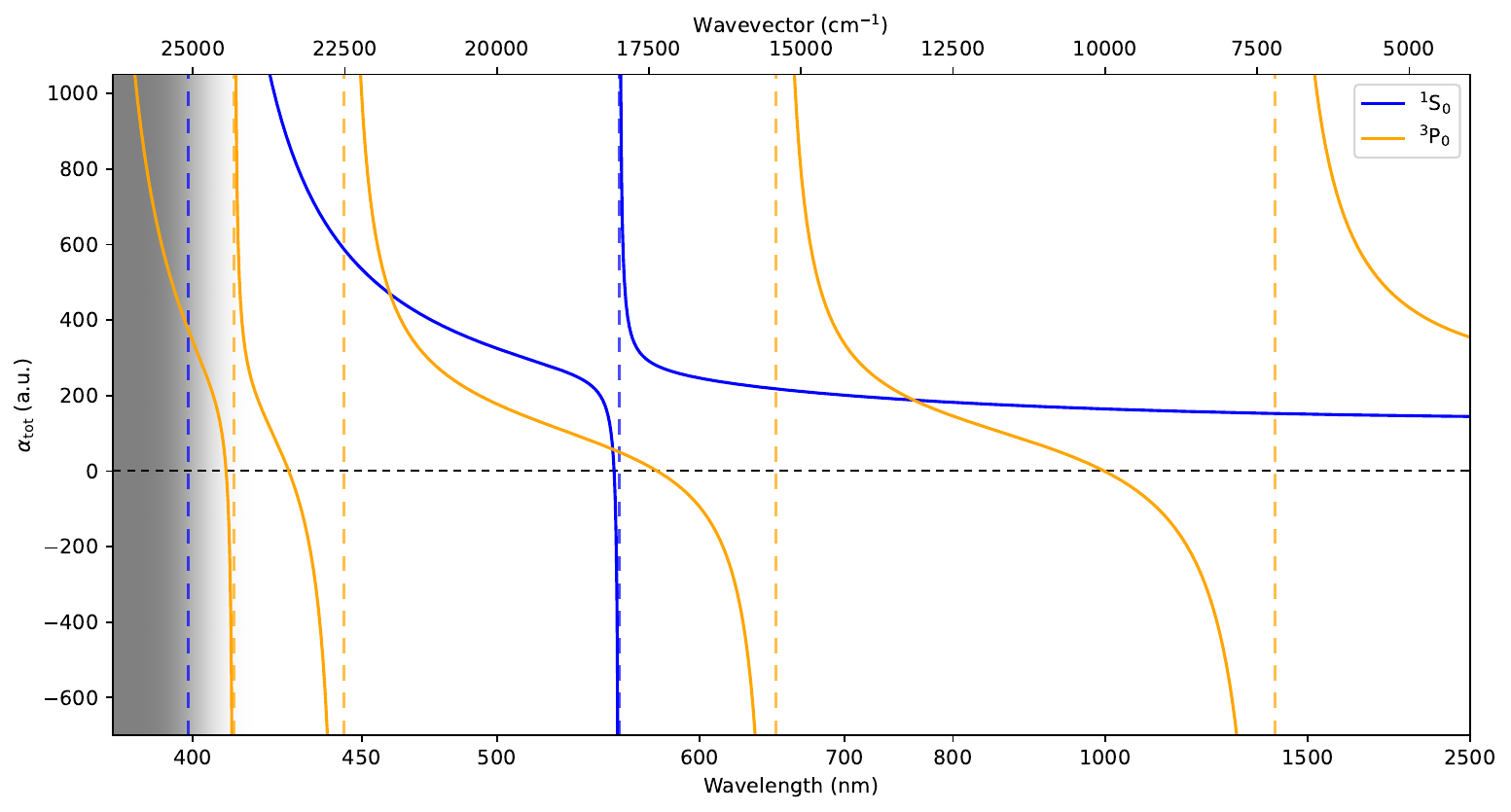}
  \caption{Polarizability of \sS{0} and \tP{0} states calculated using the semi-empirical model discussed in Refs.~\cite{Hohn_State-dependent_2023, Hohn_State-dependent_2024}.
  Note the nonlinearity of the wavelength axis.
  At short wavelengths, as indicated by the gray shaded area, the accuracy of the semi-empirical model starts to break down.}
  \label{fig:alpha}
\end{figure}

\Cref{fig:alpha} shows the dynamic polarizability of the \sS{0} and \tP{0} states.
Here, $\alpha_{\rm tot}$ is evaluated using the semi-empirical model described in Refs.~\cite{Hohn_State-dependent_2023, Hohn_State-dependent_2024}.
Rather than summing over all allowed transitions, which requires precise knowledge of a variety of highly excited states, the shortest-wavelength transitions are absorbed through a coarse-graining procedure.
For \sS{0} this entails a frequency-independent contribution 
to the polarizability.
For \tP{0} these transitions are instead captured by a single, \textit{effective} transition.
The frequency and linewidth of this effective transition, as well as the polarizability offset for \sS{0} are fitted to reproduce known state-(in)dependent wavelengths, discussed below.
The polarizability is stated in atomic units (\unit{\au}), which can be converted to SI through
\begin{equation}
    \alpha[\unit{\ampere\squared\second^4\per\kilo\gram}] = \frac{m_\text{e} e^2a_0^4}{\hbar^2}\alpha[\unit{\au}]~,
\end{equation}
or directly into light shift as
\begin{equation}
    V_\text{ac}[h\,\unit{\hertz\per(\watt\per\centi\meter\squared)}] = -\frac{2\pi^2 m_\text{e} e^2a_0^4}{c\varepsilon_0h^3}\times10^{4}\,\alpha[\unit{\au}] \approx \num{-0.0469}\,\alpha[\unit{\au}]~.
\end{equation}

\subsection{State-(in)dependent potentials}

The optical dipole potential $U$ experienced by an atom in a light field with optical frequency $\omega$ is given, in the far‐off‐resonant, low‐saturation limit, by \cref{eq:lightshift}.
A positive (negative) total polarizability produces an attractive (repulsive) potential.
At certain ``special'' wavelengths the polarizabilities of two states satisfy particular relations, giving rise to state‐(in)dependent potentials:
\begin{itemize}
    \item A \emph{magic wavelength} $\lambda_{\rm m}$ is defined by
    \[
        \alpha_{\rm tot}^g(\omega_{\rm m}) = \alpha_{\rm tot}^e(\omega_{\rm m})~,
    \]
    so that ground ($g$) and excited ($e$) states experience identical potentials.
    This cancels differential Stark shifts and preserves the unperturbed transition frequency.
    \item An \emph{anti‐magic wavelength} $\lambda_{\rm a}$ satisfies
    \[
        \alpha_{\rm tot}^g(\omega_{\rm a}) = -\alpha_{\rm tot}^e(\omega_{\rm a})~,
    \]
    so that the ground- and excited-state dipole potentials have equal magnitude but opposite sign, $U_g(\mathbf{r}) = -U_e(\mathbf{r})$.
    \item A \emph{tune‐out wavelength} $\lambda_{\rm t}$ occurs when the polarizability of one state vanishes:
    \[
        \alpha_{\rm tot}^g(\omega_{\rm t}^g)=0  \quad\text{or}\quad  \alpha_{\rm tot}^e(\omega_{\rm t}^e)=0~.
    \]
    At $\lambda_{\rm t}$, an atom in one internal state experiences no optical potential while in the other it remains subject to one, enabling state-selective mobility and precise tunneling control.
\end{itemize}

Values for magic, anti-magic, and tune-out wavelengths have been theoretically calculated in, e.g., Refs.~\cite{Hohn_State-dependent_2024, Tang_Magic_2018, Dzuba_Testing_2018, Dzuba_Dynamic_2010}.
\Cref{tab:special_wavelengths} presents experimentally measured magic and tune-out wavelengths for \yb.
We note that measurements using \ybb are included for wavelengths where hyperfine effects play a negligible role, i.e., those involving solely \sS{0} and/or \tP{0}, as well as stretched states (see the discussion above).
Special wavelengths measured in \ybb for these states can be used for \yb at the level of accuracy relevant for \cref{tab:special_wavelengths}.
All values inferred from \ybb are indicated with an asterisk.
We also note additional observations of (nearly) magic wavelengths for other transitions of \ybb in Refs.~\cite{Saskin_Narrow-Line_2019, Yamamoto_Ytterbium_2016, Yamaguchi_High-resolution_2010, Hohn_Determining_2026, Ishiyama_Excluding_2026}.

The following comments apply, as pertaining to \cref{tab:special_wavelengths}:
\begin{itemize}
    \item The $\sim$ \SI{759}{\nano\meter} magic wavelength for the \sS{0} $\to$ \tP{0} transition has been extensively studied because of its importance to optical lattice clocks.
    Various results have been obtained~\cite{Aeppli_Atomic_2025, Bothwell_Lattice_2025, Kim_Absolute_2021, Pizzocaro_Absolute_2020, Kobayashi_Demonstration_2020, Luo_Absolute_2020, Nemitz_Modeling_2019, Gao_Systematic_2018, McGrew_Atomic_2018, Kobayashi_Uncertainty_2018, Brown_Hyperpolarizability_2017, Pizzocaro_Absolute_2017, Kim_Improved_2017, Nemitz_Frequency_2016, Park_Absolute_2013, Lemke_Spin-1/2_2009, Kohno_One-Dimensional_2009}, and isotope shifts can be inferred by comparison to measurements of the magic wavelength in other isotopes~\cite{Barber_Direct_2006}.
    We note that these references frequently contain the \textit{operational} magic wavelength, which is subtly different from the electric dipole (E1) magic wavelength, see, e.g., Ref.~\cite{Bothwell_Lattice_2025} for an extended discussion.
    Nonetheless, we simply report in \cref{tab:special_wavelengths} the average of all results listed in the references above.
    The stated angle of $\sim$ \SI{17}{\degree} only applies to the triple-magic condition including \tP{1}, since \sS{0} and \tP{0} have no tensor polarizability and hence realize the magic condition under any angle.
    \item Ref.~\cite{Ma_Universal_2022} does not specify the angle between quantization axis and polarization of the light used to observe the magic condition for the \sS{0} $\to$ \tP{1} $\ket{F'=3/2, m_{F'}=\pm1/2}$ transition.
    \item For the magic wavelength reported in Ref.~\cite{Li_Fast_2025} we were likewise unable to retrieve the angle for which this magic condition was achieved.
    Additionally, it highlights that the stated number is an \textit{operational} magic wavelength, and that due to its proximity to the \tP{1} $\to$ \sD{2} transition the precise magic condition depends on residual broadband content of the laser light.
    \item The last column lists the polarizability of the \sS{0} state at each wavelength, calculated using the semi-empirical model described above.
\end{itemize}

\subsection{Calculation of Rabi Frequencies}\label{sec:Rabi}

\begin{figure}[ht]
  \centering
  \includegraphics[width=0.6\textwidth]{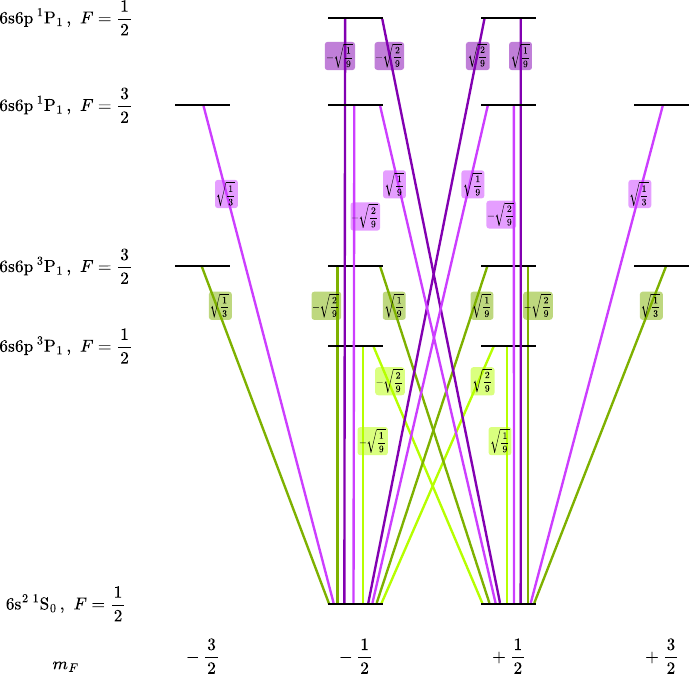}
  \caption{Dipole matrix elements for $\ket{\sS{0}, F, m_F} \rightarrow \ket{\tP{1}, F', m_{F'}}$ and $\ket{\sS{0}, F, m_F} \rightarrow \ket{\sP{1}, F', m_{F'}}$ transitions as multiples of the reduced matrix element $\braket{J \| e \vec{r} \| J'}_R$~.}
  \label{fig:prefac_1s0}
\end{figure}

\begin{figure}[ht]
  \centering
  \includegraphics[width=0.6\textwidth]{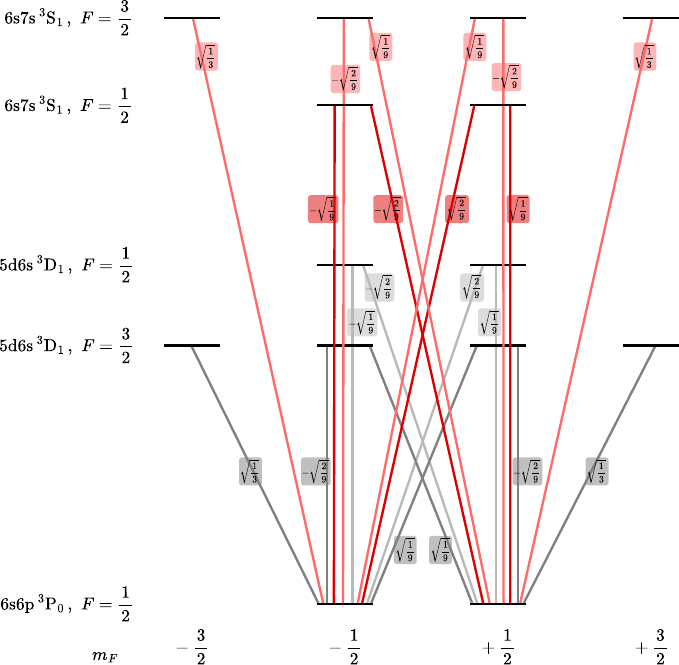}
  \caption{Dipole matrix elements for $\ket{\tP{0}, F, m_F} \rightarrow \ket{\tD{1}, F', m_{F'}}$ and $\ket{\tP{0}, F, m_F} \rightarrow \ket{\tS{1}, F', m_{F'}}$ transitions as multiples of the reduced matrix element $\braket{J \| e \vec{r} \| J'}_R$~.}
  \label{fig:prefac_3p0}
\end{figure}

\begin{figure}[ht]
  \centering
  \includegraphics[width=0.8\textwidth]{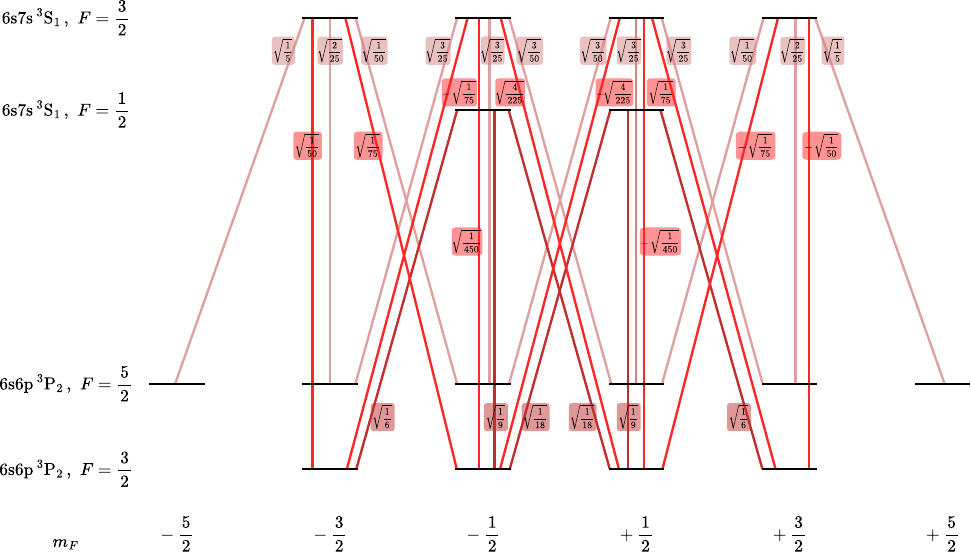}
  \caption{Dipole matrix elements for $\ket{\tP{2}, F, m_F} \rightarrow \ket{\tS{1}, F', m_{F'}}$ transitions as multiples of the reduced matrix element $\braket{J \| e \vec{r} \| J'}_R$~.}
  \label{fig:prefac_3p2}
\end{figure}

Consider an atom with a ground‐state manifold $\ket{g} =  \ket{F_g,\,m_{F_g}}$ and an excited‐state manifold $\ket{e} =  \ket{F_e,\,m_{F_e}}$.
A near‐resonant laser with an angular frequency $\omega_L$ drives the $\ket{g}  \leftrightarrow \ket{e}$ transition via the electric‐dipole Hamiltonian:
\begin{equation}
    H_{\mathrm{int}}(t) \;=\; -\,\hat{\mathbf{d}}\;\cdot\;\mathbf{E}(t) \;\approx\; -\,\braket{e | \hat{\mathbf{d}}\cdot\hat{\boldsymbol{\epsilon}}_{\,q} | g}\;E_{0}\,\cos(\omega_{L} t)\;\bigl(\ket{e} \bra{g} + \ket{g}\bra{e}\bigr)~,
\end{equation}
where $\braket{e | \hat{\mathbf{d}}\cdot\hat{\boldsymbol{\epsilon}}_{\,q} | g}$ is the matrix element of the dipole operator between the two levels, including the polarization component.
Here, $q$ labels the spherical polarization component.
$q=0$ corresponds to $\pi$-polarization, i.e., electric field along the quantization axis, while $q=+1$ and $q=-1$ correspond to right-circular $\sigma^{+}$ and left-circular $\sigma^{-}$ polarization, respectively.
Under the rotating‐wave approximation, the Rabi frequency $\Omega_{q}^{\,g\to e}$ is
\begin{equation}\label{eq:Omega_basic_general}
    \Omega_{q}^{\,g\to e} \;=\; \frac{1}{\hbar}\,\braket{e | \hat{\mathbf{d}}\cdot\hat{\boldsymbol{\epsilon}}_{\,q} | g} \;E_{0}~.
\end{equation}
This matrix element can be reduced using the Wigner-Eckart theorem~\cite{SteckBook}
\begin{equation}
    \braket{F_g m_{F_g} | d_q | F_e m_{F_e}} = \frac{\braket{F_g \| \mathbf{d} \| F_e}_R}{\sqrt{2F_g+1}} \braket{F_g m_{F_g} | F_e m_{F_e}; 1 q},
\end{equation}
separating it into a product of a reduced dipole matrix element which depends on $F_g$ and $F_e$ but not on $m_{F_g}$ or $m_{F_e}$, and a Clebsch–Gordan factor that accounts for the coupling between the specific Zeeman sublevels $(m_{F_g},\,m_{F_e})$ under polarization $q$.
The Clebsch–Gordan coefficient can be calculated using the Wigner $3$‐$j$ symbol as~\cite{SteckBook}
\begin{equation}
    \braket{F_g m_{F_g} | F_e m_{F_e}; 1 q} = \braket{F_e m_{F_e}; 1 q | F_g m_{F_g}} = \sqrt{2F_g+1}(-1)^{F_e-1+m_{F_g}}
    \begin{pmatrix}
        F_e & 1 & F_g\\
        m_{F_e} & q & -m_{F_g}
    \end{pmatrix}.
\end{equation}
This separation greatly simplifies practical calculations, since all the dependence on magnetic quantum numbers and polarization is contained in the known tabulated Clebsch–Gordan factor.
Further, using \cref{eq:reduced_dipole_matrix_element_J}, the matrix element between hyperfine substates can be related to the reduced matrix elements tabulated in \crefrange{tab:blue_MOT}{tab:magnetic_clock_repumper}.
Explicitly, this gives
\begin{align}
    \braket{F_g m_{F_g} | d_q | F_e m_{F_e}} &= (-1)^{2F_e+J_g+I+m_{F_g}}\sqrt{(2F_g+1)(2F_e+1)}
    \begin{pmatrix}
        F_e & 1 & F_g\\
        m_{F_e} & q & -m_{F_g}
    \end{pmatrix}
    \begin{Bmatrix}
        J_g & J_e & 1 \\
        F_e & F_g & I
    \end{Bmatrix}
    \braket{J_g \| \mathbf{d} \| J_e}_R  \\
    &\equiv\; w_{q}^{\,g\to e}\braket{J_g \| \mathbf{d} \| J_e}_R~.
    \label{eq:wigner_factor_def}
\end{align}
The values of the factor $w_{q}^{\,g\to e}$ for various transitions can be found in \crefrange{fig:prefac_1s0}{fig:prefac_3p2}.
Substituting into \cref{eq:Omega_basic_general} yields the (real, positive) Rabi frequency
\begin{equation}\label{eq:Omega_separated}
    \Omega^{\,g\to e}_q = \lvert w^{g\to e}_q\rvert\,\frac{\lvert\braket{J_g \| \hat{\mathbf{d}} \| J_e}_R\rvert}{\hbar}\;E_0~.
\end{equation}

We consider an atom driven by a Gaussian laser beam, whose waist is much larger than the spatial extent of the atomic wavefunction.
In this case, the atom experiences the peak intensity of the Gaussian beam, given by
\begin{equation}
    I_{0} = \;\frac{2P}{\pi w_{0}^{2}} \;=\; \frac{8P}{\pi D^{2}}~,
\end{equation}
where $P$ is the total optical power and $w_{0}$ is the $1/e^{2}$ beam radius so that the full $1/e^{2}$ diameter is $D=2w_{0}$.
Using \cref{eq:intensity_efield}, the peak electric‐field amplitude is
\begin{equation}\label{eq:E0_gaussian_peak}
    I_{0} = \;\frac12\,c\,\varepsilon_{0}\,E_{0}^{2}  \quad\Longrightarrow\quad  E_{0} = \;\sqrt{\frac{2I_{0}}{\varepsilon_{0}\,c}} = \;\sqrt{\frac{16\,P}{\varepsilon_{0}\,c\,\pi\,D^{2}}}~.
\end{equation}

Using the separated form of the Rabi frequency (see \cref{eq:Omega_separated}) together with the peak electric‐field amplitude (see \cref{eq:E0_gaussian_peak}), we find  
\begin{equation}\label{eq:Omega_final_gaussian_clean}
    \Omega_{q}^{\,g\to e} = \;\bigl|w_{q}^{\,g\to e}\bigr|\;\frac{\lvert\braket{J_g \| \hat{\mathbf{d}} \| J_e}_R\rvert}{\hbar}\;\sqrt{\frac{16\,P}{\varepsilon_{0}\,c\,\pi\,D^{2}}}~.
\end{equation}
In \cref{tab:blue_MOT}-\ref{tab:magnetic_clock_repumper}, we list numerical values of $\Omega_{q}^{\,g\to e}$ obtained from \cref{eq:Omega_final_gaussian_clean} under the following common experimental parameters: a laser power of $P=1\:\mathrm{mW}$ with a Gaussian beam diameter of $D=1\:\mathrm{mm}$, and $\pi$‐polarization ($q=0$).
Specifically, we calculate the Rabi frequency for $\ket{F_g,m_{F_g}}\to \ket{F_e,m_{F_e}}$ transitions as specified.

\section[Fluorescence]{Resonance Fluorescence}\label{sec:resonance_fluorescence}

In this \namecref{sec:resonance_fluorescence}, we analyze the interaction of a two‐level atom with a monochromatic classical electromagnetic field.
Under the electric‐dipole and rotating‐wave approximations, the atom behaves as an ideal two‐level system whose dynamics are governed by the optical Bloch equations.
We derive these equations, solve for the steady‐state excited‐state population, and then obtain expressions for the total photon scattering rate, the saturation intensity, and the on‐resonance scattering cross section.

\subsection{Optical Bloch Equations}

Consider an atom with ground state $\ket{g}$ and excited state $\ket{e}$, separated by an energy $\hbar\omega_0$.
We drive the $\ket{g} \leftrightarrow \ket{e}$ transition with a classical, linearly polarized, monochromatic field of the form
\begin{equation}
    \mathbf{E}(t) \;=\; E_0 \, \hat{\boldsymbol{\epsilon}} \, \cos(\omega_L t) \,,
\end{equation}
where $E_0$ is the real field amplitude, $\hat{\boldsymbol{\epsilon}}$ is the unit polarization vector, and $\omega_L$ is the laser frequency.
The atom–field interaction Hamiltonian in the electric‐dipole approximation is
\begin{equation}
    H_{\rm int} \;=\; -\,\hat{\mathbf{d}} \,\cdot\, \mathbf{E}(t) \;=\; -\,\bigl(d\,\hat{\sigma}_+ + d^*\,\hat{\sigma}_-\bigr)\,E_0 \cos(\omega_L t)\,,
\end{equation}
where $\hat{\mathbf{d}}$ is the atomic dipole moment, $d = \braket{e | \hat{\mathbf{d}}\cdot \hat{\boldsymbol{\epsilon}} | g}$ is the (complex) dipole matrix element, and $\hat{\sigma}_- = \ket{g} \bra{e}$ (with $\hat{\sigma}_+ = \hat{\sigma}_-^\dagger$) is the lowering operator for the two‐level system.
We define the Rabi frequency as
\begin{equation}\label{eq:rabi}
    \Omega \;=\; \frac{d \, E_0}{\hbar}\,.
\end{equation}
We move into a frame rotating at the laser frequency $\omega_L$ and apply the rotating‐wave approximation (RWA).
Then, the density operator $\rho$ of the two‐level atom is described by the optical Bloch equations~\cite{Steck2024CesiumDLine}
\begin{subequations}\label{eq:optical_bloch_raw}
\begin{align}
    \dot{\rho}_{gg} \; &= \; \frac{i \Omega}{2} \bigl(\rho_{eg} - \rho_{ge}\bigr) \;+\; \Gamma\,\rho_{ee} \,,  \\
    \dot{\rho}_{ee} \; &= \; -\,\frac{i \Omega}{2} \bigl(\rho_{eg} - \rho_{ge}\bigr) \;-\; \Gamma\,\rho_{ee} \,,  \\
    \dot{\rho}_{ge} \; &= \; -\bigl(\gamma + i \Delta\bigr)\,\rho_{ge} \;-\; \frac{i \Omega}{2}\,\bigl(\rho_{ee} - \rho_{gg}\bigr) \,,
\end{align}
\end{subequations}
where $\rho_{ij} = \braket{i | \rho | j}$ are the matrix elements of the density operator in the $\{\ket{g}, \ket{e}\}$ basis, $\gamma = \tfrac{\Gamma}{2} + \gamma_c$ is the total coherence‐decay (``transverse'') rate, with $\gamma_c$ a pure‐dephasing rate (e.g., due to collisions), and $\Delta = \omega_L - \omega_0$ is the detuning of the driving field from atomic resonance.
In writing \cref{eq:optical_bloch_raw}, we have neglected any additional couplings to auxiliary levels or motional effects.
The term $\Gamma\,\rho_{ee}$ in the population equations accounts for radiative decay from $\ket{e}$ to $\ket{g}$, while the coherence $\rho_{ge}$ and its conjugate $\rho_{eg} = \rho_{ge}^*$ evolve under both dephasing and the coherent drive $\Omega$.

\subsection{Steady‐State Excited‐State Population}

Here, we are primarily interested in the long‐time (steady‐state) solution, where $\dot{\rho}_{ij} = 0$ for all $i,j$.
In the steady state and assuming the purely radiative case ($\gamma = \Gamma/2$, i.e., $\gamma_c = 0$), one finds
\begin{equation}\label{eq:rho_ee_ss}
    \rho_{ee}^{(\infty)} \;=\; \frac{\displaystyle \bigl(\tfrac{\Omega}{\Gamma}\bigr)^2}{\displaystyle 1 + 4\bigl(\tfrac{\Delta}{\Gamma}\bigr)^2 + 2\bigl(\tfrac{\Omega}{\Gamma}\bigr)^2}~.
\end{equation}
We can rewrite this equation
\begin{equation}
    \rho_{ee}^{(\infty)} \;=\; \frac{s/2}{1 + s + 4(\Delta/\Gamma)^2}\,, 
\end{equation}
where $s$ is the on‐resonance saturation parameter, given by
\begin{equation}
    s \;=\; \frac{2\Omega^2}{\Gamma^2}~.
\end{equation}
\Cref{eq:rho_ee_ss} shows that, on resonance ($\Delta=0$), the excited‐state population saturates to $1/2$ when $\Omega \gg \Gamma$.

\subsection{Total Photon Scattering Rate}

The total photon scattering (or fluorescence) rate is the rate at which population decays from the excited state, given by
\begin{equation}
    R_{\rm sc} \;=\; \Gamma \,\rho_{ee}~.
\end{equation}
In the steady state, we find
\begin{equation}\label{eq:R_sc_raw}
    R_{\rm sc} \;=\; \Gamma \,\rho_{ee}^{(\infty)} \;=\; \frac{\Gamma}{2} \; \frac{\,s\,}{\,1 + s + 4(\Delta/\Gamma)^2\,} \,,
\end{equation}
where we have set $s = 2\Omega^2/\Gamma^2$ as above.
Rewriting \cref{eq:R_sc_raw} in terms of the incident intensity $I$ leads naturally to the definition of the saturation intensity $\Isat$.

\subsection{Saturation Intensity}

The average intensity of a plane electromagnetic wave in free space is 
\begin{equation}\label{eq:intensity_efield}
    I \;=\; \frac{1}{2}\,c \,\varepsilon_0 \, E_0^2\,,
\end{equation}
where $c$ is the speed of light and $\varepsilon_0$ is the vacuum permittivity.
In terms of the Rabi frequency (see \cref{eq:rabi}), one finds
\begin{equation}
    I = \frac{1}{2} \, c \,\varepsilon_0 \,\Bigl(\hbar \,\frac{\Omega}{\lvert d \rvert}\Bigr)^2 = \frac{c\,\varepsilon_0\,\hbar^2}{2 \lvert d \rvert^2}\;\Omega^2 \,.
\end{equation}
We define the (on‐resonance) saturation intensity $\Isat$ using the condition that $s=1$ when $I = \Isat$.
Since $s = 2\Omega^2 / \Gamma^2$, setting $s=1$ yields
\begin{equation}\label{eq:I_sat}
    \Isat \;=\; \frac{c\,\varepsilon_0\,\hbar^2}{2 \lvert d \rvert^2}\;\frac{\Gamma^2}{2} \;=\; \frac{c\,\varepsilon_0\,\hbar^2\,\Gamma^2}{4\,\lvert d \rvert^2}\,.
\end{equation}
Using \cref{eq:gamma_matrix_element}, we find the saturation intensity for a cycling transition
\begin{equation}
    \Isat = \frac{\hbar\,\omega_0^3\,\Gamma}{12\pi\,c^2} = \frac{\pi\,h\,c\,\Gamma}{3\,\lambda^3}\,.
\end{equation}
Equivalently, $\Isat$ and $\Omega$ are often defined as
\begin{equation}\label{eq:saturation_parameter}
    \frac{I}{\Isat} = \frac{2\,\Omega^2}{\Gamma^2} = s\quad\Longrightarrow\quad\Omega = \frac{\Gamma}{\sqrt2}\sqrt{\frac{I}{\Isat}}~.
\end{equation}
Using this definition of $\Isat$, we can write the scattering rate (see \cref{eq:R_sc_raw}) as
\begin{equation}\label{eq:gain_equation}
    R_{\text{sc}} = \left( \frac{\Gamma}{2} \right) \frac{\left( \frac{I}{I_{\text{sat}}} \right)}{1 + 4 \left( \frac{\Delta}{\Gamma} \right)^2 + \left( \frac{I}{I_{\text{sat}}} \right)}~.
\end{equation}
When the laser is exactly on resonance ($\Delta = 0$), the scattering rate reduces to
\begin{equation}\label{eq:R_sc_on_res}
    R_{\rm sc}(\Delta=0) = \frac{\Gamma}{2} \;\frac{\,I / \Isat\,}{\,1 + I / \Isat\,}\,.
\end{equation}
Thus, on resonance, the scattering rate on a fully saturated transition ($I \gg \Isat$) approaches $\Gamma/2$, consistent with the fact that at saturation $\rho_{ee}^{(\infty)} \to 1/2$.

\subsection{On‐Resonance Scattering Cross Section}

It is often convenient to express the scattering properties in terms of an effective cross section $\sigma(\Delta, I)$, defined as the power radiated by the atom divided by the incident energy flux so that the power scattered by the atom is $\sigma(\Delta, I)\,I$.
Using \cref{eq:gain_equation}, we can calculate the scattering cross section as
\begin{equation}
    R_{\rm sc} \;=\; \sigma(\Delta, I)\, I \quad\Longrightarrow\quad\sigma(\Delta, I) \;=\; \frac{R_{\rm sc}}{\,I\,} \;=\; \frac{\Gamma}{2\,I}\;\frac{I / \Isat}{\,1 + I/\Isat + 4(\Delta/\Gamma)^2\,}\,.
\end{equation}
We define the low-intensity on‐resonance cross section as
\begin{equation}\label{eq:sigma_0}
    \sigma_0 \;=\; \frac{\hbar \omega_0\,\Gamma}{2\,\Isat}~.
\end{equation}
For a cycling transition, this reduces to~\cite{Foot_Atomic_2004}
\begin{equation}\label{eq:sigma_0_two_level}
    \sigma_0 \;=\; \frac{3 \lambda^2}{2\pi}~.
\end{equation}
Then, we can write the scattering cross section as
\begin{equation}\label{eq:sigma_general}
    \sigma(\Delta, I) = \;\frac{\sigma_0}{\,1 + 4 (\Delta / \Gamma)^2 + I / \Isat\,}\,.
\end{equation}
On resonance ($\Delta = 0$), this reduces to
\begin{equation}\label{eq:sigma_on_res}
    \sigma(0, I) = \frac{\sigma_0}{\,1 + I / \Isat\,}\,,
\end{equation}
showing that at $I = \Isat$ the scattering cross section is half of its low‐intensity value $\sigma_0$.

\clearpage

\section{Data Tables}
\makegapedcells

\input{tables/constants}

\input{tables/physical_properties}

\input{tables/isotopes}

\input{tables/scattering_lengths}

\input{tables/zeeman_and_hyperfine_coefs}

\input{tables/isotope_shifts_part1}
\input{tables/isotope_shifts_part2}

\input{tables/magic_and_tuneout}

\input{tables/1S0_1P1}

\input{tables/1S0_3P0}

\input{tables/1S0_3P1}

\input{tables/1S0_3P2}

\input{tables/3P0_3D1}

\input{tables/3P0_3S1}

\input{tables/3P2_3S1}

\clearpage

\begin{acknowledgments}
We thank Monika Aidelsburger, William Cairncross, Jacob Covey, Wentao Fan, Frederic Hummel, Alessandro Muzi Falconi, Riccardo Pennetta, and Tao Zheng for stimulating discussions, comments, corrections, and suggestions.
R.M.K.~acknowledges support from the Alexander von Humboldt Stiftung.
S.L.K.~acknowledges support from the Independent Research Fund Denmark.
\end{acknowledgments}

\setcounter{secnumdepth}{0}
\section{Release notes}
\noindent\textbf{v2}
\begin{itemize}
    \item Include isotope shifts for the \tP{2} state.
    Accordingly update isotope shifts for the \tS{1} state, noting that this, however, yields large Birge ratios.
    \item Update the hyperfine structure constant for \tP{2} with one additional reference.
    \item Update the \SI{759}{\nm} magic wavelength with one additional reference.
    \item Add stretched-state magic-wavelength measurements obtained in \ybb to \cref{tab:special_wavelengths}, along with supporting justification in the main text for applying them to fermionic isotopes.
    \item Fixed an error in \cref{tab:special_wavelengths} where two polarizability entries were swapped.
    \item Replace preprint citations with journal-published references, where available.
\end{itemize}

\noindent\textbf{v1}

\noindent Original release.

\bibliography{171yb_line_data, miscellaneous}

\end{document}

%% file: preamble.tex
\usepackage{babel}

\usepackage[letterpaper,top=2cm,bottom=2cm,left=2cm,right=2cm,marginparwidth=1.75cm]{geometry}

\usepackage{amsmath}
\usepackage{mathtools}
\usepackage{graphicx}

\usepackage{caption}
\usepackage{subcaption}
\usepackage{xspace}
\usepackage[per-mode=symbol]{siunitx}
\DeclareSIUnit{\eaBohr}{$ea_0$}
\DeclareSIUnit{\gauss}{G}
\DeclareSIUnit{\amu}{u}
\DeclareSIUnit{\torr}{Torr}
\DeclareSIUnit{\au}{a.u.}

\usepackage{makecell}
\setcellgapes{1.5pt}
\usepackage{braket}
\usepackage{multirow}
\usepackage{orcidlink}
\usepackage[capitalize]{cleveref}

\usepackage{fancyhdr}
\DeclareCaptionLabelFormat{continued}{#1~#2~(\textit{continued})}
\renewcommand{\arraystretch}{1.25}  

%% file: commands.tex
\newcommand{\yb}{\ensuremath{{^\text{171}\text{Yb}}}\xspace}
\newcommand{\ybi}[1]{\ensuremath{{^\text{#1}\text{Yb}}}\xspace}
\newcommand{\ybb}{\ybi{174}}
\newcommand{\ybs}{\ybi{173}}

\newcommand{\birge}{\ensuremath{{R_\text{B}}}}
\newcommand{\muB}{\mu_\text{B}}
\newcommand{\muN}{\mu_\text{N}}
\newcommand{\Isat}{I_\text{sat}}

\newcommand{\sS}[1]{\ensuremath{{^1\text{S}_{#1}}}}
\newcommand{\sP}[1]{\ensuremath{{^1\text{P}_{#1}}}}
\newcommand{\sPb}[1]{\ensuremath{{^1\text{P}_{#1}^{(0)}}}}
\newcommand{\sPso}[1]{\ensuremath{{^1\text{P}_{#1}^{(\text{so})}}}}
\newcommand{\sD}[1]{\ensuremath{{^1\text{D}_{#1}}}}
\newcommand{\tS}[1]{\ensuremath{{^3\text{S}_{#1}}}}
\newcommand{\tP}[1]{\ensuremath{{^3\text{P}_{#1}}}}
\newcommand{\tPb}[1]{\ensuremath{{^3\text{P}_{#1}^{(0)}}}}
\newcommand{\tPp}[1]{\ensuremath{{^3\text{P}_{#1}'}}}
\newcommand{\tPso}[1]{\ensuremath{{^3\text{P}_{#1}^{(\text{so})}}}}
\newcommand{\tD}[1]{\ensuremath{{^3\text{D}_{#1}}}}

%% file: tables/constants.tex
\begin{table}[!ht]
\centering
\caption{Fundamental Physical Constants (2022 CODATA recommended values)~\cite{Mohr_CODATA_2025}.}
\label{tab:fundamental_constants}
\begin{tabular}{|l|c|c|}
\hline
\textbf{Constant}            & \textbf{Symbol} & \textbf{Value}                                                   \\ \hline
Speed of Light               & $c$             & \SI{2.99792458e8}{\meter\per\second} (exact)                    \\ \hline
Permeability of Vacuum       & $\mu_{0}$       & \SI{1.25663706127(20)e-6}{\newton\per\ampere\squared}                \\ \hline
Permittivity of Vacuum       & $\varepsilon_{0}$ & $(\mu_{0}c^{2})^{-1}\approx$ \SI{8.8541878188(14)e-12}{\farad\per\meter}                 \\ \hline
Planck’s Constant            & $h$             & \SI{6.62607015e-34}{\joule\second} (exact)                     \\ \hline
Reduced Planck Constant      & $\hbar$         & $\sim$ \SI{1.054571817e-34}{\joule\second}                    \\ \hline
Elementary Charge            & $e$             & \SI{1.602176634e-19}{\coulomb} (exact)                         \\ \hline
Bohr Magneton                & $\muB$       & \SI{9.2740100657(29)e-24}{\joule\per\tesla}                    \\ \hline
Nuclear Magneton                & $\muN$       & \SI{5.0507837393(16)e-27}{\joule\per\tesla}                    \\ \hline
Atomic Mass Unit             & $u$             & \SI{1.66053906892(52)e-27}{\kilogram}                          \\ \hline
Electron Mass                & $m_{\rm e}$         & \SI{9.1093837139(28)e-31}{\kilogram}                           \\ \hline
Bohr Radius                  & $a_{0}$         & \SI{5.29177210544(82)e-11}{\meter}                            \\ \hline
Boltzmann Constant           & $k_{\rm B}$         & \SI{1.380649e-23}{\joule\per\kelvin} (exact)                   \\ \hline
\end{tabular}
\end{table}

%% file: tables/physical_properties.tex
\begin{table}[!ht]
\centering
\caption{Physical properties of \yb.}
\label{tab:physical_properties}
\begin{tabular}{|l|c|c|c|}
\hline
\textbf{Property} & \textbf{Symbol} & \textbf{Value} & \textbf{Reference} \\
\hline
Atomic Number & $Z$ & 70 & \\
\hline
Total Nucleons & $Z + N$ & 171 & \\
\hline
Nuclear Spin & $I$ & 1/2 & \\
\hline
Particle Statistics &  & fermionic & \\
\hline
Relative Natural Abundance & $\eta$ & \SI{14.239(5)}{\percent} & \cite{Wang_absolute_2015} \\
\hline
Nuclear Lifetime & $\tau_n$ & stable & \\
\hline
Atomic Mass & $m$ & \makecell{\SI{170.936331515(14)}{\amu}\\ \SI{2.8384645678(9)e-25}{\kilo\gram}} & \cite{Wang_AME_2021} \\
\hline
RMS Nuclear Charge Radius & $R$ & \SI{5.2906(57)}{\femto\meter} & \cite{Angeli_Table_2013} \\
\hline
Electron Spin g-Factor  & $g_S$ & \num{2.00231930436092(36)} & \cite{Mohr_CODATA_2025} \\
\hline
Electron Orbital g-Factor & $g_L$ & $\sim$ \num{0.99999679} & \\
\hline
Nuclear Magnetic Moment & $\mu$ & \num{0.4923(4)}$\muN$ & \cite{Stone_Table_2019} \\
\hline
Nuclear Spin g-Factor & $g_I$ & \num{-0.0005362(4)} & \\
\hline
Density at \SI{20}{\celsius} & $\rho_m$ & \SI{6904(2)}{\kilogram\per\meter\cubed} & \cite{Arblaster_Selected_2018} \\
\hline
Molar Volume at \SI{20}{\celsius} & $V_m$ & \SI{25.066(7)}{\centi\meter\cubed\per\mol} & \cite{Arblaster_Selected_2018} \\ 
\hline
Melting Point & $T_{\rm M}$ & \SI{1097}{\kelvin} (\SI{824}{\celsius}) & \cite{Haynes_CRC_2016} \\
\hline
Boiling Point & $T_{\rm B}$ & \SI{1496}{\kelvin} (\SI{1196}{\celsius}) & \cite{Haynes_CRC_2016} \\
\hline
Specific Heat Capacity & $c_p$ & \SI{0.155}{\joule\per\gram\per\kelvin} & \cite{Haynes_CRC_2016} \\
\hline
Molar Heat Capacity & $C_p$ & \SI{26.74}{\joule\per\mol\per\kelvin} & \cite{Haynes_CRC_2016} \\
\hline
Vapor Pressure at \SI{25}{\celsius} & $P$ & \makecell{\SI{1.69(8)e-16}{\pascal} \\ \SI{1.27(6)e-18}{\torr}} & \cite{Alcock_Vapour_1984} \\
\hline
\multirow{5}{*}{Ionization Energy} & \multirow{3}{*}{$E_{\rm I}$} & \SI{50443.0676(12)}{\centi\meter^{-1}} & \multirow{3}{*}{\cite{Kaja_Characterization_2024, Lehec_Laser_2018, Chhetri_Investigation_2018, Aymar_Three-step_1984, Camus_Highly_1980, Camus_Spectre_1969, Hertel_Surface_1968, Zmbov_First_1966, Dresser_Surface_1965}}\\
& & \SI{6.25414330(15)}{\electronvolt} & \\
& & \birge: \num{4.7} & \\ \cline{2-4}
& $E_{\rm I}^{0}$ & \SI{50442.795744}{\centi\meter^{-1}} & \cite{Peper_Spectroscopy_2025} \\ \cline{2-4}
& $E_{\rm I}^{1}$ & \SI{50443.217463}{\centi\meter^{-1}} & \cite{Peper_Spectroscopy_2025} \\
\hline
Ionic hyperfine constant & $A_\text{hfs}(\rm6s\,^2S_{1/2})$ & \SI{12.642 812 124 2(12)}{\giga\hertz} & \cite{Blatt_Precise_1983} \\
\hline
\end{tabular}
\end{table}

%% file: tables/isotopes.tex
\begin{table}[!ht]
\centering
\caption{Relative natural abundance, mass, and nuclear spin of all stable ytterbium isotopes.
All values are from Refs.~\cite{Wang_absolute_2015,Wang_AME_2021}.}
\begin{tabular}{|c|c|c|c|}
\hline
\textbf{Isotope} & \textbf{Relative Natural Abundance} (\unit{\percent}) & \textbf{Atomic Mass} (\unit{u}) & \textbf{Nuclear Spin} \boldmath{$I$} \\ \hline
\ybi{168} & \num{0.12648(22)} & \num{167.93389130(10)} & 0 \\ \hline
\ybi{170} & \num{3.0280(19)} & \num{169.934767243(11)} & 0 \\ \hline
\yb & \num{14.239(5)} & \num{170.936331515(14)} & 1/2 \\ \hline
\ybi{172} & \num{21.789(5)} & \num{171.936386654(15)} & 0 \\ \hline
\ybs & \num{16.119(3)} & \num{172.938216212(12)} & 5/2 \\ \hline
\ybb & \num{31.881(8)} & \num{173.938867546(12)} & 0 \\ \hline
\ybi{176} & \num{12.817(12)} & \num{175.942574706(16)} & 0 \\ \hline
\end{tabular}
\label{tab:isotopes}
\end{table}

%% file: tables/scattering_lengths.tex
\begin{table}[!ht]
\centering
\caption{Scattering lengths between the ytterbium isotopes, given in units of~$a_0$. All values are calculated with a contribution from Ref.~\cite{Kitagawa_Two-color_2008} plus additional references if indicated.}
\begin{tabular}{|l|c|c|c|c|c|c|c|}
\hline
 & \boldmath\ybi{168} & \boldmath\ybi{170} & \boldmath\yb & \boldmath\ybi{172} & \boldmath\ybs & \boldmath\ybb & \boldmath\ybi{176} \\ \hline
\boldmath\ybi{168} & \num{252.86(6)}~\cite{Borkowski_Beyond-Born-Oppenheimer_2017} & \num{117.268(10)}~\cite{Borkowski_Beyond-Born-Oppenheimer_2017} & \num{89.2(17)} & \num{65.117(8)}~\cite{Borkowski_Beyond-Born-Oppenheimer_2017} & \num{38.6(25)} & \num{2.599(24)}~\cite{Borkowski_Beyond-Born-Oppenheimer_2017} & \num{-357.1(6)}~\cite{Borkowski_Beyond-Born-Oppenheimer_2017} \\ \hline
\boldmath\ybi{170} & & \num{63.957(8)}~\cite{Borkowski_Beyond-Born-Oppenheimer_2017, Fukuhara_Bose-Einstein_2007} & \num{36.5(25)} & \num{-1.916(27)}~\cite{Borkowski_Beyond-Born-Oppenheimer_2017} & \num{-81(7)} & \num{-512.7(1.1)}~\cite{Borkowski_Beyond-Born-Oppenheimer_2017} & \num{209.69(5)}~\cite{Borkowski_Beyond-Born-Oppenheimer_2017} \\ \hline
\boldmath\yb & & & \num{-3(4)} & \num{-84(7)} & \num{-580(60)} & \num{429(13)} & \num{141.5(15)} \\ \hline
\boldmath\ybi{172} & & & & \num{-592.3(14)}~\cite{Borkowski_Beyond-Born-Oppenheimer_2017} & \num{418(13)} & \num{200.82(4)}~\cite{Borkowski_Beyond-Born-Oppenheimer_2017} & \num{106.310(11)}~\cite{Borkowski_Beyond-Born-Oppenheimer_2017} \\ \hline
\boldmath\ybs & & & & & \makecell{\num{199(5)} \\ \birge: \num{2.3}}~\cite{ Fukuhara_Degenerate_2007} & \num{138.7(15)} & \num{79.7(19)} \\ \hline
\boldmath\ybb & & & & & & \num{105.037(11)}~\cite{Borkowski_Beyond-Born-Oppenheimer_2017, Enomoto_Determination_2007, Takasu_Spin-Singlet_2003}  & \num{54.573(18)}~\cite{Borkowski_Beyond-Born-Oppenheimer_2017} \\ \hline
\boldmath\ybi{176} & & & & & & & \num{-24.06(9)}~\cite{Borkowski_Beyond-Born-Oppenheimer_2017} \\ \hline
\end{tabular}
\label{tab:scattering_lengths}
\end{table}

%% file: tables/zeeman_and_hyperfine_coefs.tex
\begin{table}[!ht]
\centering
\caption{Land\'e $g_J$ factor, linear Zeeman coefficient, and hyperfine structure constant $A_\text{hfs}$ for selected states of \yb.
Asterisks denote theoretical values.}
\begin{tabular}{|c|c||c|c|c||c|c|}
\hline
\textbf{State} & \boldmath$F$ & \boldmath$g_J$ & \boldmath$\muB g_F / h$ (\unit{\kilo\hertz\per\gauss}) & \textbf{Ref.} & \boldmath$A_\text{hfs}/h$ (\unit{\mega\hertz}) & \textbf{Ref.} \\
\hline
\sS{0} & $1/2$ & \textendash & \num{-0.7505(6)} & \cite{Stone_Table_2019} & 0 & \textendash \\ \hline
\multirow{2}{*}{\sP{1}} & $1/2$ & \multirow{2}{*}{\num{1.035(5)}} & \num{1932(9)} & \multirow{2}{*}{\cite{Baumann_Factors_1968}} & \multirow{2}{*}{\makecell{\num{-214.06(17)}\\ \birge: \num{3.3}}} & \multirow{2}{*}{\cite{Kleinert_Measurement_2016, Wang_Novel_2010, Das_Absolute_2005, Banerjee_Precise_2003, Zinkstok_Hyperfine_2002, Loftus_Optical_2001, Deilamian_Isotope_1993, Berends_Hyperfine_1992, Liening_Level-crossing-spectroscopy_1985, Grundevik_Analysis_1979, Chaiko_Isotope_1966}} \\ \cline{2-2} \cline{4-4}
 & $3/2$ &  & \num{965(5)} &  &  &  \\ \hline
\tP{0} & $1/2$ & \textendash & \num{-1.1496(6)} & \cite{Bettermann_Clock-line_2023, Zhang_Precise_2023, Ono_Antiferromagnetic_2019, McGrew_Atomic_2018} & 0 & \textendash \\ \hline
\multirow{2}{*}{\tP{1}} & $1/2$ & \multirow{2}{*}{\num{1.492820(31)}} & \num{2786.10(6)} & \multirow{2}{*}{\cite{Baumann_Factors_1968, Budick_values_1967}} & \multirow{2}{*}{\makecell{\num{3957.85(5)}\\ \birge: \num{2.6}}} & \multirow{2}{*}{\cite{Jones_Intercombination_2023, Atkinson_Hyperfine_2019, McFerran_inverted_2016, Pandey_Isotope_2009, vanWijngaarden_Measurement_1994, Jin_Systematic_1991, Buchinger_Fast-beam_1982, Clark_Optical_1979, Broadhurst_High_1974, Wandel_Doppelresonanzexperimente_1970, Baumann_Die_1969}} \\ \cline{2-2} \cline{4-4}
 & $3/2$ &  & \num{1392.674(29)} &  &  &  \\ \hline
\multirow{2}{*}{\tP{2}} & $3/2$ & \multirow{2}{*}{\num{1.501}$^*$} & $\sim$ \num{2521} & \multirow{2}{*}{\textendash} & \multirow{2}{*}{\makecell{\num{2680.9(20)}\\ \birge: \num{6.3}}} & \multirow{2}{*}{\cite{Qiao_Frequency_2023, Singh_Hyperfine_2014, Wakui_High-Resolution_2003, Maier_Hyperfine_1991}} \\ \cline{2-2} \cline{4-4}
 & $5/2$ &  & $\sim$ \num{1681} &  &  &  \\ \hline
\multirow{2}{*}{\tD{1}} & $1/2$ & \multirow{2}{*}{\num{0.499}$^*$} & $\sim$ \num{931} &  \multirow{2}{*}{\textendash} & \multirow{2}{*}{\makecell{\num{-2038.30(11)}\\ \birge: \num{1.0}}} & \multirow{2}{*}{\cite{Zhao_Absolute_2024, Ai_Absolute_2023, Ai_Erratum_2024, Beloy_Determination_2012, Bowers_Experimental_1999}} \\ \cline{2-2} \cline{4-4}
 & $3/2$ &  & $\sim$ \num{465} &  &  &  \\ \hline
\multirow{2}{*}{\tD{2}} & $3/2$ & \multirow{2}{*}{\num{1.167}$^*$} & $\sim$ \num{1960} & \multirow{2}{*}{\textendash} & \multirow{2}{*}{\num{1313.8(33)}} & \multirow{2}{*}{\cite{Bowers_Experimental_1999, Topper_unpublished}} \\ \cline{2-2} \cline{4-4}
 & $5/2$ &  & $\sim$ \num{1307} &  &  &  \\ \hline
\multirow{2}{*}{\tS{1}} & $1/2$ & \multirow{2}{*}{\num{2.01(1)}} & \num{3751(19)} & \multirow{2}{*}{\cite{Baumann_Radiative_1985}} & \multirow{2}{*}{\makecell{\num{6838.41(26)}\\ \birge: \num{3.1}}} & \multirow{2}{*}{\cite{Qiao_Frequency_2023, Qiao_Investigation_2023, Zhou_Characterization_2020, Wakui_High-Resolution_2003, Berends_Hyperfine_1992, Schulz_Resonance_1991}} \\ \cline{2-2} \cline{4-4}
 & $3/2$ &  & \num{1875(9)} &  &  &  \\ \hline
\end{tabular}
\label{tab:zeeman_and_hyperfine}
\end{table}

%% file: tables/isotope_shifts_part1.tex
\begin{table}[!ht]
\centering
\caption{Isotope shifts in \unit{\mega\hertz} for selected states, relative to \sS{0} and compared to \ybb.}
\begin{tabular}{|c||c|c|c|}
\hline
\textbf{State} & \boldmath\ybi{168} & \boldmath\ybi{170} & \boldmath\yb \\
\hline
\sP{1} & \makecell{\num{1887.64(23)}\\ \birge: \num{5.0}\\ \cite{Kleinert_Measurement_2016, Wang_Novel_2010, Nizamani_Doppler-free_2010, Das_Absolute_2005, Banerjee_Precise_2003, Zinkstok_Hyperfine_2002,  Deilamian_Isotope_1993}} & \makecell{\num{1187.8(27)}\\ \birge: \num{47.2}\\ \cite{Utreja_Cross_2025, Laupretre_Absolute_2020, Tanabe_Frequency-stabilized_2018, Kleinert_Measurement_2016, Wang_Novel_2010, Nizamani_Doppler-free_2010, Das_Absolute_2005, Banerjee_Precise_2003, Zinkstok_Hyperfine_2002, Loftus_Optical_2001, Deilamian_Isotope_1993, Grundevik_Analysis_1979}} & \makecell{\num{940.8(9)}\\ \birge: \num{26.8} \\ \cite{Laupretre_Absolute_2020, Tanabe_Frequency-stabilized_2018, Kleinert_Measurement_2016, Wang_Novel_2010, Nizamani_Doppler-free_2010, Johanning_Resonance_2010, Das_Absolute_2005, Banerjee_Precise_2003, Loftus_Optical_2001, Deilamian_Isotope_1993, Grundevik_Analysis_1979}} \\ \hline
\tP{0} & \makecell{\num{3626.9710688(29)}\\ \cite{Ono_Observation_2022}} & \makecell{\num{2268.4865926(19)}\\ \cite{Ono_Observation_2022}} & \makecell{\num{1811.2816473(10)}\\ \cite{Ono_Observation_2022, Lemke_Spin-1/2_2009}} \\ \hline
\tP{1} & \makecell{\num{3655.88(20)}\\ \birge: \num{3.6}\\ \cite{Atkinson_Hyperfine_2019, Pandey_Isotope_2009, vanWijngaarden_Measurement_1994, Jin_Systematic_1991, Buchinger_Fast-beam_1982, Clark_Optical_1979, Broadhurst_High_1974, Ross_Isotope_1963}} & \makecell{\num{2286.55(5)}\\ \birge: \num{1.3}\\ \cite{Atkinson_Hyperfine_2019, McFerran_inverted_2016, Pandey_Isotope_2009, vanWijngaarden_Measurement_1994, Jin_Systematic_1991, Buchinger_Fast-beam_1982, Clark_Optical_1979, Broadhurst_High_1974, Chaiko_Isotope_1966, Ross_Isotope_1963}} & \makecell{\num{1826.88(19)}\\ \birge: \num{5.8}\\ \cite{Atkinson_Hyperfine_2019, McFerran_inverted_2016, Pandey_Isotope_2009, vanWijngaarden_Measurement_1994, Jin_Systematic_1991, Buchinger_Fast-beam_1982, Clark_Optical_1979, Broadhurst_High_1974, Chaiko_Isotope_1966, Ross_Isotope_1963}} \\ \hline
\tP{2} & \makecell{\num{3677.546251(7)}\\ \cite{Ishiyama_Excluding_2026}} & \makecell{\num{2300.235712(6)}\\ \cite{Ishiyama_Excluding_2026}} & ? \\ \hline
\tD{1} & ? & \makecell{\num{2542(12)}\\ \birge: \num{8.1}\\ \cite{Zhao_Absolute_2024, Bowers_Experimental_1999}} & \makecell{\num{2038.9(12)}\\ \birge: \num{2.0}\\ \cite{Zhao_Absolute_2024, Bowers_Experimental_1999}} \\ \hline
\tD{2} & ? & \num{2531(5)}~\cite{Bowers_Experimental_1999} & \num{2019(6)}~\cite{Bowers_Experimental_1999} \\ \hline
\tS{1} & \makecell{\num{2100(8)}\\ \birge: \num{3.5}\\ \cite{Wakui_High-Resolution_2003, Buchinger_Fast-beam_1982}} & \makecell{\num{1321.2(1)}\\ \birge: \num{4.1}\\ \cite{Qiao_Investigation_2023, Zhou_Characterization_2020, Yamaguchi_Metastable_2008, Wakui_High-Resolution_2003, Berends_Hyperfine_1992, Buchinger_Fast-beam_1982}} & \makecell{\num{1057.34(15)}\\ \cite{Zhou_Characterization_2020, Yamaguchi_Metastable_2008, Berends_Hyperfine_1992}} \\ \hline
\end{tabular}
\label{tab:isotope_shifts}
\end{table}

%% file: tables/isotope_shifts_part2.tex
\begin{table}[!ht]\ContinuedFloat
\centering
\caption{Isotope shifts in \unit{\mega\hertz} for selected states, relative to \sS{0} and compared to \ybb.}
\begin{tabular}{|c||c|c|c|}
\hline
\textbf{State} & \boldmath\ybi{172} & \boldmath\ybs & \boldmath\ybi{176} \\
\hline
\sP{1} & \makecell{\num{532.95(34)}\\ \birge: \num{8.0}\\ \cite{Utreja_Cross_2025, Laupretre_Absolute_2020, Tanabe_Frequency-stabilized_2018, Kleinert_Measurement_2016, Wang_Novel_2010, Nizamani_Doppler-free_2010, Johanning_Resonance_2010, Das_Absolute_2005, Banerjee_Precise_2003, Zinkstok_Hyperfine_2002, Loftus_Optical_2001, Grundevik_Analysis_1979}} & \makecell{\num{291.51(5)}\\ \cite{Laupretre_Absolute_2020, Wang_Novel_2010, Das_Absolute_2005, Banerjee_Precise_2003, Grundevik_Analysis_1979}} & \makecell{\num{-509.25(7)}\\ \birge: \num{1.7}\\ \cite{Utreja_Cross_2025, Laupretre_Absolute_2020, Tanabe_Frequency-stabilized_2018, Kleinert_Measurement_2016, Wang_Novel_2010, Nizamani_Doppler-free_2010, Johanning_Resonance_2010, Das_Absolute_2005, Banerjee_Precise_2003, Zinkstok_Hyperfine_2002, Loftus_Optical_2001, Grundevik_Analysis_1979}} \\ \hline
\tP{0} & \num{992.7145866(21)}~\cite{Ono_Observation_2022} & \num{551.536050(10)}~\cite{Clivati_Measuring_2016} & \num{-946.9217749(29)}~\cite{Ono_Observation_2022} \\ \hline
\tP{1} & \makecell{\num{1000.73(8)}\\ \birge: \num{3.0}\\ \cite{Atkinson_Hyperfine_2019, McFerran_inverted_2016, Pandey_Isotope_2009, vanWijngaarden_Measurement_1994, Jin_Systematic_1991, Buchinger_Fast-beam_1982, Clark_Optical_1979, Broadhurst_High_1974, Chaiko_Isotope_1966, Ross_Isotope_1963}} & \makecell{\num{556.23(16)}\\ \birge: \num{3.5}\\ \cite{Atkinson_Hyperfine_2019, McFerran_inverted_2016, Pandey_Isotope_2009, vanWijngaarden_Measurement_1994, Jin_Systematic_1991, Buchinger_Fast-beam_1982, Clark_Optical_1979, Broadhurst_High_1974, Chaiko_Isotope_1966, Ross_Isotope_1963}} & \makecell{\num{-954.69(4)}\\ \birge: \num{1.9}\\ \cite{Atkinson_Hyperfine_2019, McFerran_inverted_2016, Pandey_Isotope_2009, vanWijngaarden_Measurement_1994, Jin_Systematic_1991, Buchinger_Fast-beam_1982, Clark_Optical_1979, Broadhurst_High_1974, Chaiko_Isotope_1966, Ross_Isotope_1963}} \\ \hline
\tP{2} & \num{1006.735628(5)}~\cite{Ishiyama_Excluding_2026} & ? & \num{-960.303321(7)}~\cite{Ishiyama_Excluding_2026} \\ \hline
\tD{1} & \makecell{\num{1112(8)}\\ \birge: \num{11.5}\\ \cite{Zhao_Absolute_2024, Bowers_Experimental_1999}} & \makecell{\num{619.4(8)}\\ \birge: \num{1.3}\\ \cite{Zhao_Absolute_2024, Bowers_Experimental_1999}} & \makecell{\num{-1063(6)}\\ \birge: \num{9.5}\\ \cite{Zhao_Absolute_2024, Bowers_Experimental_1999}} \\ \hline
\tD{2} & \num{1100(4)}~\cite{Bowers_Experimental_1999} & \num{609(6)}~\cite{Bowers_Experimental_1999} & \num{-1059(11)}~\cite{Bowers_Experimental_1999} \\ \hline
\tS{1} & \makecell{\num{579.0(19)}\\ \birge: \num{13.1}\\ \cite{Qiao_Investigation_2023, Zhou_Characterization_2020, Yamaguchi_Metastable_2008, Wakui_High-Resolution_2003, Berends_Hyperfine_1992, Buchinger_Fast-beam_1982}} & \makecell{\num{320.97(12)}\\ \cite{Zhou_Characterization_2020, Yamaguchi_Metastable_2008, Berends_Hyperfine_1992}} & \makecell{\num{-543.1(29)}\\ \birge: \num{19.2}\\ \cite{Qiao_Investigation_2023, Zhou_Characterization_2020, Yamaguchi_Metastable_2008, Wakui_High-Resolution_2003, Berends_Hyperfine_1992, Buchinger_Fast-beam_1982}} \\ \hline
\end{tabular}
\end{table}

%% file: tables/magic_and_tuneout.tex
\begin{table}[!ht]
\centering
\caption{Experimentally measured magic and tune‐out wavelengths, supplemented with theoretical polarizability values for \sS{0}.
Asterisks denote a measurement performed with \ybb.}
\label{tab:special_wavelengths}
\begin{tabular}{|l|l|c|c|c|c|c|}
\hline
\textbf{Type} & \textbf{Transition} & $\boldsymbol{\beta}$ (\unit\degree) & $\boldsymbol{\lambda}$ (\unit{\nano\m}) & $\boldsymbol{\omega_0/2\pi}$ (\unit{\tera\hertz}) & \textbf{Ref.} & $\boldsymbol{\alpha_{\rm tot}}$ (a.u.) \\
\hline
\multirow{11}{*}{magic} & \sS{0} $-$ \sP{1} $\ket{F'=3/2, m_{F'}=\pm 3/2}$ & \num{0} &
 $\sim$ \num{532}$^*$ & $\sim$ \num{564}$^*$ & \cite{MuziFalconi_Microsecond-Scale_2025} & \num{258} \\ \cline{2-7}
& \multirow{3}{*}{\sS{0} $-$ \tP{0}} & $\forall\angle$ & $\sim$ \num{397.634} & $\sim$ \num{753.941} & \cite{McGrew_Ytterbium_2020} & \num{-15495} \\ \cline{3-7}
 &  & $\forall\angle$ & \num{459.5960(5)} & \num{652.2956(7)} & \cite{Hohn_State-dependent_2023, Norcia_Midcircuit_2023} & \num{469} \\ \cline{3-7}
 &  & $\forall\angle$ & \num{552.61074(19)}$^*$ & \num{542.50205(19)}$^*$ & \cite{Hohn_State-dependent_2023} & \num{57} \\ \cline{2-7}
 & \sS{0} $-$ \tP{0} $-$ \tP{1} $\ket{F'=3/2, m_{F'}=-1/2}$ & $\sim$ \num{17} & \num{759.3560633(31)} & \makecell{\num{394.7982672(16)}\\ \birge: \num{4.3}}  & \cite{ Lis_Midcircuit_2023}, $^a$ & \num{188} \\ \cline{2-7}
 & \multirow{2}{*}{\sS{0} $-$ \tP{1} $\ket{F'=3/2, m_{F'}=\pm1/2}$} & ? & \num{486.78(1)} & \num{615.868(13)} & \cite{Ma_Universal_2022} & \num{370} \\ \cline{3-7}
 &  & ? & \num{1036.13(1)} & \num{289.3387(28)} & \cite{Li_Fast_2025} & \num{162} \\ \cline{2-7}
 & \multirow{4}{*}{\sS{0} $-$ \tP{1} $\ket{F'=3/2, m_{F'} = +3/2}$} & \num{90} & $\sim$ \num{483} & $\sim$ \num{621} & \cite{Norcia_Midcircuit_2023} & $\sim$ \num{359} \\ \cline{3-7}
 &  & \num{90} & $\sim$ \num{783.8} & $\sim$ \num{382.5} & \cite{Norcia_Iterative_2024} & $\sim$ \num{184} \\ \cline{3-7}
  &  & \num{0}  & \num{1036.12(3)}$^*$ & \num{289.341(8)}$^*$ & \cite{Zheng_Magic_2020} & \num{162} \\ \cline{3-7}
 &  & \num{90} & \num{1035.83(3)}$^*$ & \num{289.422(8)}$^*$ & \cite{Zheng_Magic_2020} & \num{162} \\ \cline{3-7}
\hline
\multirow{2}{*}{tune‐out} & \sS{0} & $\forall\angle$ & \num{553.2936(5)}$^*$ & \num{541.8325(5)}$^*$ & \cite{Hohn_State-dependent_2023} & \num{0} \\
\cline{2-7}
 & \tP{0} & $\forall\angle$ & \num{576.613(10)}$^*$ & \num{519.920(9)}$^*$ & \cite{Hohn_Determining_2026} & \num{277} \\
\hline
\end{tabular}\\
\begin{flushleft}
\hspace{5mm}$^a$: \cite{Aeppli_Atomic_2025, Bothwell_Lattice_2025, Kim_Absolute_2021, Pizzocaro_Absolute_2020, Kobayashi_Demonstration_2020, Luo_Absolute_2020, Nemitz_Modeling_2019, Gao_Systematic_2018, McGrew_Atomic_2018, Kobayashi_Uncertainty_2018, Brown_Hyperpolarizability_2017, Pizzocaro_Absolute_2017, Kim_Improved_2017, Nemitz_Frequency_2016, Park_Absolute_2013, Lemke_Spin-1/2_2009, Kohno_One-Dimensional_2009}
\end{flushleft}
\end{table}

%% file: tables/1S0_1P1.tex
\begin{table}[!ht]
\centering
\caption{\sS{0} $\to$ \sP{1} transition (Zeeman slower, blue MOT).}
\begin{tabular}{|l|c|c|c|}
\hline
\textbf{Parameter} & \textbf{Symbol} & \textbf{Value} & \textbf{Ref.} \\ \hline
Frequency & $\omega_{0}$ & \makecell{$2\pi\times$\SI{751.5274711(8)}{\tera\hertz}\\ \birge: \num{6.3}} & \cite{Laupretre_Absolute_2020, Tanabe_Frequency-stabilized_2018, Kleinert_Measurement_2016, Nizamani_Doppler-free_2010} \\ \hline
Transition Energy & $\hbar\omega_0$ & \SI{3.1080678856(32)}{\electronvolt} & \\ \hline
Wavelength (vacuum) & $\lambda$ & \SI{398.9108443(4)}{\nano\meter} & \\ \hline
Wavelength (air) & $\lambda_{\rm air}$ & \SI{398.802070(4)}{\nano\meter} & \\ \hline
Wavenumber (vacuum) & $k_L/2\pi$ & \SI{25068.258091(26)}{\centi\meter^{-1}} & \\ \hline
Lifetime excited state & $\tau$ & \makecell{\SI{5.464(11)}{\nano\second}\\ \birge: \num{2.3}} & \cite{Takasu_Photoassociation_2004, Blagoev_1978, Rambow_Radiative_1976, Andersen_Lifetimes_1975, Lange_1970, Komarovskii_Oscillator_1969, Baumann_Lifetimes_1966} \\ \hline
Decay rate & $\Gamma$ & \SI{183.0(4)e6}{\second^{-1}} & \\ \hline
Natural Line Width & $\Delta\nu_{\rm FWHM}$ & $2\pi\times$\SI{29.13(6)}{\mega\hertz} & \\ \hline
Branching Ratio & $\beta$ & $1 - \num{6(4)e-8}$ & \cite{Porsev_Electric-dipole_1999} \\ \hline
Oscillator strength & $f$ & \num{1.3098(27)} & \\ \hline
Recoil velocity & $v_r$ & $\sim$ \SI{5.852}{\milli\meter\per\second} & \\ \hline
Recoil energy & $\omega_r$ & $\sim$ $2\pi\times$\SI{7.335}{\kilo\hertz} & \\ \hline
Recoil temperature & $T_r$ & $\sim$ \SI{704}{\nano\kelvin} & \\ \hline
Doppler shift ($v=v_r$) & $\Delta\omega_d$ & $\sim$ $2\pi\times$\SI{14.670}{\kilo\hertz} & \\ \hline
Doppler Temperature  & $T_\mathrm{D}$ & \SI{699.0(15)}{\micro\kelvin} & \\ \hline
Reduced E1 Matrix Element & $\braket{J \| er \| J'}_R$ & \num{4.147(4)} $ea_0$ & \\ \hline
Saturation Intensity & $\Isat$ & \SI{59.97(12)}{\milli\watt\per\centi\meter\squared} & \\ \hline
Resonant Cross Section & $\sigma_0$ & $\sim$ \SI{7.598e-14}{\meter\squared} & \\ \hline
\makecell[l]{
  Rabi Frequency \\
\footnotesize $P=\SI{1}{\milli\watt}$, $D=\SI{1}{\milli\meter}$
} & $\Omega_0^{\ket{\frac{1}{2}, \frac{1}{2}} \text{\scalebox{0.7}[1]{$\to$}} \ket{\frac{3}{2}, \frac{3}{2}}}$ & $2\pi\times$ \SI{42.44(4)}{\mega\hertz} &  \\ \hline
\end{tabular}
\label{tab:blue_MOT}
\end{table}

%% file: tables/1S0_3P0.tex
\begin{table}[!ht]
\centering
\caption{\sS{0} $\to$ \tP{0} transition (yellow clock).}
\begin{tabular}{|l|c|c|c|}
\hline
\textbf{Parameter} & \textbf{Symbol} & \textbf{Value} & \textbf{Ref.} \\ \hline
Frequency & $\omega_{0}$ & $2\pi\times$\SI{518.29583659086363(10)}{\tera\hertz} & \cite{Margolis_CIPM_2024} \\ \hline
Transition Energy & $\hbar\omega_0$ & \SI{2.1434993488389615(4)}{\electronvolt} & \\ \hline
Wavelength (vacuum) & $\lambda$ & \SI{578.41957591616250(11)}{\nano\meter} & \\ \hline
Wavelength (air) & $\lambda_{\rm air}$ & \SI{578.264935(6)}{\nano\meter} & \\ \hline
Wavenumber (vacuum) & $k_L/2\pi$ & \SI{17288.488177740069(3)}{\centi\meter^{-1}} & \\ \hline
Lifetime excited state & $\tau$ & \makecell{\SI{21.0(18)}{\second}\\ \birge: \num{1.3}} & \cite{Siegel_Excited-Band_2024, Xu_Measurement_2014} \\ \hline
Decay rate & $\Gamma$ & \SI{0.048(4)}{\second^{-1}} & \\ \hline
Natural Line Width & $\Delta\nu_{\rm FWHM}$ & $2\pi\times$\SI{7.6(6)}{\milli\hertz} & \\ \hline
Recoil velocity & $v_r$ & $\sim$ \SI{4.036}{\milli\meter\per\second} & \\ \hline
Recoil energy & $\omega_r$ & $\sim$ $2\pi\times$\SI{3.489}{\kilo\hertz} & \\ \hline
Recoil temperature & $T_r$ & $\sim$ \SI{335}{\nano\kelvin} & \\ \hline
Doppler shift ($v=v_r$) & $\Delta\omega_d$ & $\sim$ $2\pi\times$\SI{6.977}{\kilo\hertz} & \\ \hline
\end{tabular}
\label{tab:clock}
\end{table}

%% file: tables/1S0_3P1.tex
\begin{table}[!ht]
\centering
\caption{\sS{0} $\to$ \tP{1} transition (green MOT).}
\begin{tabular}{|l|c|c|c|}
\hline
\textbf{Parameter} & \textbf{Symbol} & \textbf{Value} & \textbf{Ref.} \\ \hline
Frequency & $\omega_{0}$ & \makecell{$2\pi\times$\SI{539.38842684(12)}{\tera\hertz}\\ \birge: \num{2.9}} & \cite{Jones_Intercombination_2023, Atkinson_Hyperfine_2019, Pandey_Isotope_2009} \\ \hline
Transition Energy & $\hbar\omega_0$ & \SI{2.2307312930(5)}{\electronvolt} & \\ \hline
Wavelength (vacuum) & $\lambda$ & \SI{555.80068663(13)}{\nano\meter} & \\ \hline
Wavelength (air) & $\lambda_{\rm air}$ & \SI{555.651877(6)}{\nano\meter} & \\ \hline
Wavenumber (vacuum) & $k_L/2\pi$ & \SI{17992.061256(4)}{\centi\meter^{-1}} & \\ \hline
Lifetime excited state & $\tau$ & \makecell{\SI{872.6(19)}{\nano\second}\\ \birge: \num{1.2}} & \cite{Beloy_Determination_2012, Bowers_Experimental_1996, Golub_Radiative_1988, Gustavsson_Lifetime_1979, Blagoev_1978, Rambow_Radiative_1976, Burshtein_Lifetimes_1974, Gornik_Quantum_1972, Budick_Hyperfine-Structure_1970, Komarovskii_Oscillator_1969, Baumann_Lifetimes_1966} \\ \hline
Decay rate & $\Gamma$ & \SI{1.1460(25)e6}{\second^{-1}} & \\ \hline
Natural Line Width & $\Delta\nu_{\rm FWHM}$ & $2\pi\times$\SI{182.4(4)}{\kilo\hertz} & \\ \hline
Oscillator strength & $f$ & \num{0.015923(35)} & \\ \hline
Recoil velocity & $v_r$ & $\sim$ \SI{4.200}{\milli\meter\per\second} & \\ \hline
Recoil energy & $\omega_r$ & $\sim$ $2\pi\times$\SI{3.778}{\kilo\hertz} & \\ \hline
Recoil temperature & $T_r$ & $\sim$ \SI{363}{\nano\kelvin} & \\ \hline
Doppler shift ($v=v_r$) & $\Delta\omega_d$ & $\sim$ $2\pi\times$\SI{7.557}{\kilo\hertz} & \\ \hline
Doppler Temperature  & $T_\mathrm{D}$ & \SI{4.377(10)}{\micro\kelvin} & \\ \hline
Reduced E1 Matrix Element & $\braket{J \| er \| J'}_R$ & \num{0.5398(6)} $ea_0$ & \\ \hline
Saturation Intensity & $\Isat$ & \SI{138.85(30)}{\micro\watt\per\centi\meter\squared} & \\ \hline
Resonant Cross Section & $\sigma_0$ & $\sim$ \SI{1.475e-13}{\meter\squared} & \\ \hline
\makecell[l]{
  Rabi Frequency \\
\footnotesize $P=\SI{1}{\milli\watt}$, $D=\SI{1}{\milli\meter}$
} & $\Omega_0^{\ket{\frac{1}{2}, \frac{1}{2}} \text{\scalebox{0.7}[1]{$\to$}} \ket{\frac{3}{2}, \frac{3}{2}}}$ & $2\pi\times$ \SI{5.523(6)}{\mega\hertz} &  \\ \hline
\end{tabular}
\label{tab:green_MOT}
\end{table}

%% file: tables/1S0_3P2.tex
\begin{table}[!ht]
\centering
\caption{\sS{0} $\to$ \tP{2} transition (green clock).}
\begin{tabular}{|l|c|c|c|}
\hline
\textbf{Parameter} & \textbf{Symbol} & \textbf{Value} & \textbf{Ref.} \\ \hline
Frequency & $\omega_{0}$ & $2\pi\times$\SI{590.90433(4)}{\tera\hertz} & \cite{Yamaguchi_Metastable_2008} \\ \hline
Transition Energy & $\hbar\omega_0$ & \SI{2.44378393(17)}{\electronvolt} & \\ \hline
Wavelength (vacuum) & $\lambda$ & \SI{507.34517(4)}{\nano\meter} & \\ \hline
Wavelength (air) & $\lambda_{\rm air}$ & \SI{507.20882(4)}{\nano\meter} & \\ \hline
Wavenumber (vacuum) & $k_L/2\pi$ & \SI{19710.4467(14)}{\centi\meter^{-1}} & \\ \hline
Lifetime excited state & $\tau$ & $\sim$ \SI{8.9}{\second} & ~\cite{Mishra_Radiative_2001} \\ \hline
Decay rate & $\Gamma$ & $\sim$ \SI{0.112}{\second^{-1}} & \\ \hline
Natural Line Width & $\Delta\nu_{\rm FWHM}$ & $\sim2\pi\times$\SI{17.9}{\milli\hertz} & \\ \hline
Recoil velocity & $v_r$ & $\sim$ \SI{4.601}{\milli\meter\per\second} & \\ \hline
Recoil energy & $\omega_r$ & $\sim$ $2\pi\times$\SI{4.535}{\kilo\hertz} & \\ \hline
Recoil temperature & $T_r$ & $\sim$ \SI{435}{\nano\kelvin} & \\ \hline
Doppler shift ($v=v_r$) & $\Delta\omega_d$ & $\sim$ $2\pi\times$\SI{9.069}{\kilo\hertz} & \\ \hline
\end{tabular}
\label{tab:magnetic_clock}
\end{table}

%% file: tables/3P0_3D1.tex
\begin{table}[!ht]
\centering
\caption{\tP{0} $\to$ \tD{1} transition (infrared repumper for yellow clock).}
\begin{tabular}{|l|c|c|c|}
\hline
\textbf{Parameter} & \textbf{Symbol} & \textbf{Value} & \textbf{Ref.} \\ \hline
Frequency & $\omega_{0}$ & $2\pi\times$\SI{215.87014457(9)}{\tera\hertz} & \cite{Zhao_Absolute_2024, Ai_Absolute_2023, Ai_Erratum_2024} \\ \hline
Transition Energy & $\hbar\omega_0$ & \SI{0.8927671836(4)}{\electronvolt} & \\ \hline
Wavelength (vacuum) & $\lambda$ & \SI{1388.7629464(6)}{\nano\meter} & \\ \hline
Wavelength (air) & $\lambda_{\rm air}$ & \SI{1388.396915(14)}{\nano\meter} & \\ \hline
Wavenumber (vacuum) & $k_L/2\pi$ & \SI{7200.6529453(29)}{\centi\meter^{-1}} & \\ \hline
Lifetime excited state & $\tau$ & \makecell{\SI{332(11)}{\nano\second}\\ \birge: \num{1.6}} & \cite{Beloy_Determination_2012, Bowers_Experimental_1996} \\ \hline
Decay rate & $\Gamma$ & \SI{3.01(10)e6}{\second^{-1}} & \\ \hline
Natural Line Width & $\Delta\nu_{\rm FWHM}$ & $2\pi\times$\SI{479(16)}{\kilo\hertz} & \\ \hline
Branching Ratios & \makecell{$\beta_{\rm 3D1 \to 3P2}$ \\ $\beta_{\rm 3D1 \to 3P1}$ \\ $\beta_{\rm 3D1 \to 3P0}$} & \makecell{\num{0.0099(39)} \\ \num{0.352(27)} \\ \num{0.638(27)}}& \cite{Porsev_Electric-dipole_1999} \\ \hline
Partial decay rate & $\Gamma_{\rm partial}$ & \SI{1.92(10)e6}{\second^{-1}} & \\ \hline
Recoil velocity & $v_r$ & $\sim$ \SI{1.681}{\milli\meter\per\second} & \\ \hline
Recoil energy & $\omega_r$ & $\sim$ $2\pi\times$\SI{605.2}{\hertz} & \\ \hline
Recoil temperature & $T_r$ & $\sim$ \SI{58.1}{\nano\kelvin} & \\ \hline
Doppler shift ($v=v_r$) & $\Delta\omega_d$ & $\sim$ $2\pi\times$\SI{1.210}{\kilo\hertz} & \\ \hline
Doppler Temperature  & $T_\mathrm{D}$ & \SI{11.5(4)}{\micro\kelvin} & \\ \hline
Reduced E1 Matrix Element & $\braket{J \| er \| J'}_R$ & \num{2.76(7)} $ea_0$ & \\ \hline
\makecell[l]{
  Rabi Frequency \\
\footnotesize $P=\SI{1}{\milli\watt}$, $D=\SI{1}{\milli\meter}$
} & $\Omega_0^{\ket{\frac{1}{2}, \frac{1}{2}} \text{\scalebox{0.7}[1]{$\to$}} \ket{\frac{3}{2}, \frac{3}{2}}}$ & $2\pi\times$ \SI{28.3(8)}{\mega\hertz} &  \\ \hline
\end{tabular}
\label{tab:clock_repumper}
\end{table}

%% file: tables/3P0_3S1.tex
\begin{table}[!ht]
\centering
\caption{\tP{0} $\to$ \tS{1} transition (red repumper for yellow clock).}
\begin{tabular}{|l|c|c|c|}
\hline
\textbf{Parameter} & \textbf{Symbol} & \textbf{Value} & \textbf{Ref.} \\ \hline
Frequency & $\omega_{0}$ & $2\pi\times$\SI{461.86817672(22)}{\tera\hertz} & \cite{Qiao_Frequency_2023, Zhou_Characterization_2020} \\ \hline
Transition Energy & $\hbar\omega_0$ & \SI{1.9101332987(9)}{\electronvolt} & \\ \hline
Wavelength (vacuum) & $\lambda$ & \SI{649.08662928(31)}{\nano\meter} & \\ \hline
Wavelength (air) & $\lambda_{\rm air}$ & \SI{648.913718(6)}{\nano\meter} & \\ \hline
Wavenumber (vacuum) & $k_L/2\pi$ & \SI{15406.264047(7)}{\centi\meter^{-1}} & \\ \hline
Lifetime excited state & $\tau$ & \makecell{\SI{13.8(17)}{\nano\second}\\ \birge: \num{1.4}} & \cite{Baumann_Radiative_1985, Lange_1970} \\ \hline
Decay rate & $\Gamma$ & \SI{7.2(9)e7}{\second^{-1}} & \\ \hline
Natural Line Width & $\Delta\nu_{\rm FWHM}$ & $2\pi\times$\SI{11.5(14)}{\mega\hertz} & \\ \hline
Branching Ratios & \makecell{$\beta_{\rm 3S1 \to 3P2}$ \\ $\beta_{\rm 3S1 \to 3P1}$ \\ $\beta_{\rm 3S1 \to 3P0}$ \\ $\beta_{\rm 3S1 \to 1P1}$} & \makecell{\num{0.508(26)} \\ \num{0.360(25)} \\ \num{0.130(13)} \\ \num{0.0021(9)}} & \cite{Porsev_Electric-dipole_1999} \\ \hline
Partial decay rate & $\Gamma_{\rm partial}$ & \SI{9.4(15)e6}{\per\second} & \\ \hline
Recoil velocity & $v_r$ & $\sim$ \SI{3.596}{\milli\meter\per\second} & \\ \hline
Recoil energy & $\omega_r$ & $\sim$ $2\pi\times$\SI{2.770}{\kilo\hertz} & \\ \hline
Recoil temperature & $T_r$ & $\sim$ \SI{266}{\nano\kelvin} & \\ \hline
Doppler shift ($v=v_r$) & $\Delta\omega_d$ & $\sim$ $2\pi\times$\SI{5.541}{\kilo\hertz} & \\ \hline
Doppler Temperature  & $T_\mathrm{D}$ & \SI{277(33)}{\micro\kelvin} & \\ \hline
Reduced E1 Matrix Element & $\braket{J \| er \| J'}_R$ & \num{1.96(15)} $ea_0$ & \\ \hline
\makecell[l]{
  Rabi Frequency \\
\footnotesize $P=\SI{1}{\milli\watt}$, $D=\SI{1}{\milli\meter}$
} & $\Omega_0^{\ket{\frac{1}{2}, \frac{1}{2}} \text{\scalebox{0.7}[1]{$\to$}} \ket{\frac{3}{2}, \frac{3}{2}}}$ & $2\pi\times$ \SI{20.0(16)}{\mega\hertz} &  \\ \hline
\end{tabular}
\label{tab:clock_alt_repumper}
\end{table}

%% file: tables/3P2_3S1.tex
\begin{table}[!ht]
\centering
\caption{\tP{2} $\to$ \tS{1} transition (red repumper for green clock).}
\begin{tabular}{|l|c|c|c|}
\hline
\textbf{Parameter} & \textbf{Symbol} & \textbf{Value} & \textbf{Ref.} \\ \hline
Frequency & $\omega_{0}$ & $2\pi\times$\SI{389.2598337(6)}{\tera\hertz} & \cite{Qiao_Frequency_2023} \\ \hline
Transition Energy & $\hbar\omega_0$ & \SI{1.6098493199(24)}{\electronvolt} & \\ \hline
Wavelength (vacuum) & $\lambda$ & \SI{770.160284(11)}{\nano\meter} & \\ \hline
Wavelength (air) & $\lambda_{\rm air}$ & \SI{769.955919(8)}{\nano\meter} & \\ \hline
Wavenumber (vacuum) & $k_L/2\pi$ & \SI{12984.310422(19)}{\centi\meter^{-1}} & \\ \hline
Lifetime excited state & $\tau$ & \makecell{\SI{13.8(17)}{\nano\second}\\ \birge: \num{1.4}} & \cite{Baumann_Radiative_1985, Lange_1970} \\ \hline
Decay rate & $\Gamma$ & \SI{7.2(9)e7}{\second^{-1}} & \\ \hline
Natural Line Width & $\Delta\nu_{\rm FWHM}$ & $2\pi\times$\SI{11.5(14)}{\mega\hertz} & \\ \hline
Branching Ratios & \makecell{$\beta_{\rm 3S1 \to 3P2}$ \\ $\beta_{\rm 3S1 \to 3P1}$ \\ $\beta_{\rm 3S1 \to 3P0}$ \\ $\beta_{\rm 3S1 \to 1P1}$} & \makecell{\num{0.508(26)} \\ \num{0.360(25)} \\ \num{0.130(13)} \\ \num{0.0021(9)}} & \cite{Porsev_Electric-dipole_1999} \\ \hline
Partial decay rate & $\Gamma_{\rm partial}$ & \SI{3.7(5)e7}{\second^{-1}} & \\ \hline
Recoil velocity & $v_r$ & $\sim$ \SI{3.031}{\milli\meter\per\second} & \\ \hline
Recoil energy & $\omega_r$ & $\sim$ $2\pi\times$\SI{1.968}{\kilo\hertz} & \\ \hline
Recoil temperature & $T_r$ & $\sim$ \SI{189}{\nano\kelvin} & \\ \hline
Doppler shift ($v=v_r$) & $\Delta\omega_d$ & $\sim$ $2\pi\times$\SI{3.936}{\kilo\hertz} & \\ \hline
Doppler Temperature  & $T_\mathrm{D}$ & \SI{277(33)}{\micro\kelvin} & \\ \hline
Reduced E1 Matrix Element & $\braket{J \| er \| J'}_R$ & \num{4.99(32)} $ea_0$ & \\ \hline
\makecell[l]{
  Rabi Frequency \\
\footnotesize $P=\SI{1}{\milli\watt}$, $D=\SI{1}{\milli\meter}$
} & $\Omega_0^{\ket{\frac{5}{2}, \frac{5}{2}} \text{\scalebox{0.7}[1]{$\to$}} \ket{\frac{3}{2}, \frac{3}{2}}}$ & $2\pi\times$ \SI{39.5(26)}{\mega\hertz} &  \\ \hline
\end{tabular}
\label{tab:magnetic_clock_repumper}
\end{table}

%% file: 171yb_line_data.bib
@article{Ai_Absolute_2023,
	author = {Ai, Di and Jin, Taoyun and Zhang, Tao and Luo, Limeng and Liu, Luhua and Zhou, Min and Xu, Xinye},
	doi = {10.1103/PhysRevA.107.063107},
	journal = {Phys. Rev. A},
	month = {6},
	number = {6},
	pages = {063107},
	title = {{Absolute frequency measurement of the $6s6p$ $^3$P$_0$ $\rightarrow$ $5d6s$ $^3$D$_1$ transition based on ultracold ytterbium atoms}},
	url = {https://link.aps.org/doi/10.1103/PhysRevA.107.063107},
	volume = {107},
	year = {2023}
}

@article{Ai_Erratum_2024,
	author = {Ai, Di and Jin, Taoyun and Zhang, Tao and Luo, Limeng and Liu, Luhua and Zhou, Min and Xu, Xinye},
	doi = {10.1103/PhysRevA.109.039902},
	journal = {Phys. Rev. A},
	month = {3},
	number = {3},
	pages = {039902},
	title = {{Erratum: Absolute frequency measurement of the $6s6p$ $^3$P$_0$ $\rightarrow$ $5d6s$ $^3$D$_1$ transition based on ultracold ytterbium atoms [Phys. Rev. A 107, 063107 (2023)]}},
	url = {https://link.aps.org/doi/10.1103/PhysRevA.109.039902},
	volume = {109},
	year = {2024}
}

@article{Alcock_Vapour_1984,
	author = {Alcock, C. B. and Itkin, V. P. and Horrigan, M. K.},
	doi = {10.1179/cmq.1984.23.3.309},
	journal = {Can. Met. Q.},
	month = {7},
	number = {3},
	pages = {309--313},
	title = {{Vapour Pressure Equations for the Metallic Elements: 298-2500 K}},
	url = {http://www.tandfonline.com/doi/full/10.1179/cmq.1984.23.3.309},
	volume = {23},
	year = {1984}
}

@article{Andersen_Lifetimes_1975,
	author = {Andersen, T. and Poulsen, O. and Ramanujam, P. S. and Petkov, A. Petrakiev},
	doi = {10.1007/BF00153206},
	journal = {Sol. Phys.},
	month = {10},
	number = {2},
	pages = {257--267},
	title = {{Lifetimes of some excited states in the rare earths: La ii, Ce ii, Pr ii, Nd ii, Sm ii, Yb i, Yb ii, and Lu ii}},
	url = {http://link.springer.com/10.1007/BF00153206},
	volume = {44},
	year = {1975}
}

@article{Angeli_Table_2013,
	author = {Angeli, I. and Marinova, K. P.},
	doi = {10.1016/j.adt.2011.12.006},
	journal = {Data Nucl. Data Tables},
	number = {1},
	pages = {69--95},
	title = {{Table of experimental nuclear ground state charge radii: An update}},
	url = {http://dx.doi.org/10.1016/j.adt.2011.12.006},
	volume = {99},
	year = {2013}
}

@book{Arblaster_Selected_2018,
	address = {Materials Park, Ohio},
	author = {Arblaster, John W.},
	isbn = {9781627081542},
	publisher = {ASM International},
	title = {{Selected Values of the Crystallographic Properties of the Elements}},
	url = {https://www.asminternational.org/results/-/journal_content/56/39867022/PUBLICATION/?srsltid=AfmBOopb92FTcO385HjIdrp9yOoBwe_ikPX5sJTYxzqcv7WU2S05ZJ3U},
	year = {2018}
}

@article{Atkinson_Hyperfine_2019,
	author = {Atkinson, Peggy E. and Schelfhout, Jesse S. and McFerran, John J.},
	doi = {10.1103/PhysRevA.100.042505},
	journal = {Phys. Rev. A},
	month = {10},
	number = {4},
	pages = {042505},
	title = {{Hyperfine constants and line separations for the $^1$S$_0$-$^3$P$_1$ intercombination line in neutral ytterbium with sub-Doppler resolution}},
	url = {https://link.aps.org/doi/10.1103/PhysRevA.100.042505},
	volume = {100},
	year = {2019}
}

@article{Aymar_Three-step_1984,
	author = {Aymar, M. and Champeau, R. J. and Delsart, C. and Robaux, O.},
	doi = {10.1088/0022-3700/17/18/006},
	journal = {J. Phys. B Mol. Phys.},
	month = {9},
	number = {18},
	pages = {3645--3661},
	title = {{Three-step laser spectroscopy and multichannel quantum defect analysis of odd-parity Rydberg states of neutral ytterbium}},
	url = {https://iopscience.iop.org/article/10.1088/0022-3700/17/18/006},
	volume = {17},
	year = {1984}
}

@article{Banerjee_Precise_2003,
	author = {Banerjee, A. and Rapol, U. D. and Das, D. and Krishna, A. and Natarajan, V.},
	doi = {10.1209/epl/i2003-00543-x},
	journal = {Eur. Lett.},
	month = {8},
	number = {3},
	pages = {340--346},
	title = {{Precise measurements of UV atomic lines: Hyperfine structure and isotope shifts in the 398.8 nm line of Yb}},
	url = {https://iopscience.iop.org/article/10.1209/epl/i2003-00543-x},
	volume = {63},
	year = {2003}
}

@article{Barber_Direct_2006,
	author = {Barber, Z. and Hoyt, C. and Oates, C. and Hollberg, L. and Taichenachev, A. and Yudin, V.},
	doi = {10.1103/PhysRevLett.96.083002},
	journal = {Phys. Rev. Lett.},
	month = {3},
	number = {8},
	pages = {083002},
	title = {{Direct Excitation of the Forbidden Clock Transition in Neutral $^{174}$Yb Atoms Confined to an Optical Lattice}},
	url = {https://link.aps.org/doi/10.1103/PhysRevLett.96.083002},
	volume = {96},
	year = {2006}
}

@article{Baumann_Die_1969,
	author = {Baumann, M and Liening, H. and Wandel, G.},
	doi = {10.1007/BF01392176},
	journal = {Zeitschrift f{\"{u}}r Phys. A Hadron. Nucl.},
	month = {6},
	number = {3},
	pages = {245--256},
	title = {{Die Hyperfeinstruktur des $6s6p$ $^3$P$_1$-Zustandes von $^{171}$Yb und $^{173}$Yb}},
	url = {http://link.springer.com/10.1007/BF01392176},
	volume = {221},
	year = {1969}
}

@article{Baumann_Factors_1968,
	author = {Baumann, M. and Wandel, G.},
	doi = {10.1016/0375-9601(68)90200-4},
	journal = {Phys. Lett. A},
	month = {11},
	number = {3},
	pages = {200--201},
	title = {{$g_J$ Factors of the $6s6p$ $^3$P$_1$ and $6s6p$ $^1$P$_1$ states of ytterbium}},
	url = {https://linkinghub.elsevier.com/retrieve/pii/0375960168902004},
	volume = {28},
	year = {1968}
}

@article{Baumann_Lifetimes_1966,
	author = {Baumann, M. and Wandel, G.},
	doi = {10.1016/0031-9163(66)90614-7},
	journal = {Phys. Lett.},
	month = {8},
	number = {3},
	pages = {283--285},
	title = {{Lifetimes of the excited states $(6s6p)$ $^1$P$_1$ and $(6s6p)$ $^3$P$_1$ of ytterbium}},
	url = {https://linkinghub.elsevier.com/retrieve/pii/0031916366906147},
	volume = {22},
	year = {1966}
}

@article{Baumann_Radiative_1985,
	author = {Baumann, M. and Braun, M. and Gaiser, A. and Liening, H.},
	doi = {10.1088/0022-3700/18/17/001},
	journal = {J. Phys. B Mol. Phys.},
	month = {9},
	number = {17},
	pages = {L601--L604},
	title = {{Radiative lifetimes and $g_J$ factors of low-lying even-parity levels in the Yb I spectrum}},
	url = {https://iopscience.iop.org/article/10.1088/0022-3700/18/17/001},
	volume = {18},
	year = {1985}
}

@article{Beloy_Determination_2012,
	author = {Beloy, K. and Sherman, J. A. and Lemke, N. D. and Hinkley, N. and Oates, C. W. and Ludlow, A. D.},
	doi = {10.1103/PhysRevA.86.051404},
	journal = {Phys. Rev. A},
	month = {11},
	number = {5},
	pages = {051404},
	title = {{Determination of the $5d6s$ $^3$D$_1$ state lifetime and blackbody-radiation clock shift in Yb}},
	url = {https://link.aps.org/doi/10.1103/PhysRevA.86.051404},
	volume = {86},
	year = {2012}
}

@article{Berends_Hyperfine_1992,
	author = {Berends, R. W. and Maleki, L.},
	doi = {10.1364/JOSAB.9.000332},
	journal = {J. Opt. Soc. Am. B},
	month = {3},
	number = {3},
	pages = {332},
	title = {{Hyperfine structure and isotope shifts of transitions in neutral and singly ionized ytterbium}},
	url = {https://opg.optica.org/abstract.cfm?URI=josab-9-3-332},
	volume = {9},
	year = {1992}
}

@article{Berengut_Probing_2018,
	author = {Berengut, Julian C. and Budker, Dmitry and Delaunay, C{\'{e}}dric and Flambaum, Victor V. and Frugiuele, Claudia and Fuchs, Elina and Grojean, Christophe and Harnik, Roni and Ozeri, Roee and Perez, Gilad and Soreq, Yotam},
	doi = {10.1103/PhysRevLett.120.091801},
	journal = {Phys. Rev. Lett.},
	month = {2},
	number = {9},
	pages = {091801},
	title = {{Probing New Long-Range Interactions by Isotope Shift Spectroscopy}},
	url = {https://link.aps.org/doi/10.1103/PhysRevLett.120.091801},
	volume = {120},
	year = {2018}
}

@book{Bethe_Quantum_1957,
	address = {Heidelberg},
	author = {Bethe, Hans A. and Salpeter, Edwin E.},
	doi = {10.1007/978-3-662-12869-5},
	isbn = {978-3-662-12871-8},
	publisher = {Springer-Verlag Berlin},
	title = {{Quantum Mechanics of One- and Two-Electron Atoms}},
	url = {http://link.springer.com/10.1007/978-3-662-12869-5},
	year = {1957}
}

@article{Bettermann_Clock-line_2023,
	author = {Bettermann, O. and {Darkwah Oppong}, N. and Pasqualetti, G. and Riegger, L. and Bloch, I. and F{\"{o}}lling, S.},
	doi = {10.1103/PhysRevA.108.L041302},
	journal = {Phys. Rev. A},
	month = {10},
	number = {4},
	pages = {L041302},
	title = {{Clock-line photoassociation of strongly bound dimers in a magic-wavelength lattice}},
	url = {https://link.aps.org/doi/10.1103/PhysRevA.108.L041302},
	volume = {108},
	year = {2023}
}

@article{Birge_Calculation_1932,
	author = {Birge, Raymond T.},
	doi = {10.1103/PhysRev.40.207},
	journal = {Phys. Rev.},
	month = {4},
	number = {2},
	pages = {207--227},
	title = {{The Calculation of Errors by the Method of Least Squares}},
	url = {https://link.aps.org/doi/10.1103/PhysRev.40.207},
	volume = {40},
	year = {1932}
}

@article{Blagoev_Lifetimes_1994,
	author = {Blagoev, K.B. and Komarovskii, V.A.},
	doi = {10.1006/adnd.1994.1001},
	journal = {Data Nucl. Data Tables},
	month = {1},
	number = {1},
	pages = {1--40},
	title = {{Lifetimes of Levels of Neutral and Singly Ionized Lanthanide Atoms}},
	url = {https://linkinghub.elsevier.com/retrieve/pii/S0092640X84710011},
	volume = {56},
	year = {1994}
}

@article{Blatt_Precise_1983,
	author = {Blatt, R. and Schnatz, H. and Werth, G.},
	doi = {10.1007/BF01412156},
	journal = {Zeitschrift f{\"{u}}r Phys. A Atoms Nucl.},
	month = {9},
	number = {3},
	pages = {143--147},
	title = {{Precise Determination of the $^{171}$Yb$^+$ Ground State Hyperfine Separation}},
	url = {http://link.springer.com/10.1007/BF01412156},
	volume = {312},
	year = {1983}
}

@article{Borkowski_Beyond-Born-Oppenheimer_2017,
	author = {Borkowski, Mateusz and Buchachenko, Alexei A. and Ciury{\l}o, Roman and Julienne, Paul S. and Yamada, Hirotaka and Kikuchi, Yuu and Takahashi, Kakeru and Takasu, Yosuke and Takahashi, Yoshiro},
	doi = {10.1103/PhysRevA.96.063405},
	journal = {Phys. Rev. A},
	month = {12},
	number = {6},
	pages = {063405},
	title = {{Beyond-Born-Oppenheimer effects in sub-kHz-precision photoassociation spectroscopy of ytterbium atoms}},
	url = {https://link.aps.org/doi/10.1103/PhysRevA.96.063405},
	volume = {96},
	year = {2017}
}

@article{Bothwell_Lattice_2025,
	author = {Bothwell, Tobias and Hunt, Benjamin D. and Siegel, Jacob L. and Hassan, Youssef S. and Grogan, Tanner and Kobayashi, Takumi and Gibble, Kurt and Porsev, Sergey G. and Safronova, Marianna S. and Brown, Roger C. and Beloy, Kyle and Ludlow, Andrew D.},
	doi = {10.1103/PhysRevLett.134.033201},
	journal = {Phys. Rev. Lett.},
	month = {1},
	number = {3},
	pages = {033201},
	title = {{Lattice Light Shift Evaluations in a Dual-Ensemble Yb Optical Lattice Clock}},
	url = {https://doi.org/10.1103/PhysRevLett.134.033201 https://link.aps.org/doi/10.1103/PhysRevLett.134.033201},
	volume = {134},
	year = {2025}
}

@article{Bouganne_Clock_2017,
	author = {Bouganne, R. and Aguilera, M. Bosch and Dareau, A. and Soave, E. and Beugnon, J. and Gerbier, F.},
	doi = {10.1088/1367-2630/aa8c45},
	journal = {New J. Phys.},
	month = {11},
	number = {11},
	pages = {113006},
	title = {{Clock spectroscopy of interacting bosons in deep optical lattices}},
	url = {https://iopscience.iop.org/article/10.1088/1367-2630/aa8c45},
	volume = {19},
	year = {2017}
}

@article{Bowers_Experimental_1996,
	author = {Bowers, C. J. and Budker, D. and Commins, E. D. and DeMille, D. and Freedman, S. J. and Nguyen, A.-T. and Shang, S.-Q. and Zolotorev, M.},
	doi = {10.1103/PhysRevA.53.3103},
	journal = {Phys. Rev. A},
	month = {5},
	number = {5},
	pages = {3103--3109},
	title = {{Experimental investigation of excited-state lifetimes in atomic ytterbium}},
	url = {https://link.aps.org/doi/10.1103/PhysRevA.53.3103},
	volume = {53},
	year = {1996}
}

@article{Bowers_Experimental_1999,
	author = {Bowers, C. J. and Budker, D. and Freedman, S. J. and Gwinner, G. and Stalnaker, J. E. and DeMille, D.},
	doi = {10.1103/PhysRevA.59.3513},
	journal = {Phys. Rev. A},
	month = {5},
	number = {5},
	pages = {3513--3526},
	title = {{Experimental investigation of the $6s^2$ $^1$S$_0$$\rightarrow$$5d6s$ $^3$D$_{1,2}$ forbidden transitions in atomic ytterbium}},
	url = {https://link.aps.org/doi/10.1103/PhysRevA.59.3513},
	volume = {59},
	year = {1999}
}

@article{Boyd_Nuclear_2007,
	author = {Boyd, Martin M. and Zelevinsky, Tanya and Ludlow, Andrew D. and Blatt, Sebastian and Zanon-Willette, Thomas and Foreman, Seth M. and Ye, Jun},
	doi = {10.1103/PhysRevA.76.022510},
	journal = {Phys. Rev. A},
	month = {8},
	number = {2},
	pages = {022510},
	title = {{Nuclear spin effects in optical lattice clocks}},
	url = {https://link.aps.org/doi/10.1103/PhysRevA.76.022510},
	volume = {76},
	year = {2007}
}

@article{Braaten_Universality_2006,
	author = {Braaten, Eric and Hammer, H. W.},
	doi = {10.1016/j.physrep.2006.03.001},
	journal = {Phys. Rep.},
	number = {5-6},
	pages = {259--390},
	title = {{Universality in few-body systems with large scattering length}},
	volume = {428},
	year = {2006}
}

@article{Breit_Hyperfine_1933,
	author = {Breit, G. and Wills, Lawrence A.},
	doi = {10.1103/PhysRev.44.470},
	journal = {Phys. Rev.},
	month = {9},
	number = {6},
	pages = {470--490},
	title = {{Hyperfine Structure in Intermediate Coupling}},
	url = {https://link.aps.org/doi/10.1103/PhysRev.44.470},
	volume = {44},
	year = {1933}
}

@book{Brink_Angular_1968,
	author = {Brink, D. M. and Satchler, G. R.},
	title = {{Angular Momentum}},
	series = {Oxford Library of the Physical Sciences},
	publisher = {Clarendon Press},
	address = {Oxford},
	year = {1968}
}

@article{Broadhurst_High_1974,
	author = {Broadhurst, J. H. and Cage, M. E. and Clark, D. L. and Greenlees, G. W. and Griffith, J A R and Isaak, G. R.},
	doi = {10.1088/0022-3700/7/18/001},
	journal = {J. Phys. B Mol. Phys.},
	month = {12},
	number = {18},
	pages = {L513--L517},
	title = {{High resolution measurements of isotope shifts and hyperfine splittings for ytterbium using a CW tunable laser}},
	url = {https://iopscience.iop.org/article/10.1088/0022-3700/7/18/001},
	volume = {7},
	year = {1974}
}

@article{Brown_Hyperpolarizability_2017,
	author = {Brown, R. C. and Phillips, N. B. and Beloy, K. and McGrew, W. F. and Schioppo, M. and Fasano, R. J. and Milani, G. and Zhang, X. and Hinkley, N. and Leopardi, H. and Yoon, T. H. and Nicolodi, D. and Fortier, T. M. and Ludlow, A. D.},
	doi = {10.1103/PhysRevLett.119.253001},
	journal = {Phys. Rev. Lett.},
	month = {12},
	number = {25},
	pages = {253001},
	title = {{Hyperpolarizability and Operational Magic Wavelength in an Optical Lattice Clock}},
	url = {https://link.aps.org/doi/10.1103/PhysRevLett.119.253001},
	volume = {119},
	year = {2017}
}

@article{Buchinger_Fast-beam_1982,
	author = {Buchinger, F. and Mueller, A.C. and Schinzler, B. and Wendt, K. and Ekstr{\"{o}}m, C. and Klempt, W. and Neugart, R.},
	doi = {10.1016/0167-5087(82)90390-8},
	journal = {Nucl. Instruments Methods Phys. Res.},
	month = {11},
	number = {1-2},
	pages = {159--165},
	title = {{Fast-beam laser spectroscopy on metastable atoms applied to neutron-deficient ytterbium isotopes}},
	url = {https://linkinghub.elsevier.com/retrieve/pii/0167508782903908},
	volume = {202},
	year = {1982}
}

@article{Buck_New_1981,
	author = {Buck, Arden L.},
	doi = {10.1175/1520-0450(1981)020<1527:NEFCVP>2.0.CO;2},
	journal = {J. Appl. Meteorol.},
	month = {12},
	number = {12},
	pages = {1527--1532},
	title = {{New Equations for Computing Vapor Pressure and Enhancement Factor}},
	url = {http://journals.ametsoc.org/doi/10.1175/1520-0450(1981)020{\%}3C1527:NEFCVP{\%}3E2.0.CO;2},
	volume = {20},
	year = {1981}
}

@article{Budick_Hyperfine-Structure_1970,
	author = {Budick, B. and Snir, J.},
	doi = {10.1103/PhysRevA.1.545},
	journal = {Phys. Rev. A},
	month = {3},
	number = {3},
	pages = {545--551},
	title = {{Hyperfine-Structure Anomalies of Stable Ytterbium Isotopes}},
	url = {https://link.aps.org/doi/10.1103/PhysRevA.1.545},
	volume = {1},
	year = {1970}
}

@article{Budick_values_1967,
	author = {Budick, B. and Snir, J.},
	doi = {10.1016/0375-9601(67)91031-6},
	journal = {Phys. Lett. A},
	month = {6},
	number = {12},
	pages = {689--690},
	title = {{$g_J$ values of excited states of the ytterbium atom}},
	url = {https://linkinghub.elsevier.com/retrieve/pii/0375960167910316},
	volume = {24},
	year = {1967}
}

@article{Camus_Highly_1980,
	author = {Camus, P and Debarre, A and Morillon, C},
	doi = {10.1088/0022-3700/13/6/015},
	journal = {J. Phys. B Mol. Phys.},
	month = {3},
	number = {6},
	pages = {1073--1087},
	title = {{Highly excited levels of neutral ytterbium. I. Two-photon and two-step spectroscopy of even spectra}},
	url = {https://iopscience.iop.org/article/10.1088/0022-3700/13/6/015},
	volume = {13},
	year = {1980}
}

@article{Camus_Spectre_1969,
	author = {Camus, Pierre and Tomkins, Frank S.},
	doi = {10.1051/jphys:01969003007054500},
	journal = {J. Phys.},
	number = {7},
	pages = {545--550},
	title = {{Spectre d'absorption de Yb I}},
	url = {http://www.edpsciences.org/10.1051/jphys:01969003007054500},
	volume = {30},
	year = {1969}
}

@article{Chhetri_Investigation_2018,
	author = {Chhetri, P. and Moodley, C.S. and Raeder, S. and Block, M. and Giacoppo, F. and G{\"{o}}tz, S. and He{\ss}berger, F.P. and Eibach, M. and Kaleja, O. and Laatiaoui, M. and Mistry, A.K. and Murb{\"{o}}ck, T. and Walther, Th},
	doi = {10.5506/APhysPolB.49.599},
	journal = {Acta Phys. Pol. B},
	number = {3},
	pages = {599},
	title = {{Investigation of the First Ionization Potential of Ytterbium in Argon Buffer Gas}},
	url = {http://www.actaphys.uj.edu.pl/findarticle?series=Reg{\&}vol=49{\&}page=599},
	volume = {49},
	year = {2018}
}

@article{Chin_Feshbach_2010,
	author = {Chin, Cheng and Grimm, Rudolf and Julienne, Paul and Tiesinga, Eite},
	doi = {10.1103/RevModPhys.82.1225},
	journal = {Rev. Mod. Phys.},
	number = {2},
	pages = {1225--1286},
	title = {{Feshbach resonances in ultracold gases}},
	volume = {82},
	year = {2010}
}

@article{Clark_Optical_1979,
	author = {Clark, D. L. and Cage, M. E. and Lewis, D. A. and Greenlees, G. W.},
	doi = {10.1103/PhysRevA.20.239},
	journal = {Phys. Rev. A},
	month = {7},
	number = {1},
	pages = {239--253},
	title = {{Optical isotopic shifts and hyperfine splittings for Yb}},
	url = {https://link.aps.org/doi/10.1103/PhysRevA.20.239},
	volume = {20},
	year = {1979}
}

@article{Clivati_Coherent_2022,
	author = {Clivati, C. and Pizzocaro, M. and Bertacco, E.K. and Condio, S. and Costanzo, G.A. and Donadello, S. and Goti, I. and Gozzelino, M. and Levi, F. and Mura, A. and Risaro, M. and Calonico, D. and T{\o}nnes, M. and Pointard, B. and Mazouth-Laurol, M. and {Le Targat}, R. and Abgrall, M. and Lours, M. and {Le Goff}, H. and Lorini, L. and Pottie, P.-E. and Cantin, E. and Lopez, O. and Chardonnet, C. and Amy-Klein, A.},
	doi = {10.1103/PhysRevApplied.18.054009},
	journal = {Phys. Rev. Appl.},
	month = {11},
	number = {5},
	pages = {054009},
	title = {{Coherent Optical-Fiber Link Across Italy and France}},
	url = {https://doi.org/10.1103/PhysRevApplied.18.054009 https://link.aps.org/doi/10.1103/PhysRevApplied.18.054009},
	volume = {18},
	year = {2022}
}

@article{Clivati_Measuring_2016,
	author = {Clivati, Cecilia and Cappellini, Giacomo and Livi, Lorenzo F. and Poggiali, Francesco and de Cumis, Mario Siciliani and Mancini, Marco and Pagano, Guido and Frittelli, Matteo and Mura, Alberto and Costanzo, Giovanni A. and Levi, Filippo and Calonico, Davide and Fallani, Leonardo and Catani, Jacopo and Inguscio, Massimo},
	doi = {10.1364/OE.24.011865},
	journal = {Opt. Express},
	month = {5},
	number = {11},
	pages = {11865},
	title = {{Measuring absolute frequencies beyond the GPS limit via long-haul optical frequency dissemination}},
	url = {https://opg.optica.org/abstract.cfm?URI=oe-24-11-11865},
	volume = {24},
	year = {2016}
}

@book{Corney_Atomic_1977,
	address = {Oxford, United Kingdom},
	author = {Corney, Alan},
	doi = {10.1093/acprof:oso/9780199211456.001.0001},
	isbn = {9780199211456},
	month = {10},
	publisher = {Clarendon Press, Oxford University Press},
	title = {{Atomic and Laser Spectroscopy}},
	url = {https://academic.oup.com/book/8068},
	year = {1977}
}

@article{Das_Absolute_2005,
	author = {Das, Dipankar and Barthwal, Sachin and Banerjee, Ayan and Natarajan, Vasant},
	doi = {10.1103/PhysRevA.72.032506},
	journal = {Phys. Rev. A},
	month = {9},
	number = {3},
	pages = {032506},
	title = {{Absolute frequency measurements in Yb with $0.08$ ppb uncertainty: Isotope shifts and hyperfine structure in the 399-nm $^1$S$_0\rightarrow^1$P$_1$ line}},
	url = {https://link.aps.org/doi/10.1103/PhysRevA.72.032506},
	volume = {72},
	year = {2005}
}

@article{Deilamian_Isotope_1993,
	author = {Deilamian, K. and Gillaspy, J. D. and Kelleher, D. E.},
	doi = {10.1364/JOSAB.10.000789},
	journal = {J. Opt. Soc. Am. B},
	month = {5},
	number = {5},
	pages = {789},
	title = {{Isotope shifts and hyperfine splittings of the 398.8-nm Yb I line}},
	url = {https://opg.optica.org/abstract.cfm?URI=josab-10-5-789},
	volume = {10},
	year = {1993}
}

@book{Demtroder_Laser_2008,
	address = {Heidelberg, Germany},
	author = {Demtr{\"{o}}der, Wolfgang},
	doi = {10.1007/978-3-540-73418-5},
	isbn = {978-3-540-73415-4},
	publisher = {Springer-Verlag Berlin},
	title = {{Laser Spectroscopy}},
	url = {http://link.springer.com/10.1007/978-3-540-73418-5},
	year = {2008}
}

@article{Dorscher_Creation_2013,
	author = {D{\"{o}}rscher, S{\"{o}}ren and Thobe, Alexander and Hundt, Bastian and Kochanke, Andr{\'{e}} and {Le Targat}, Rodolphe and Windpassinger, Patrick and Becker, Christoph and Sengstock, Klaus},
	doi = {10.1063/1.4802682},
	journal = {Rev. Sci. Instrum.},
	month = {4},
	number = {4},
	title = {{Creation of quantum-degenerate gases of ytterbium in a compact 2D-/3D-magneto-optical trap setup}},
	url = {https://pubs.aip.org/rsi/article/84/4/043109/357904/Creation-of-quantum-degenerate-gases-of-ytterbium},
	volume = {84},
	year = {2013}
}

@article{Dresser_Surface_1965,
	author = {Dresser, M. J. and Hudson, D. E.},
	doi = {10.1103/PhysRev.137.A673},
	journal = {Phys. Rev.},
	month = {1},
	number = {2A},
	pages = {A673--A682},
	title = {{Surface Ionization of Some Rare Earths on Tungsten}},
	url = {https://link.aps.org/doi/10.1103/PhysRev.137.A673},
	volume = {137},
	year = {1965}
}

@article{Dzuba_Dynamic_2010,
	author = {Dzuba, V. A. and Derevianko, A.},
	doi = {10.1088/0953-4075/43/7/074011},
	journal = {J. Phys. B Mol. Opt. Phys.},
	month = {4},
	number = {7},
	pages = {074011},
	title = {{Dynamic polarizabilities and related properties of clock states of the ytterbium atom}},
	url = {https://iopscience.iop.org/article/10.1088/0953-4075/43/7/074011},
	volume = {43},
	year = {2010}
}

@article{Dzuba_Testing_2018,
	author = {Dzuba, V. A. and Flambaum, V. V. and Schiller, S.},
	doi = {10.1103/PhysRevA.98.022501},
	journal = {Phys. Rev. A},
	month = {8},
	number = {2},
	pages = {022501},
	title = {{Testing physics beyond the standard model through additional clock transitions in neutral ytterbium}},
	url = {https://link.aps.org/doi/10.1103/PhysRevA.98.022501},
	volume = {98},
	year = {2018}
}

@article{Edlen_Refractive_1966,
	author = {Edl{\'{e}}n, Bengt},
	doi = {10.1088/0026-1394/2/2/002},
	journal = {Metrologia},
	month = {4},
	number = {2},
	pages = {71--80},
	title = {{The Refractive Index of Air}},
	url = {https://iopscience.iop.org/article/10.1088/0026-1394/2/2/002},
	volume = {2},
	year = {1966}
}

@article{Enomoto_Comparison_2016,
	author = {Enomoto, Katsunari and Hizawa, Nagisa and Suzuki, Takahiro and Kobayashi, Kaori and Moriwaki, Yoshiki},
	doi = {10.1007/s00340-016-6400-5},
	journal = {Appl. Phys. B},
	month = {5},
	number = {5},
	pages = {126},
	title = {{Comparison of resonance frequencies of major atomic lines in 398-423 nm}},
	url = { http://link.springer.com/10.1007/s00340-016-6400-5},
	volume = {122},
	year = {2016}
}

@article{Enomoto_Determination_2007,
	author = {Enomoto, K. and Kitagawa, M. and Kasa, K. and Tojo, S. and Takahashi, Y.},
	doi = {10.1103/PhysRevLett.98.203201},
	journal = {Phys. Rev. Lett.},
	month = {5},
	number = {20},
	pages = {203201},
	title = {{Determination of the $s$-Wave Scattering Length and the $C_6$ van der Waals Coefficient of $^{174}$Yb via Photoassociation Spectroscopy}},
	url = {https://link.aps.org/doi/10.1103/PhysRevLett.98.203201},
	volume = {98},
	year = {2007}
}

@article{Figueroa_Precision_2022,
	author = {Figueroa, N. L. and Berengut, J. C. and Dzuba, V. A. and Flambaum, V. V. and Budker, D. and Antypas, D.},
	doi = {10.1103/PhysRevLett.128.073001},
	journal = {Phys. Rev. Lett.},
	month = {2},
	number = {7},
	pages = {073001},
	title = {{Precision Determination of Isotope Shifts in Ytterbium and Implications for New Physics}},
	url = {https://doi.org/10.1103/PhysRevLett.128.073001 https://link.aps.org/doi/10.1103/PhysRevLett.128.073001},
	volume = {128},
	year = {2022}
}

@book{Foot_Atomic_2004,
	author = {Foot, Christopher J.},
	title = {{Atomic Physics}},
	publisher = {Oxford University Press},
	address = {Oxford, United Kingdom},
	year = {2004}
}

@article{Franchi_State-dependent_2017,
	author = {Franchi, L. and Livi, L. F. and Cappellini, G. and Binella, G. and Inguscio, M. and Catani, J. and Fallani, L.},
	doi = {10.1088/1367-2630/aa8fb4},
	journal = {New J. Phys.},
	month = {11},
	number = {10},
	pages = {103037},
	title = {{State-dependent interactions in ultracold $^{174}$Yb probed by optical clock spectroscopy}},
	url = {https://iopscience.iop.org/article/10.1088/1367-2630/aa8fb4},
	volume = {19},
	year = {2017}
}

@article{Frugiuele_Constraining_2017,
	author = {Frugiuele, Claudia and Fuchs, Elina and Perez, Gilad and Schlaffer, Matthias},
	doi = {10.1103/PhysRevD.96.015011},
	journal = {Phys. Rev. D},
	month = {7},
	number = {1},
	pages = {015011},
	title = {{Constraining new physics models with isotope shift spectroscopy}},
	url = {http://link.aps.org/doi/10.1103/PhysRevD.96.015011},
	volume = {96},
	year = {2017}
}

@article{Fry_Discovery_2013,
	author = {Fry, C. and Thoennessen, M.},
	doi = {10.1016/j.adt.2012.05.004},
	journal = {Data Nucl. Data Tables},
	month = {9},
	number = {5},
	pages = {520--544},
	title = {{Discovery of dysprosium, holmium, erbium, thulium, and ytterbium isotopes}},
	url = {https://linkinghub.elsevier.com/retrieve/pii/S0092640X13000247},
	volume = {99},
	year = {2013}
}

@article{Fukuhara_All-optical_2009,
	author = {Fukuhara, Takeshi and Sugawa, Seiji and Takasu, Yosuke and Takahashi, Yoshiro},
	doi = {10.1103/PhysRevA.79.021601},
	journal = {Phys. Rev. A},
	month = {2},
	number = {2},
	pages = {021601},
	title = {{All-optical formation of quantum degenerate mixtures}},
	url = {https://link.aps.org/doi/10.1103/PhysRevA.79.021601},
	volume = {79},
	year = {2009}
}

@article{Fukuhara_Bose-Einstein_2007,
	author = {Fukuhara, Takeshi and Sugawa, Seiji and Takahashi, Yoshiro},
	doi = {10.1103/PhysRevA.76.051604},
	journal = {Phys. Rev. A},
	month = {11},
	number = {5},
	pages = {051604},
	title = {{Bose-Einstein condensation of an ytterbium isotope}},
	url = {https://link.aps.org/doi/10.1103/PhysRevA.76.051604},
	volume = {76},
	year = {2007}
}

@article{Fukuhara_Degenerate_2007,
	author = {Fukuhara, Takeshi and Takasu, Yosuke and Kumakura, Mitsutaka and Takahashi, Yoshiro},
	doi = {10.1103/PhysRevLett.98.030401},
	journal = {Phys. Rev. Lett.},
	month = {1},
	number = {3},
	pages = {030401},
	title = {{Degenerate Fermi Gases of Ytterbium}},
	url = {https://link.aps.org/doi/10.1103/PhysRevLett.98.030401},
	volume = {98},
	year = {2007}
}

@article{Fukuhara_Quantum_2007,
	author = {Fukuhara, T. and Takasu, Y. and Sugawa, S. and Takahashi, Y.},
	doi = {10.1007/s10909-007-9411-0},
	journal = {J. Low Temp. Phys.},
	month = {6},
	number = {3-4},
	pages = {441--445},
	title = {{Quantum Degenerate Fermi Gases of Ytterbium Atoms}},
	url = {http://link.springer.com/10.1007/s10909-007-9411-0},
	volume = {148},
	year = {2007}
}

@article{Fuller_Nuclear_1976,
	author = {Fuller, Gladys H.},
	doi = {10.1063/1.555544},
	journal = {J. Phys. Chem. Ref. Data},
	month = {10},
	number = {4},
	pages = {835--1092},
	title = {{Nuclear Spins and Moments}},
	url = {https://pubs.aip.org/jpr/article/5/4/835/242157/Nuclear-Spins-and-Moments},
	volume = {5},
	year = {1976}
}

@article{Gao_Systematic_2018,
	author = {Gao, Qi and Zhou, Min and Han, Chengyin and Li, Shangyan and Zhang, Shuang and Yao, Yuan and Li, Bo and Qiao, Hao and Ai, Di and Lou, Ge and Zhang, Mengya and Jiang, Yanyi and Bi, Zhiyi and Ma, Longsheng and Xu, Xinye},
	doi = {10.1038/s41598-018-26365-w},
	journal = {Sci. Rep.},
	month = {5},
	number = {1},
	pages = {8022},
	title = {{Systematic evaluation of a $^{171}$Yb optical clock by synchronous comparison between two lattice systems}},
	url = {https://www.nature.com/articles/s41598-018-26365-w},
	volume = {8},
	year = {2018}
}

@article{Golub_Radiative_1988,
	author = {Golub, J. E. and Bai, Y. S. and Mossberg, T. W.},
	doi = {10.1103/PhysRevA.37.119},
	journal = {Phys. Rev. A},
	month = {1},
	number = {1},
	pages = {119--124},
	title = {{Radiative and dynamical properties of homogeneously prepared atomic samples}},
	url = {https://link.aps.org/doi/10.1103/PhysRevA.37.119},
	volume = {37},
	year = {1988}
}

@article{Gornik_Quantum_1972,
	author = {Gornik, W. and Kaiser, D. and Lange, W. and Luther, J. and Schulz, H.-H.},
	doi = {10.1016/0030-4018(72)90147-2},
	journal = {Opt. Commun.},
	month = {12},
	number = {4},
	pages = {327--328},
	title = {{Quantum beats under pulsed dye laser excitation}},
	url = {https://linkinghub.elsevier.com/retrieve/pii/0030401872901472},
	volume = {6},
	year = {1972}
}

@article{Gossard_Ytterbium_1964,
	author = {Gossard, A. C. and Jaccarino, V. and Wernick, J. H.},
	doi = {10.1103/PhysRev.133.A881},
	journal = {Phys. Rev.},
	month = {2},
	number = {3A},
	pages = {A881--A884},
	title = {{Ytterbium NMR: Yb$^{171}$ Nuclear Moment and Yb Metal Knight Shift}},
	url = {https://link.aps.org/doi/10.1103/PhysRev.133.A881},
	volume = {133},
	year = {1964}
}

@article{Goti_Absolute_2023,
	author = {Goti, Irene and Condio, Stefano and Clivati, Cecilia and Risaro, Matias and Gozzelino, Michele and Costanzo, Giovanni A. and Levi, Filippo and Calonico, Davide and Pizzocaro, Marco},
	doi = {10.1088/1681-7575/accbc5},
	journal = {Metrologia},
	number = {3},
	title = {{Absolute frequency measurement of a Yb optical clock at the limit of the Cs fountain}},
	volume = {60},
	year = {2023}
}

@article{Gribakin_Calculation_1993,
	author = {Gribakin, G. F. and Flambaum, V. V.},
	doi = {10.1103/PhysRevA.48.546},
	journal = {Phys. Rev. A},
	month = {7},
	number = {1},
	pages = {546--553},
	title = {{Calculation of the scattering length in atomic collisions using the semiclassical approximation}},
	url = {https://link.aps.org/doi/10.1103/PhysRevA.48.546},
	volume = {48},
	year = {1993}
}

@incollection{Grimm_Optical_2000,
	author = {Grimm, Rudolf and Weidem{\"u}ller, Matthias and Ovchinnikov, Yurii B.},
	title = {{Optical Dipole Traps for Neutral Atoms}},
	booktitle = {{Advances in Atomic, Molecular, and Optical Physics}},
	volume = {42},
	pages = {95--170},
	publisher = {Academic Press},
	year = {2000},
	doi = {10.1016/S1049-250X(08)60186-X},
	url = {https://doi.org/10.1016/S1049-250X(08)60186-X}
}

@article{Grundevik_Analysis_1979,
	author = {Grundevik, P. and Gustavsson, M. and Ros{\'e}n, A. and Rydberg, S.},
	doi = {10.1007/BF01546427},
	journal = {Zeitschrift f{\"{u}}r Phys. A Atoms Nucl.},
	month = {12},
	number = {4},
	pages = {307--310},
	title = {{Analysis of the isotope shifts and hyperfine structure in the 3 988 \r{A} ($6s6p$ $^1$P$_1\leftrightarrow6s^2$ $^1$S$_0$) YbI line}},
	url = {http://link.springer.com/10.1007/BF01546427},
	volume = {292},
	year = {1979}
}

@article{Gustavsson_Lifetime_1979,
	author = {Gustavsson, M. and Lundberg, H. and Nilsson, L. and Svanberg, S.},
	doi = {10.1364/JOSA.69.000984},
	journal = {J. Opt. Soc. Am.},
	month = {7},
	number = {7},
	pages = {984},
	title = {{Lifetime measurements for excited states of rare-earth atoms using pulse modulation of a cw dye-laser beam}},
	url = {https://opg.optica.org/abstract.cfm?URI=josa-69-7-984},
	volume = {69},
	year = {1979}
}

@article{Guttridge_Direct_2016,
	author = {Guttridge, A. and Hopkins, S. A. and Kemp, S. L. and Boddy, D. and Freytag, R. and Jones, M P A and Tarbutt, M. R. and Hinds, E. A. and Cornish, S. L.},
	doi = {10.1088/0953-4075/49/14/145006},
	journal = {J. Phys. B Mol. Opt. Phys.},
	month = {7},
	number = {14},
	pages = {145006},
	title = {{Direct loading of a large Yb MOT on the $^1$S$_0$ $\rightarrow$ $^3$P$_1$ transition}},
	url = {https://iopscience.iop.org/article/10.1088/0953-4075/49/14/145006},
	volume = {49},
	year = {2016}
}

@book{Haynes_CRC_2016,
	address = {Boca Raton, Florida},
	author = {Haynes, W. M.},
	doi = {10.1201/9781315380476},
	isbn = {9781315380476},
	month = {6},
	publisher = {CRC Press},
	title = {{CRC Handbook of Chemistry and Physics}},
	url = {http://dx.doi.org/10.1016/j.bpj.2015.06.056{\%}0Ahttps://academic.oup.com/bioinformatics/article-abstract/34/13/2201/4852827{\%}0Ainternal-pdf://semisupervised-3254828305/semisupervised.ppt{\%}0Ahttp://dx.doi.org/10.1016/j.str.2013.02.005{\%}0Ahttp://dx.doi.org/10.10 https://www.taylorfrancis.com/books/9781498754293},
	year = {2016}
}

@article{Hertel_Surface_1968,
	author = {Hertel, G. R.},
	doi = {10.1063/1.1669015},
	journal = {J. Chem. Phys.},
	month = {3},
	number = {5},
	pages = {2053--2058},
	title = {{Surface Ionization. III. The First Ionization Potentials of the Lanthanides}},
	url = {https://pubs.aip.org/jcp/article/48/5/2053/772650/Surface-Ionization-III-The-First-Ionization},
	volume = {48},
	year = {1968}
}

@article{Hohn_Determining_2026,
	author = {H{\"{o}}hn, Tim O and Villela, Ren{\'{e}} A and Zu, Er and Bezzo, Leonardo and Kroeze, Ronen M and Aidelsburger, Monika},
	doi = {10.1103/32q9-j82c},
	journal = {PRX Quantum},
	month = {1},
	number = {1},
	pages = {010303},
	title = {{Determining the $^3$P$_0$ Excited-State Tune-Out Wavelength of $^{174}$Yb in a Triple-Magic Lattice}},
	url = {https://link.aps.org/doi/10.1103/32q9-j82c},
	volume = {7},
	year = {2026}
}

@article{Hohn_State-dependent_2023,
	author = {H{\"{o}}hn, Tim O and Staub, Etienne and Brochier, Guillaume and {Darkwah Oppong}, Nelson and Aidelsburger, Monika},
	doi = {10.1103/PhysRevA.108.053325},
	journal = {Phys. Rev. A},
	month = {11},
	number = {5},
	pages = {053325},
	title = {{State-dependent potentials for the $^1$S$_0$ and $^3$P$_0$ clock states of neutral ytterbium atoms}},
	url = {https://link.aps.org/doi/10.1103/PhysRevA.108.053325},
	volume = {108},
	year = {2023}
}

@phdthesis{Hohn_State-dependent_2024,
	author = {H{\"{o}}hn, Tim O},
	doi = {10.5282/edoc.34662},
	school = {Ludwig-Maximilians-Universit{\"{a}}t M{\"{u}}nchen},
	title = {{State-dependent potentials and clock ground-state cooling in an ytterbium quantum simulator}},
	year = {2024}
}

@article{Honda_Magneto-optical_1999,
	author = {Honda, K. and Takahashi, Y. and Kuwamoto, T. and Fujimoto, M. and Toyoda, K. and Ishikawa, K. and Yabuzaki, T.},
	doi = {10.1103/PhysRevA.59.R934},
	journal = {Phys. Rev. A},
	month = {2},
	number = {2},
	pages = {R934--R937},
	title = {{Magneto-optical trapping of Yb atoms and a limit on the branching ratio of the $^1$P$_1$ state}},
	url = {https://link.aps.org/doi/10.1103/PhysRevA.59.R934},
	volume = {59},
	year = {1999}
}

@article{Hoyt_Observation_2005,
	author = {Hoyt, C. and Barber, Z. and Oates, C. and Fortier, T. and Diddams, S. and Hollberg, L.},
	doi = {10.1103/PhysRevLett.95.083003},
	journal = {Phys. Rev. Lett.},
	month = {8},
	number = {8},
	pages = {083003},
	title = {{Observation and Absolute Frequency Measurements of the $^1$S$_0$-$^3$P$_0$ Optical Clock Transition in Neutral Ytterbium}},
	url = {https://link.aps.org/doi/10.1103/PhysRevLett.95.083003},
	volume = {95},
	year = {2005}
}

@article{Ishiyama_Observation_2023,
	author = {Ishiyama, Taiki and Ono, Koki and Takano, Tetsushi and Sunaga, Ayaki and Takahashi, Yoshiro},
	doi = {10.1103/PhysRevLett.130.153402},
	journal = {Phys. Rev. Lett.},
	month = {4},
	number = {15},
	pages = {153402},
	title = {{Observation of an Inner-Shell Orbital Clock Transition in Neutral Ytterbium Atoms}},
	url = {https://doi.org/10.1103/PhysRevLett.130.153402 https://link.aps.org/doi/10.1103/PhysRevLett.130.153402},
	volume = {130},
	year = {2023}
}

@article{Jenkins_Ytterbium_2022,
	author = {Jenkins, Alec and Lis, Joanna W. and Senoo, Aruku and McGrew, William F. and Kaufman, Adam M.},
	doi = {10.1103/PhysRevX.12.021027},
	journal = {Phys. Rev. X},
	number = {2},
	pages = {21027},
	title = {{Ytterbium Nuclear-Spin Qubits in an Optical Tweezer Array}},
	url = {https://doi.org/10.1103/PhysRevX.12.021027},
	volume = {12},
	year = {2022}
}

@article{Jin_Systematic_1991,
	author = {Jin, Wei-Guo and Horiguchi, Takayoshi and Wakasugi, Masanori and Hasegawa, Takashi and Yang, Wei},
	doi = {10.1143/JPSJ.60.2896},
	journal = {J. Phys. Soc. Jpn.},
	month = {9},
	number = {9},
	pages = {2896--2906},
	title = {{Systematic Study of Isotope Shiftsand Hyperfine Structures in Yb I by Atomic-Beam Laser Spectroscopy}},
	url = {http://journals.jps.jp/doi/10.1143/JPSJ.60.2896},
	volume = {60},
	year = {1991}
}

@article{Jones_Intercombination_2023,
	author = {Jones, Daniel M. and van Kann, Frank and McFerran, John J.},
	doi = {10.1364/AO.488653},
	journal = {Appl. Opt.},
	month = {5},
	number = {15},
	pages = {3932},
	title = {{Intercombination line frequencies in $^{171}$Yb validated with the clock transition}},
	url = {https://opg.optica.org/abstract.cfm?URI=ao-62-15-3932},
	volume = {62},
	year = {2023}
}

@article{Judd_Theory_1961,
	author = {Judd, B. R. and Lindgren, I.},
	doi = {10.1103/PhysRev.122.1802},
	journal = {Phys. Rev.},
	month = {6},
	number = {6},
	pages = {1802--1812},
	title = {{Theory of Zeeman Effect in the Ground Multiplets of Rare-Earth Atoms}},
	url = {https://link.aps.org/doi/10.1103/PhysRev.122.1802},
	volume = {122},
	year = {1961}
}

@article{Kaja_Characterization_2024,
	author = {Kaja, Magdalena and Studer, Dominik and Heinke, Reinhard and Kieck, Tom and Wendt, Klaus},
	doi = {10.1016/j.nimb.2023.165213},
	journal = {Nucl. Instruments Methods Phys. Res. Sect. B Beam Interact. with Mater. Atoms},
	month = {2},
	number = {December 2023},
	pages = {165213},
	title = {{Characterization of the field ionization extension for the laser ion source and trap: Measurement of the ionization potential of ytterbium}},
	url = {https://doi.org/10.1016/j.nimb.2023.165213 https://linkinghub.elsevier.com/retrieve/pii/S0168583X23004548},
	volume = {547},
	year = {2024}
}

@article{Kato_Control_2013,
	author = {Kato, Shinya and Sugawa, Seiji and Shibata, Kosuke and Yamamoto, Ryuta and Takahashi, Yoshiro},
	doi = {10.1103/PhysRevLett.110.173201},
	journal = {Phys. Rev. Lett.},
	month = {4},
	number = {17},
	pages = {173201},
	title = {{Control of Resonant Interaction between Electronic Ground and Excited States}},
	url = {https://link.aps.org/doi/10.1103/PhysRevLett.110.173201},
	volume = {110},
	year = {2013}
}

@article{Khramov_Ultracold_2014,
	author = {Khramov, Alexander and Hansen, Anders and Dowd, William and Roy, Richard J. and Makrides, Constantinos and Petrov, Alexander and Kotochigova, Svetlana and Gupta, Subhadeep},
	doi = {10.1103/PhysRevLett.112.033201},
	journal = {Phys. Rev. Lett.},
	month = {1},
	number = {3},
	pages = {033201},
	title = {{Ultracold Heteronuclear Mixture of Ground and Excited State Atoms}},
	url = {https://link.aps.org/doi/10.1103/PhysRevLett.112.033201},
	volume = {112},
	year = {2014}
}

@article{Kim_Absolute_2021,
	author = {Kim, Huidong and Heo, Myoung Sun and Park, Chang Yong and Yu, Dai Hyuk and Lee, Won Kyu},
	doi = {10.1088/1681-7575/ac1950},
	journal = {Metrologia},
	number = {5},
	title = {{Absolute frequency measurement of the $^{171}$Yb optical lattice clock at KRISS using TAI for over a year}},
	volume = {58},
	year = {2021}
}

@article{Kim_Improved_2017,
	author = {Kim, Huidong and Heo, Myoung-Sun and Lee, Won-Kyu and Park, Chang Yong and Hong, Hyun-Gue and Hwang, Sang-Wook and Yu, Dai-Hyuk},
	doi = {10.7567/JJAP.56.050302},
	journal = {Jpn. J. Appl. Phys.},
	month = {5},
	number = {5},
	pages = {050302},
	title = {{Improved absolute frequency measurement of the $^{171}$Yb optical lattice clock at KRISS relative to the SI second}},
	url = {https://iopscience.iop.org/article/10.7567/JJAP.56.050302},
	volume = {56},
	year = {2017}
}

@article{Kitagawa_Two-color_2008,
	author = {Kitagawa, Masaaki and Enomoto, Katsunari and Kasa, Kentaro and Takahashi, Yoshiro and Ciury{\l}o, Roman and Naidon, Pascal and Julienne, Paul S.},
	doi = {10.1103/PhysRevA.77.012719},
	journal = {Phys. Rev. A},
	month = {1},
	number = {1},
	pages = {012719},
	title = {{Two-color photoassociation spectroscopy of ytterbium atoms and the precise determinations of $s$-wave scattering lengths}},
	url = {https://link.aps.org/doi/10.1103/PhysRevA.77.012719},
	volume = {77},
	year = {2008}
}

@article{Kleinert_Measurement_2016,
	author = {Kleinert, Michaela and {Gold Dahl}, M. E. and Bergeson, Scott},
	doi = {10.1103/PhysRevA.94.052511},
	journal = {Phys. Rev. A},
	month = {11},
	number = {5},
	pages = {052511},
	title = {{Measurement of the Yb I $^1$S$_0$-$^1$P$_1$ transition frequency at 399 nm using an optical frequency comb}},
	url = {https://link.aps.org/doi/10.1103/PhysRevA.94.052511},
	volume = {94},
	year = {2016}
}

@article{Kobayashi_Demonstration_2020,
	author = {Kobayashi, Takumi and Akamatsu, Daisuke and Hosaka, Kazumoto and Hisai, Yusuke and Wada, Masato and Inaba, Hajime and Suzuyama, Tomonari and Hong, Feng-Lei and Yasuda, Masami},
	doi = {10.1088/1681-7575/ab9f1f},
	journal = {Metrologia},
	month = {12},
	number = {6},
	pages = {065021},
	title = {{Demonstration of the nearly continuous operation of an $^{171}$Yb optical lattice clock for half a year}},
	url = {https://iopscience.iop.org/article/10.1088/1681-7575/ab9f1f},
	volume = {57},
	year = {2020}
}

@article{Kobayashi_Improved_2025,
	author = {Kobayashi, Takumi and Nishiyama, Akiko and Hosaka, Kazumoto and Akamatsu, Daisuke and Kawasaki, Akio and Wada, Masato and Inaba, Hajime and Tanabe, Takehiko and Yasuda, Masami},
	doi = {10.1088/1681-7575/adb754},
	journal = {Metrologia},
	month = {4},
	number = {2},
	pages = {025006},
	title = {{Improved absolute frequency measurement of $^{171}$Yb at NMIJ with uncertainty below $2\times10^{-16}$}},
	url = {https://iopscience.iop.org/article/10.1088/1681-7575/adb754},
	volume = {62},
	year = {2025}
}

@article{Kobayashi_Search_2022,
	author = {Kobayashi, Takumi and Takamizawa, Akifumi and Akamatsu, Daisuke and Kawasaki, Akio and Nishiyama, Akiko and Hosaka, Kazumoto and Hisai, Yusuke and Wada, Masato and Inaba, Hajime and Tanabe, Takehiko and Yasuda, Masami},
	doi = {10.1103/PhysRevLett.129.241301},
	journal = {Phys. Rev. Lett.},
	month = {12},
	number = {24},
	pages = {241301},
	title = {{Search for Ultralight Dark Matter from Long-Term Frequency Comparisons of Optical and Microwave Atomic Clocks}},
	url = {https://doi.org/10.1103/PhysRevLett.129.241301 https://link.aps.org/doi/10.1103/PhysRevLett.129.241301},
	volume = {129},
	year = {2022}
}

@article{Kobayashi_Uncertainty_2018,
	author = {Kobayashi, Takumi and Akamatsu, Daisuke and Hisai, Yusuke and Tanabe, Takehiko and Inaba, Hajime and Suzuyama, Tomonari and Hong, Feng-Lei and Hosaka, Kazumoto and Yasuda, Masami},
	doi = {10.1109/TUFFC.2018.2870937},
	journal = {IEEE Trans. Ultrason. Ferroelectr. Freq. Control},
	month = {12},
	number = {12},
	pages = {2449--2458},
	title = {{Uncertainty Evaluation of an $^{171}$Yb Optical Lattice Clock at NMIJ}},
	url = {https://ieeexplore.ieee.org/document/8467344/},
	volume = {65},
	year = {2018}
}

@article{Kohno_One-Dimensional_2009,
	author = {Kohno, Takuya and Yasuda, Masami and Hosaka, Kazumoto and Inaba, Hajime and Nakajima, Yoshiaki and Hong, Feng-Lei},
	doi = {10.1143/APEX.2.072501},
	journal = {Appl. Phys. Express},
	month = {6},
	number = {7},
	pages = {072501},
	title = {{One-Dimensional Optical Lattice Clock with a Fermionic $^{171}$Yb Isotope}},
	url = {https://iopscience.iop.org/article/10.1143/APEX.2.072501},
	volume = {2},
	year = {2009}
}

@article{Kondev_NUBASE2020_2021,
	author = {Kondev, F.G. and Wang, M. and Huang, W.J. and Naimi, S. and Audi, G.},
	doi = {10.1088/1674-1137/abddae},
	journal = {Chin. Phys. C},
	month = {3},
	number = {3},
	pages = {030001},
	title = {{The NUBASE2020 evaluation of nuclear physics properties}},
	url = {https://iopscience.iop.org/article/10.1088/1674-1137/abddae},
	volume = {45},
	year = {2021}
}

@article{Kuwamoto_Magneto-optical_1999,
	author = {Kuwamoto, T. and Honda, K. and Takahashi, Y. and Yabuzaki, T.},
	doi = {10.1103/PhysRevA.60.R745},
	journal = {Phys. Rev. A},
	month = {8},
	number = {2},
	pages = {R745--R748},
	title = {{Magneto-optical trapping of Yb atoms using an intercombination transition}},
	url = {https://link.aps.org/doi/10.1103/PhysRevA.60.R745},
	volume = {60},
	year = {1999}
}

@article{Labzowsky_Estimates_1999,
	author = {Labzowsky, Leonti and Goidenko, Igor and Pyykk{\"{o}}, Pekka},
	doi = {10.1016/S0375-9601(99)00335-7},
	journal = {Phys. Lett. A},
	month = {7},
	number = {1},
	pages = {31--37},
	title = {{Estimates of the bound-state QED contributions to the $g$-factor of valence $ns$ electrons in alkali metal atoms}},
	url = {https://linkinghub.elsevier.com/retrieve/pii/S0375960199003357},
	volume = {258},
	year = {1999}
}

@article{Laupretre_Absolute_2020,
	author = {Laupr{\^{e}}tre, Thomas and Groult, Lucas and Achi, Bachir and Petersen, Michael and Kersal{\'{e}}, Yann and Delehaye, Marion and Lacro{\^{u}}te, Cl{\'{e}}ment},
	doi = {10.1364/OSAC.3.000050},
	journal = {OSA Contin.},
	month = {1},
	number = {1},
	pages = {50},
	title = {{Absolute frequency measurements of the $^1$S$_0\rightarrow^1$P$_1$ transition in ytterbium}},
	url = {https://opg.optica.org/abstract.cfm?URI=osac-3-1-50},
	volume = {3},
	year = {2020}
}

@article{LeKien_Dynamical_2013,
	author = {Le Kien, Fam and Schneeweiss, Philipp and Rauschenbeutel, Arno},
	doi = {10.1140/epjd/e2013-30729-x},
	journal = {Eur. Phys. J. D},
	month = {5},
	number = {5},
	pages = {92},
	title = {{Dynamical polarizability of atoms in arbitrary light fields: general theory and application to cesium}},
	url = {http://link.springer.com/10.1140/epjd/e2013-30729-x},
	volume = {67},
	year = {2013}
}

@article{Lee_Core-shell_2015,
	author = {Lee, Jeongwon and Lee, Jae Hoon and Noh, Jiho and Mun, Jongchul},
	doi = {10.1103/PhysRevA.91.053405},
	journal = {Phys. Rev. A},
	month = {5},
	number = {5},
	pages = {053405},
	title = {{Core-shell magneto-optical trap for alkaline-earth-metal-like atoms}},
	url = {https://link.aps.org/doi/10.1103/PhysRevA.91.053405},
	volume = {91},
	year = {2015}
}

@article{Lehec_Laser_2018,
	author = {Lehec, H. and Zuliani, A. and Maineult, W. and Luc-Koenig, E. and Pillet, P. and Cheinet, P. and Niyaz, F. and Gallagher, T. F.},
	doi = {10.1103/PhysRevA.98.062506},
	journal = {Phys. Rev. A},
	month = {12},
	number = {6},
	pages = {062506},
	title = {{Laser and microwave spectroscopy of even-parity Rydberg states of neutral ytterbium and multichannel-quantum-defect-theory analysis}},
	url = {https://link.aps.org/doi/10.1103/PhysRevA.98.062506},
	volume = {98},
	year = {2018}
}

@article{Lemke_Spin-1/2_2009,
	author = {Lemke, N. and Ludlow, A. and Barber, Z. and Fortier, T. and Diddams, S. and Jiang, Y. and Jefferts, S. and Heavner, T. and Parker, T. and Oates, C.},
	doi = {10.1103/PhysRevLett.103.063001},
	journal = {Phys. Rev. Lett.},
	month = {8},
	number = {6},
	pages = {063001},
	title = {{Spin-1/2 Optical Lattice Clock}},
	url = {https://link.aps.org/doi/10.1103/PhysRevLett.103.063001},
	volume = {103},
	year = {2009}
}

@article{Letellier_Loading_2023,
	author = {Letellier, Hector and {Mitchell Galv{\~{a}}o de Melo}, {\'{A}}lvaro and Dorne, Ana{\"{i}}s and Kaiser, Robin},
	doi = {10.1063/5.0169772},
	journal = {Rev. Sci. Instrum.},
	month = {12},
	number = {12},
	pages = {0--13},
	title = {{Loading of a large Yb MOT on the $^1$S$_0$ $\rightarrow$ $^1$P$_1$ transition}},
	url = {https://doi.org/10.1063/5.0169772 https://pubs.aip.org/rsi/article/94/12/123203/2930404/Loading-of-a-large-Yb-MOT-on-the-1S0-1P1},
	volume = {94},
	year = {2023}
}

@article{Lett_Observation_1988,
	author = {Lett, Paul D. and Watts, Richard N. and Westbrook, Christoph I. and Phillips, William D. and Gould, Phillip L. and Metcalf, Harold J.},
	doi = {10.1103/PhysRevLett.61.169},
	journal = {Phys. Rev. Lett.},
	month = {7},
	number = {2},
	pages = {169--172},
	title = {{Observation of Atoms Laser Cooled below the Doppler Limit}},
	url = {https://link.aps.org/doi/10.1103/PhysRevLett.61.169},
	volume = {61},
	year = {1988}
}

@article{Liening_Level-crossing-spectroscopy_1985,
	author = {Liening, H.},
	doi = {10.1007/BF01415711},
	journal = {Zeitschrift f{\"{u}}r Phys. A Atoms Nucl.},
	month = {9},
	number = {3},
	pages = {363--368},
	title = {{Level-crossing-spectroscopy with additional optical pumping in the ground state}},
	url = {http://link.springer.com/10.1007/BF01415711},
	volume = {320},
	year = {1985}
}

@article{Lis_Midcircuit_2023,
	author = {Lis, Joanna W. and Senoo, Aruku and McGrew, William F. and R{\"{o}}nchen, Felix and Jenkins, Alec and Kaufman, Adam M.},
	doi = {10.1103/PhysRevX.13.041035},
	journal = {Phys. Rev. X},
	month = {11},
	number = {4},
	pages = {041035},
	title = {{Midcircuit Operations Using the omg Architecture in Neutral Atom Arrays}},
	url = {http://arxiv.org/abs/2305.19266 https://doi.org/10.1103/PhysRevX.13.041035 https://link.aps.org/doi/10.1103/PhysRevX.13.041035},
	volume = {13},
	year = {2023}
}

@article{Loftus_Optical_2001,
	author = {Loftus, T. and Bochinski, J. R. and Mossberg, T. W.},
	doi = {10.1103/PhysRevA.63.023402},
	journal = {Phys. Rev. A},
	month = {1},
	number = {2},
	pages = {023402},
	title = {{Optical double-resonance cooled-atom spectroscopy}},
	url = {https://link.aps.org/doi/10.1103/PhysRevA.63.023402},
	volume = {63},
	year = {2001}
}

@article{Ludlow_Cold-collision-shift_2011,
	author = {Ludlow, A. D. and Lemke, N. D. and Sherman, J. A. and Oates, C. W. and Qu{\'{e}}m{\'{e}}ner, G. and von Stecher, J. and Rey, A. M.},
	doi = {10.1103/PhysRevA.84.052724},
	journal = {Phys. Rev. A},
	month = {11},
	number = {5},
	pages = {052724},
	title = {{Cold-collision-shift cancellation and inelastic scattering in a Yb optical lattice clock}},
	url = {https://link.aps.org/doi/10.1103/PhysRevA.84.052724},
	volume = {84},
	year = {2011}
}

@article{Luo_Absolute_2020,
	author = {Luo, Limeng and Qiao, Hao and Ai, Di and Zhou, Min and Zhang, Shuang and Zhang, Sheng and Sun, Changyue and Qi, Qichao and Peng, Chengquan and Jin, Taoyun and Fang, Wei and Yang, Zhiqiang and Li, Tianchu and Liang, Kun and Xu, Xinye},
	doi = {10.1088/1681-7575/abb879},
	journal = {Metrologia},
	month = {12},
	number = {6},
	pages = {065017},
	title = {{Absolute frequency measurement of an Yb optical clock at the $10^{-16}$ level using International Atomic Time}},
	url = {https://iopscience.iop.org/article/10.1088/1681-7575/abb879},
	volume = {57},
	year = {2020}
}

@article{Lurio_Second-Order_1962,
	author = {Lurio, A. and Mandel, M. and Novick, R.},
	doi = {10.1103/PhysRev.126.1758},
	journal = {Phys. Rev.},
	month = {6},
	number = {5},
	pages = {1758--1767},
	title = {{Second-Order Hyperfine and Zeeman Corrections for an ($sl$) Configuration}},
	url = {https://link.aps.org/doi/10.1103/PhysRev.126.1758},
	volume = {126},
	year = {1962}
}

@article{Ma_Universal_2022,
	author = {Ma, Shuo and Burgers, Alex P. and Liu, Genyue and Wilson, Jack and Zhang, Bichen and Thompson, Jeff D.},
	doi = {10.1103/PhysRevX.12.021028},
	journal = {Phys. Rev. X},
	month = {5},
	number = {2},
	pages = {021028},
	title = {{Universal Gate Operations on Nuclear Spin Qubits in an Optical Tweezer Array of $^{171}$Yb Atoms}},
	url = {https://link.aps.org/doi/10.1103/PhysRevX.12.021028},
	volume = {12},
	year = {2022}
}

@article{Maier_Hyperfine_1991,
	author = {Maier, J. and Kischkel, C. S. and Baumann, M.},
	doi = {10.1007/BF01425593},
	journal = {Zeitschrift f{\"{u}}r Phys. D Atoms Mol. Clust.},
	month = {6},
	number = {2},
	pages = {145--151},
	title = {{Hyperfine structure and isotope shift of some even parity levels in the YbI spectrum}},
	url = {https://link.springer.com/10.1007/BF01425593},
	volume = {21},
	year = {1991}
}

@article{Margolis_CIPM_2024,
	author = {Margolis, H. S. and Panfilo, G. and Petit, G. and Oates, C. and Ido, T. and Bize, S.},
	doi = {10.1088/1681-7575/ad3afc},
	journal = {Metrologia},
	month = {6},
	number = {3},
	pages = {035005},
	title = {{The CIPM list 'Recommended values of standard frequencies': 2021 update}},
	url = {https://iopscience.iop.org/article/10.1088/1681-7575/ad3afc},
	volume = {61},
	year = {2024}
}

@article{Masson_Dicke_2024,
	author = {Masson, Stuart J. and Covey, Jacob P. and Will, Sebastian and Asenjo-Garcia, Ana},
	doi = {10.1103/PRXQuantum.5.010344},
	journal = {PRX Quantum},
	month = {3},
	number = {1},
	pages = {010344},
	title = {{Dicke Superradiance in Ordered Arrays of Multilevel Atoms}},
	url = {https://doi.org/10.1103/PRXQuantum.5.010344 https://link.aps.org/doi/10.1103/PRXQuantum.5.010344},
	volume = {5},
	year = {2024}
}

@article{McFerran_inverted_2016,
	author = {McFerran, J. J.},
	doi = {10.1364/josab.33.001278},
	journal = {J. Opt. Soc. Am. B},
	number = {6},
	pages = {1278},
	title = {{An inverted crossover resonance aiding laser cooling of $^{171}$Yb}},
	volume = {33},
	year = {2016}
}

@article{McGrew_Atomic_2018,
	author = {McGrew, W. F. and Zhang, X. and Fasano, R. J. and Sch{\"{a}}ffer, S. A. and Beloy, K. and Nicolodi, D. and Brown, R. C. and Hinkley, N. and Milani, G. and Schioppo, M. and Yoon, T. H. and Ludlow, A. D.},
	doi = {10.1038/s41586-018-0738-2},
	journal = {Nature},
	month = {12},
	number = {7734},
	pages = {87--90},
	title = {{Atomic clock performance enabling geodesy below the centimetre level}},
	url = {http://dx.doi.org/10.1038/s41586-018-0738-2 https://www.nature.com/articles/s41586-018-0738-2},
	volume = {564},
	year = {2018}
}

@article{McGrew_Towards_2019,
	author = {McGrew, W. F. and Zhang, X. and Leopardi, H. and Fasano, R. J. and Nicolodi, D. and Beloy, K. and Yao, J. and Sherman, J. A. and Sch{\"{a}}ffer, S. A. and Savory, J. and Brown, R. C. and R{\"{o}}misch, S. and Oates, C. W. and Parker, T. E. and Fortier, T. M. and Ludlow, A. D.},
	doi = {10.1364/OPTICA.6.000448},
	journal = {Optica},
	month = {4},
	number = {4},
	pages = {448},
	title = {{Towards the optical second: verifying optical clocks at the SI limit}},
	url = {https://opg.optica.org/abstract.cfm?URI=optica-6-4-448},
	volume = {6},
	year = {2019}
}

@phdthesis{McGrew_Ytterbium_2020,
	author = {McGrew, W F},
	school = {University of Colorado},
	title = {{An Ytterbium Optical Lattice Clock with Eighteen Digits of Uncertainty, Instability, and Reproducibility}},
	year = {2020}
}

@article{Mishra_Radiative_2001,
	author = {Mishra, Adya Prasad and Balasubramanian, T.K},
	doi = {10.1016/S0022-4073(00)00117-5},
	journal = {J. Quant. Spectrosc. Radiat. Transf.},
	month = {6},
	number = {6},
	pages = {769--780},
	title = {{Radiative lifetimes of the first excited $^3$P$_{2,0}^o$ metastable levels in Kr I, Xe I, Yb I and Hg-like atoms}},
	url = {https://linkinghub.elsevier.com/retrieve/pii/S0022407300001175},
	volume = {69},
	year = {2001}
}

@article{Mohr_CODATA_2025,
	author = {Mohr, Peter J. and Newell, David B. and Taylor, Barry N. and Tiesinga, Eite},
	doi = {10.1103/RevModPhys.97.025002},
	journal = {Rev. Mod. Phys.},
	month = {4},
	number = {2},
	pages = {025002},
	title = {{CODATA recommended values of the fundamental physical constants: 2022}},
	url = {https://doi.org/10.1103/RevModPhys.97.025002 https://link.aps.org/doi/10.1103/RevModPhys.97.025002},
	volume = {97},
	year = {2025}
}

@article{MuziFalconi_Microsecond-Scale_2025,
	author = {Muzi Falconi, A. and Panza, R. and Sbernardori, S. and Forti, R. and Klemt, R. and Abdel Karim, O. and Marinelli, M. and Scazza, F.},
	doi = {10.1103/n3bg-7yw7},
	journal = {Phys. Rev. Lett.},
	number = {20},
	pages = {203402},
	title = {{Microsecond-Scale High-Survival and Number-Resolved Detection of Ytterbium Atom Arrays}},
	url = {https://doi.org/10.1103/n3bg-7yw7},
	volume = {135},
	year = {2025}
}

@phdthesis{MuziFalconi_Trapping_2025,
	author = {Muzi Falconi, Alessandro Thomas},
	school = {University of Trieste},
	title = {{Trapping, loading and imaging fermionic ytterbium atom arrays}},
	year = {2025}
}

@article{Nemitz_Frequency_2016,
	author = {Nemitz, Nils and Ohkubo, Takuya and Takamoto, Masao and Ushijima, Ichiro and Das, Manoj and Ohmae, Noriaki and Katori, Hidetoshi},
	doi = {10.1038/nphoton.2016.20},
	journal = {Nat. Photonics},
	month = {4},
	number = {4},
	pages = {258--261},
	title = {{Frequency ratio of Yb and Sr clocks with $5\times10^{-17}$ uncertainty at 150 seconds averaging time}},
	url = {https://www.nature.com/articles/nphoton.2016.20},
	volume = {10},
	year = {2016}
}

@article{Nemitz_Modeling_2019,
	author = {Nemitz, Nils and J{\o}rgensen, Asbj{\o}rn Arvad and Yanagimoto, Ryotatsu and Bregolin, Filippo and Katori, Hidetoshi},
	doi = {10.1103/PhysRevA.99.033424},
	journal = {Phys. Rev. A},
	month = {3},
	number = {3},
	pages = {033424},
	title = {{Modeling light shifts in optical lattice clocks}},
	url = {https://link.aps.org/doi/10.1103/PhysRevA.99.033424},
	volume = {99},
	year = {2019}
}

@article{Nenadovic_Clock_2016,
	author = {Nenadovi{\'{c}}, L. and McFerran, J. J.},
	doi = {10.1088/0953-4075/49/6/065004},
	journal = {J. Phys. B Mol. Opt. Phys.},
	month = {3},
	number = {6},
	pages = {065004},
	title = {{Clock and inter-combination line frequency separation in $^{171}$Yb}},
	url = {https://iopscience.iop.org/article/10.1088/0953-4075/49/6/065004},
	volume = {49},
	year = {2016}
}

@article{Nizamani_Doppler-free_2010,
	author = {Nizamani, Altaf H. and McLoughlin, James J. and Hensinger, Winfried K.},
	doi = {10.1103/PhysRevA.82.043408},
	journal = {Phys. Rev. A},
	month = {10},
	number = {4},
	pages = {043408},
	title = {{Doppler-free Yb spectroscopy with the fluorescence spot technique}},
	url = {https://link.aps.org/doi/10.1103/PhysRevA.82.043408},
	volume = {82},
	year = {2010}
}

@article{Norcia_Iterative_2024,
	author = {Norcia, M A and Kim, H and Cairncross, W B and Stone, M and Ryou, A and Jaffe, M and Brown, M. O. and Barnes, K and Battaglino, P and Bohdanowicz, T C and Brown, A and Cassella, K and Chen, C.-A. and Coxe, R and Crow, D and Epstein, J and Griger, C and Halperin, E and Hummel, F and Jones, A M W and Kindem, J M and King, J and Kotru, K and Lauigan, J and Li, M and Lu, M and Megidish, E and Marjanovic, J and McDonald, M. and Mittiga, T and Muniz, J A and Narayanaswami, S and Nishiguchi, C and Paule, T and Pawlak, K A and Peng, L S and Pudenz, K L and {Rodr{\'{i}}guez P{\'{e}}rez}, D. and Smull, A and Stack, D and Urbanek, M and van de Veerdonk, R. J. M. and Vendeiro, Z and Wadleigh, L and Wilkason, T and Wu, T.-Y. and Xie, X. and Zalys-Geller, E. and Zhang, X. and Bloom, B. J.},
	doi = {10.1103/PRXQuantum.5.030316},
	journal = {PRX Quantum},
	month = {7},
	number = {3},
	pages = {030316},
	title = {{Iterative Assembly of $^{171}$Yb Atom Arrays with Cavity-Enhanced Optical Lattices}},
	url = {https://link.aps.org/doi/10.1103/PRXQuantum.5.030316 http://arxiv.org/abs/2401.16177},
	volume = {5},
	year = {2024}
}

@article{Norcia_Midcircuit_2023,
	author = {Norcia, M. A. and Cairncross, W. B. and Barnes, K. and Battaglino, P. and Brown, A. and Brown, M. O. and Cassella, K. and Chen, C.-A. and Coxe, R. and Crow, D. and Epstein, J. and Griger, C. and Jones, A. M. W. and Kim, H. and Kindem, J. M. and King, J. and Kondov, S. S. and Kotru, K. and Lauigan, J. and Li, M. and Lu, M. and Megidish, E. and Marjanovic, J. and McDonald, M. and Mittiga, T. and Muniz, J. A. and Narayanaswami, S. and Nishiguchi, C. and Notermans, R. and Paule, T. and Pawlak, K. A. and Peng, L. S. and Ryou, A. and Smull, A. and Stack, D. and Stone, M. and Sucich, A. and Urbanek, M. and van de Veerdonk, R. J. M. and Vendeiro, Z. and Wilkason, T. and Wu, T.-Y. and Xie, X. and Zhang, X. and Bloom, B. J.},
	doi = {10.1103/PhysRevX.13.041034},
	journal = {Phys. Rev. X},
	month = {11},
	number = {4},
	pages = {041034},
	title = {{Midcircuit Qubit Measurement and Rearrangement in a $^{171}$Yb Atomic Array}},
	url = {https://doi.org/10.1103/PhysRevX.13.041034 https://link.aps.org/doi/10.1103/PhysRevX.13.041034 http://arxiv.org/abs/2305.19119},
	volume = {13},
	year = {2023}
}

@article{Olschewski_Bestimmung_1967,
	author = {Olschewski, L. and Otten, E. -W.},
	doi = {10.1007/BF01328937},
	journal = {Zeitschrift f{\"{u}}r Phys.},
	month = {4},
	number = {2},
	pages = {224--226},
	title = {{Bestimmung der Kerndipolmomente von $^{171}$Yb und $^{173}$Yb durch optisches Pumpen}},
	url = {http://link.springer.com/10.1007/BF01328937},
	volume = {200},
	year = {1967}
}

@article{Olschewski_Messung_1972,
	author = {Olschewski, L.},
	doi = {10.1007/BF01400226},
	journal = {Zeitschrift f{\"{u}}r Phys.},
	month = {6},
	number = {3},
	pages = {205--227},
	title = {{Messung der magnetischen Kerndipolmomente an freien $^{43}$Ca-, $^{87}$Sr-, $^{135}$Ba-, $^{137}$Ba-, $^{171}$Yb- und $^{173}$Yb-Atomen mit optischem Pumpen}},
	url = {http://link.springer.com/10.1007/BF01400226},
	volume = {249},
	year = {1972}
}

@article{Ono_Antiferromagnetic_2019,
	author = {Ono, Koki and Kobayashi, Jun and Amano, Yoshiki and Sato, Koji and Takahashi, Yoshiro},
	doi = {10.1103/PhysRevA.99.032707},
	journal = {Phys. Rev. A},
	month = {3},
	number = {3},
	pages = {032707},
	title = {{Antiferromagnetic interorbital spin-exchange interaction of $^{171}$Yb}},
	url = {https://link.aps.org/doi/10.1103/PhysRevA.99.032707},
	volume = {99},
	year = {2019}
}

@article{Ono_Observation_2022,
	author = {Ono, Koki and Saito, Yugo and Ishiyama, Taiki and Higomoto, Toshiya and Takano, Tetsushi and Takasu, Yosuke and Yamamoto, Yasuhiro and Tanaka, Minoru and Takahashi, Yoshiro},
	doi = {10.1103/PhysRevX.12.021033},
	journal = {Phys. Rev. X},
	month = {5},
	number = {2},
	pages = {021033},
	title = {{Observation of Nonlinearity of Generalized King Plot in the Search for New Boson}},
	url = {https://doi.org/10.1103/PhysRevX.12.021033 https://link.aps.org/doi/10.1103/PhysRevX.12.021033},
	volume = {12},
	year = {2022}
}

@article{Pandey_Isotope_2009,
	author = {Pandey, Kanhaiya and Singh, Alok K. and Kumar, P. V. Kiran and Suryanarayana, M. V. and Natarajan, Vasant},
	doi = {10.1103/PhysRevA.80.022518},
	journal = {Phys. Rev. A},
	month = {8},
	number = {2},
	pages = {022518},
	title = {{Isotope shifts and hyperfine structure in the 555.8-nm $^1$S$_0\rightarrow^3$P$_1$ line of Yb}},
	url = {https://link.aps.org/doi/10.1103/PhysRevA.80.022518},
	volume = {80},
	year = {2009}
}

@article{Park_Absolute_2013,
	author = {Park, Chang Yong and Yu, Dai-Hyuk and Lee, Won-Kyu and Park, Sang Eon and Kim, Eok Bong and Lee, Sun Kyung and Cho, Jun Woo and Yoon, Tai Hyun and Mun, Jongchul and Park, Sung Jong and Kwon, Taeg Yong and Lee, Sang-Bum},
	doi = {10.1088/0026-1394/50/2/119},
	journal = {Metrologia},
	month = {4},
	number = {2},
	pages = {119--128},
	title = {{Absolute frequency measurement of $^1$S$_0(F = 1/2)$-$^3$P$_0(F = 1/2)$ transition of $^{171}$Yb atoms in a one-dimensional optical lattice at KRISS}},
	url = {https://iopscience.iop.org/article/10.1088/0026-1394/50/2/119},
	volume = {50},
	year = {2013}
}

@article{Park_Efficient_2003,
	author = {Park, Chang Yong and Yoon, Tai Hyun},
	doi = {10.1103/PhysRevA.68.055401},
	journal = {Phys. Rev. A},
	month = {11},
	number = {5},
	pages = {055401},
	title = {{Efficient magneto-optical trapping of Yb atoms with a violet laser diode}},
	url = {https://link.aps.org/doi/10.1103/PhysRevA.68.055401},
	volume = {68},
	year = {2003}
}

@article{Peck_Dispersion_1972,
	author = {Peck, Edson R. and Reeder, Kaye},
	doi = {10.1364/JOSA.62.000958},
	journal = {J. Opt. Soc. Am.},
	month = {8},
	number = {8},
	pages = {958},
	title = {{Dispersion of Air}},
	url = {https://opg.optica.org/abstract.cfm?URI=josa-62-8-958},
	volume = {62},
	year = {1972}
}

@article{Peper_Spectroscopy_2025,
	author = {Peper, Michael and Li, Yiyi and Knapp, Daniel Y. and Bileska, Mila and Ma, Shuo and Liu, Genyue and Peng, Pai and Zhang, Bichen and Horvath, Sebastian P. and Burgers, Alex P. and Thompson, Jeff D.},
	doi = {10.1103/PhysRevX.15.011009},
	journal = {Phys. Rev. X},
	month = {1},
	number = {1},
	pages = {011009},
	title = {{Spectroscopy and Modeling of $^{171}$Yb Rydberg States for High-Fidelity Two-Qubit Gates}},
	url = {https://doi.org/10.1103/PhysRevX.15.011009 https://link.aps.org/doi/10.1103/PhysRevX.15.011009},
	volume = {15},
	year = {2025}
}

@article{Pizzocaro_Absolute_2017,
	author = {Pizzocaro, Marco and Thoumany, Pierre and Rauf, Benjamin and Bregolin, Filippo and Milani, Gianmaria and Clivati, Cecilia and Costanzo, Giovanni A. and Levi, Filippo and Calonico, Davide},
	doi = {10.1088/1681-7575/aa4e62},
	journal = {Metrologia},
	month = {2},
	number = {1},
	pages = {102--112},
	title = {{Absolute frequency measurement of the $^1$S$_0$ - $^3$P$_0$ transition of $^{171}$Yb}},
	url = {https://iopscience.iop.org/article/10.1088/1681-7575/aa4e62},
	volume = {54},
	year = {2017}
}

@article{Pizzocaro_Absolute_2020,
	author = {Pizzocaro, Marco and Bregolin, Filippo and Barbieri, Piero and Rauf, Benjamin and Levi, Filippo and Calonico, Davide},
	doi = {10.1088/1681-7575/ab50e8},
	journal = {Metrologia},
	number = {3},
	title = {{Absolute frequency measurement of the $^1$S$_0$-$^3$P$_0$ transition of $^{171}$Yb with a link to international atomic time}},
	volume = {57},
	year = {2020}
}

@article{Poli_Frequency_2008,
	author = {Poli, N. and Barber, Z. W. and Lemke, N. D. and Oates, C. W. and Ma, L. S. and Stalnaker, J. E. and Fortier, T. M. and Diddams, S. A. and Hollberg, L. and Bergquist, J. C. and Brusch, A. and Jefferts, S. and Heavner, T. and Parker, T.},
	doi = {10.1103/PhysRevA.77.050501},
	journal = {Phys. Rev. A},
	month = {5},
	number = {5},
	pages = {050501},
	title = {{Frequency evaluation of the doubly forbidden $^1$S$_0$$\rightarrow$$^3$P$_0$ transition in bosonic $^{174}$Yb}},
	url = {https://link.aps.org/doi/10.1103/PhysRevA.77.050501},
	volume = {77},
	year = {2008}
}

@article{Porsev_Electric-dipole_1999,
	author = {Porsev, S. G. and Rakhlina, Yu. G. and Kozlov, M. G.},
	doi = {10.1103/PhysRevA.60.2781},
	journal = {Phys. Rev. A},
	month = {10},
	number = {4},
	pages = {2781--2785},
	title = {{Electric-dipole amplitudes, lifetimes, and polarizabilities of the low-lying levels of atomic ytterbium}},
	url = {https://link.aps.org/doi/10.1103/PhysRevA.60.2781},
	volume = {60},
	year = {1999}
}

@article{Porsev_Hyperfine_2004,
	author = {Porsev, Sergey G. and Derevianko, Andrei},
	doi = {10.1103/PhysRevA.69.042506},
	journal = {Phys. Rev. A},
	month = {4},
	number = {4},
	pages = {042506},
	title = {{Hyperfine quenching of the metastable $^3$P$_{0,2}$ states in divalent atoms}},
	url = {https://link.aps.org/doi/10.1103/PhysRevA.69.042506},
	volume = {69},
	year = {2004}
}

@article{Prohaska_Standard_2022,
	author = {Prohaska, Thomas and Irrgeher, Johanna and Benefield, Jacqueline and B{\"{o}}hlke, John K. and Chesson, Lesley A. and Coplen, Tyler B. and Ding, Tiping and Dunn, Philip J. H. and Gr{\"{o}}ning, Manfred and Holden, Norman E. and Meijer, Harro A. J. and Moossen, Heiko and Possolo, Antonio and Takahashi, Yoshio and Vogl, Jochen and Walczyk, Thomas and Wang, Jun and Wieser, Michael E. and Yoneda, Shigekazu and Zhu, Xiang-Kun and Meija, Juris},
	doi = {10.1515/pac-2019-0603},
	journal = {Pure Appl. Chem.},
	month = {5},
	number = {5},
	pages = {573--600},
	title = {{Standard atomic weights of the elements 2021 (IUPAC Technical Report)}},
	url = {https://www.degruyter.com/document/doi/10.1515/pac-2019-0603/html},
	volume = {94},
	year = {2022}
}

@article{Qiao_Frequency_2023,
	author = {Qiao, Hao and Sun, Chang-Yue and Peng, Cheng-Quan and Qi, Qi-Chao and Zhao, Cheng-Cheng and Zhou, Min and Xu, Xin-Ye},
	doi = {10.1016/j.rinp.2023.106439},
	journal = {Results Phys.},
	month = {5},
	number = {April},
	pages = {106439},
	title = {{Frequency measurement of $6s6p$ $^3$P$_{0,2}$-$6s7s$ $^3$S$_1$ transitions in ultracold $^{171}$Yb atoms referenced to local optical clock}},
	volume = {48},
	year = {2023}
}

@article{Qiao_Investigation_2023,
	author = {Qiao, Hao and Liu, Luhua and Zhou, Min and Luo, Limeng and Xu, Xinye},
	doi = {10.1063/5.0155776},
	journal = {Appl. Phys. Lett.},
	month = {5},
	number = {22},
	pages = {3--9},
	title = {{Investigation of the $6s6p$ $^3$P$_2$-$6s7s$ $^s$S$_1$ transition of ytterbium atoms with modulation-transfer spectroscopy}},
	url = {https://doi.org/10.1063/5.0155776 https://pubs.aip.org/apl/article/122/22/224002/2893691/Investigation-of-the-6s6p-3P2-6s7s-3S1-transition},
	volume = {122},
	year = {2023}
}

@article{Raghavan_Table_1989,
	author = {Raghavan, Pramila},
	doi = {10.1016/0092-640X(89)90008-9},
	journal = {Data Nucl. Data Tables},
	month = {7},
	number = {2},
	pages = {189--291},
	title = {{Table of nuclear moments}},
	url = {https://linkinghub.elsevier.com/retrieve/pii/0092640X89900089},
	volume = {42},
	year = {1989}
}

@article{Rambow_Radiative_1976,
	author = {Rambow, F. H. K. and Schearer, L. D.},
	doi = {10.1103/PhysRevA.14.738},
	journal = {Phys. Rev. A},
	month = {8},
	number = {2},
	pages = {738--743},
	title = {{Radiative lifetimes and alignment depolarization cross sections for YbI and II by the Hanle effect in a flowing helium system}},
	url = {https://link.aps.org/doi/10.1103/PhysRevA.14.738},
	volume = {14},
	year = {1976}
}

@phdthesis{Riegger_Interorbital_2019,
	author = {Riegger, Luis},
	doi = {10.5282/edoc.23994},
	school = {Ludwig-Maximilians-Universit{\"{a}}t M{\"{u}}nchen},
	title = {{Interorbital spin exchange in a state-dependent optical lattice}},
	year = {2019}
}

@article{Ross_Isotope_1963,
	author = {Ross, John S.},
	doi = {10.1364/JOSA.53.000299},
	journal = {J. Opt. Soc. Am.},
	month = {2},
	number = {2},
	pages = {299},
	title = {{Isotope Shift in $\lambda$5556 \r{A} of Ytterbium I}},
	url = {https://opg.optica.org/abstract.cfm?URI=josa-53-2-299},
	volume = {53},
	year = {1963}
}

@article{Russell_New_1925,
	author = {Russell, H. N. and Saunders, F. A.},
	doi = {10.1086/142872},
	journal = {Astrophys. J.},
	month = {1},
	pages = {38},
	title = {{New Regularities in the Spectra of the Alkaline Earths}},
	url = {http://adsabs.harvard.edu/doi/10.1086/142872},
	volume = {61},
	year = {1925}
}

@article{Saffman_Quantum_2016,
	author = {Saffman, M.},
	doi = {10.1088/0953-4075/49/20/202001},
	journal = {J. Phys. B Mol. Opt. Phys.},
	month = {10},
	number = {20},
	pages = {202001},
	title = {{Quantum computing with atomic qubits and Rydberg interactions: progress and challenges}},
	url = {https://iopscience.iop.org/article/10.1088/0953-4075/49/20/202001},
	volume = {49},
	year = {2016}
}

@article{Safronova_Two_2018,
	author = {Safronova, Marianna S. and Porsev, Sergey G. and Sanner, Christian and Ye, Jun},
	doi = {10.1103/PhysRevLett.120.173001},
	journal = {Phys. Rev. Lett.},
	month = {4},
	number = {17},
	pages = {173001},
	title = {{Two Clock Transitions in Neutral Yb for the Highest Sensitivity to Variations of the Fine-Structure Constant}},
	url = {https://doi.org/10.1103/PhysRevLett.120.173001 https://link.aps.org/doi/10.1103/PhysRevLett.120.173001},
	volume = {120},
	year = {2018}
}

@article{Saskin_Narrow-Line_2019,
	author = {Saskin, S. and Wilson, J. T. and Grinkemeyer, B. and Thompson, J. D.},
	doi = {10.1103/PhysRevLett.122.143002},
	journal = {Phys. Rev. Lett.},
	month = {4},
	number = {14},
	pages = {143002},
	title = {{Narrow-Line Cooling and Imaging of Ytterbium Atoms in an Optical Tweezer Array}},
	url = {https://doi.org/10.1103/PhysRevLett.122.143002 https://link.aps.org/doi/10.1103/PhysRevLett.122.143002},
	volume = {122},
	year = {2019}
}

@article{Scazza_Observation_2014,
	author = {Scazza, F and Hofrichter, C and H{\"{o}}fer, M. and {De Groot}, P. C. and Bloch, I and F{\"{o}}lling, S.},
	doi = {10.1038/nphys3061},
	journal = {Nat. Phys.},
	month = {10},
	number = {10},
	pages = {779--784},
	title = {{Observation of two-orbital spin-exchange interactions with ultracold SU(N)-symmetric fermions}},
	url = {https://www.nature.com/articles/nphys3061},
	volume = {10},
	year = {2014}
}

@article{Schulz_Resonance_1991,
	author = {Schulz, C. and Arnold, E. and Borchers, W. and Neu, W. and Neugart, R. and Neuroth, M. and Otten, E. W. and Scherf, M. and Wendt, K. and Lievens, P. and Kudryavtsev, Y. A. and Letokhov, V. S. and Mishin, V. I. and Petrunin, V. V.},
	doi = {10.1088/0953-4075/24/22/020},
	journal = {J. Phys. B Mol. Opt. Phys.},
	month = {11},
	number = {22},
	pages = {4831--4844},
	title = {{Resonance ionization spectroscopy on a fast atomic ytterbium beam}},
	url = {https://iopscience.iop.org/article/10.1088/0953-4075/24/22/020},
	volume = {24},
	year = {1991}
}

@article{Siegel_Excited-Band_2024,
	author = {Siegel, J. L. and McGrew, W. F. and Hassan, Y. S. and Chen, C.-C. and Beloy, K. and Grogan, T. and Zhang, X. and Ludlow, A. D.},
	doi = {10.1103/PhysRevLett.132.133201},
	journal = {Phys. Rev. Lett.},
	month = {3},
	number = {13},
	pages = {133201},
	title = {{Excited-Band Coherent Delocalization for Improved Optical Lattice Clock Performance}},
	url = {https://doi.org/10.1103/PhysRevLett.132.133201 https://link.aps.org/doi/10.1103/PhysRevLett.132.133201},
	volume = {132},
	year = {2024}
}

@phdthesis{Singh_Hyperfine_2014,
	author = {Singh, Alok Kumar},
	url = {https://etd.iisc.ac.in/handle/2005/3106},
	school = {Indian Institute of Science, Bengaluru},
	title = {{Hyperfine Structure-Measurement in Alkali-metal Atoms and Ytterbium Atom}},
	year = {2014}
}

@article{Stone_Table_2005,
	author = {Stone, N.J.},
	doi = {10.1016/j.adt.2005.04.001},
	journal = {Data Nucl. Data Tables},
	month = {5},
	number = {1},
	pages = {75--176},
	title = {{Table of nuclear magnetic dipole and electric quadrupole moments}},
	url = {https://linkinghub.elsevier.com/retrieve/pii/S0092640X05000239},
	volume = {90},
	year = {2005}
}

@techreport{Stone_Table_2019,
	address = {Vienna, Austria},
	author = {Stone, N.J.},
	doi = {10.61092/iaea.yjpc-cns6},
	institution = {International Atomic Energy Agency},
	title = {{Table of Recommended Nuclear Magnetic Dipole Moments: Part I - Long-lived States}},
	url = {https://nds.iaea.org/publications/indc/indc-nds-0794.pdf},
	year = {2019}
}

@incollection{Sugawa_ULTRACOLD_2013,
	author = {Sugawa, S. and Takasu, Y. and Enomoto, K. and Takahashi, Y.},
	booktitle = {{Annu. Rev. Cold Atoms Mol.}},
	doi = {10.1142/9789814440400_0001},
	editor = {Madison, Kirk W and Wang, Yiqiu and Rey, Ana Maria and Bongs, Kai},
	month = {2},
	pages = {3--51},
	publisher = {WORLD SCIENTIFIC},
	title = {{ULTRACOLD YTTERBIUM: GENERATION, MANY-BODY PHYSICS, AND MOLECULES}},
	url = {http://www.worldscientific.com/doi/abs/10.1142/9789814440400{\_}0001},
	volume = {1},
	year = {2013}
}

@article{Taichenachev_Magnetic_2006,
	author = {Taichenachev, A. and Yudin, V. and Oates, C. and Hoyt, C. and Barber, Z. and Hollberg, L.},
	doi = {10.1103/PhysRevLett.96.083001},
	journal = {Phys. Rev. Lett.},
	month = {3},
	number = {8},
	pages = {083001},
	title = {{Magnetic Field-Induced Spectroscopy of Forbidden Optical Transitions with Application to Lattice-Based Optical Atomic Clocks}},
	url = {https://link.aps.org/doi/10.1103/PhysRevLett.96.083001},
	volume = {96},
	year = {2006}
}

@article{Taie_Realization_2010,
	author = {Taie, Shintaro and Takasu, Yosuke and Sugawa, Seiji and Yamazaki, Rekishu and Tsujimoto, Takuya and Murakami, Ryo and Takahashi, Yoshiro},
	doi = {10.1103/PhysRevLett.105.190401},
	journal = {Phys. Rev. Lett.},
	month = {11},
	number = {19},
	pages = {190401},
	title = {{Realization of a SU(2) $\times$ SU(6) System of Fermions in a Cold Atomic Gas}},
	url = {https://link.aps.org/doi/10.1103/PhysRevLett.105.190401},
	volume = {105},
	year = {2010}
}

@article{Takasu_Magnetoassociation_2017,
	author = {Takasu, Yosuke and Fukushima, Yoshiaki and Nakamura, Yusuke and Takahashi, Yoshiro},
	doi = {10.1103/PhysRevA.96.023602},
	journal = {Phys. Rev. A},
	month = {8},
	number = {2},
	pages = {023602},
	title = {{Magnetoassociation of a Feshbach molecule and spin-orbit interaction between the ground and electronically excited states}},
	url = {https://link.aps.org/doi/10.1103/PhysRevA.96.023602},
	volume = {96},
	year = {2017}
}

@article{Takasu_Photoassociation_2004,
	author = {Takasu, Y. and Komori, K. and Honda, K. and Kumakura, M. and Yabuzaki, T. and Takahashi, Y.},
	doi = {10.1103/PhysRevLett.93.123202},
	journal = {Phys. Rev. Lett.},
	month = {9},
	number = {12},
	pages = {123202},
	title = {{Photoassociation Spectroscopy of Laser-Cooled Ytterbium Atoms}},
	url = {https://link.aps.org/doi/10.1103/PhysRevLett.93.123202},
	volume = {93},
	year = {2004}
}

@article{Takasu_Spin-Singlet_2003,
	author = {Takasu, Yosuke and Maki, Kenichi and Komori, Kaduki and Takano, Tetsushi and Honda, Kazuhito and Kumakura, Mitsutaka and Yabuzaki, Tsutomu and Takahashi, Yoshiro},
	doi = {10.1103/PhysRevLett.91.040404},
	journal = {Phys. Rev. Lett.},
	month = {7},
	number = {4},
	pages = {040404},
	title = {{Spin-Singlet Bose-Einstein Condensation of Two-Electron Atoms}},
	url = {https://link.aps.org/doi/10.1103/PhysRevLett.91.040404},
	volume = {91},
	year = {2003}
}

@article{Tanabe_frequency-stabilized_2018,
	author = {Tanabe, Takehiko and Akamatsu, Daisuke and Inaba, Hajime and Okubo, Sho and Kobayashi, Takumi and Yasuda, Masami and Hosaka, Kazumoto and Hong, Feng-Lei},
	doi = {10.7567/JJAP.57.062501},
	journal = {Jpn. J. Appl. Phys.},
	month = {6},
	number = {6},
	pages = {062501},
	title = {{A frequency-stabilized light source at 399 nm using an Yb hollow-cathode lamp}},
	url = {https://iopscience.iop.org/article/10.7567/JJAP.57.062501},
	volume = {57},
	year = {2018}
}

@article{Tang_Magic_2018,
	author = {Tang, Zhi-Ming and Yu, Yan-Mei and Jiang, Jun and Dong, Chen-Zhong},
	doi = {10.1088/1361-6455/aac181},
	journal = {J. Phys. B Mol. Opt. Phys.},
	month = {6},
	number = {12},
	pages = {125002},
	title = {{Magic wavelengths for the $6s^2$ $^1$S$_0$-$6s6p$ $^3$P$_1^o$ transition in ytterbium atom}},
	url = {https://iopscience.iop.org/article/10.1088/1361-6455/aac181},
	volume = {51},
	year = {2018}
}

@article{Tomita_Dissipative_2019,
	author = {Tomita, Takafumi and Nakajima, Shuta and Takasu, Yosuke and Takahashi, Yoshiro},
	doi = {10.1103/PhysRevA.99.031601},
	journal = {Phys. Rev. A},
	month = {3},
	number = {3},
	pages = {031601},
	title = {{Dissipative Bose-Hubbard system with intrinsic two-body loss}},
	url = {https://link.aps.org/doi/10.1103/PhysRevA.99.031601},
	volume = {99},
	year = {2019}
}

@article{Tsyganok_Scalar_2019,
	author = {Tsyganok, V. V. and Pershin, D. A. and Davletov, E. T. and Khlebnikov, V. A. and Akimov, A. V.},
	doi = {10.1103/PhysRevA.100.042502},
	journal = {Phys. Rev. A},
	month = {10},
	number = {4},
	pages = {042502},
	title = {{Scalar, tensor, and vector polarizability of Tm atoms in a 532-nm dipole trap}},
	url = {https://link.aps.org/doi/10.1103/PhysRevA.100.042502},
	volume = {100},
	year = {2019}
}

@article{Ueda_Ionization_1969,
	author = {Ueda, Nozomu},
	doi = {10.5702/massspec1953.17.643},
	journal = {J. Mass Spectrom. Soc. Jpn.},
	number = {2},
	pages = {643--647},
	title = {{Ionization Energy Difference between Isotopes and its Effect on Isotope Abundance Measurement by Surface Ionization Method}},
	url = {http://www.jstage.jst.go.jp/article/massspec1953/17/2/17{\_}2{\_}643/{\_}article},
	volume = {17},
	year = {1969}
}

@article{Uetake_Spin-dependent_2012,
	author = {Uetake, Satoshi and Murakami, Ryo and Doyle, John M. and Takahashi, Yoshiro},
	doi = {10.1103/PhysRevA.86.032712},
	journal = {Phys. Rev. A},
	month = {9},
	number = {3},
	pages = {032712},
	title = {{Spin-dependent collision of ultracold metastable atoms}},
	url = {https://link.aps.org/doi/10.1103/PhysRevA.86.032712},
	volume = {86},
	year = {2012}
}

@article{Utreja_Cross_2025,
	author = {Utreja, Shubham and Roy, Pallab and Rathore, Harish and Choudhury, Sourin and Das, Manoj and Panja, Subhasis},
	doi = {10.1016/j.optcom.2025.131870},
	journal = {Opt. Commun.},
	number = {March},
	pages = {131870},
	title = {{Cross Beam Saturated Absorption Spectroscopy: A novel technique for optimization of power broadening}},
	url = {https://doi.org/10.1016/j.optcom.2025.131870},
	volume = {585},
	year = {2025}
}

@article{Verhaar_Predicting_2009,
	author = {Verhaar, B. J. and van Kempen, E. G. M. and Kokkelmans, S. J. J. M. F.},
	doi = {10.1103/PhysRevA.79.032711},
	journal = {Phys. Rev. A},
	month = {3},
	number = {3},
	pages = {032711},
	title = {{Predicting scattering properties of ultracold atoms: Adiabatic accumulated phase method and mass scaling}},
	url = {https://link.aps.org/doi/10.1103/PhysRevA.79.032711},
	volume = {79},
	year = {2009}
}

@article{Wakui_High-Resolution_2003,
	author = {Wakui, Takashi and Jin, Wei-Guo and Hasegawa, Kenji and Uematsu, Haruko and Minowa, Tatsuya and Katsuragawa, Hidetsugu},
	doi = {10.1143/JPSJ.72.2219},
	journal = {J. Phys. Soc. Jpn.},
	month = {9},
	number = {9},
	pages = {2219--2223},
	title = {{High-Resolution Diode-Laser Spectroscopy of the Rare-Earth Elements}},
	url = {http://journals.jps.jp/doi/10.1143/JPSJ.72.2219},
	volume = {72},
	year = {2003}
}

@article{Wandel_Doppelresonanzexperimente_1970,
	author = {Wandel, G},
	doi = {10.1007/BF01642534},
	journal = {Zeitschrift f{\"{u}}r Phys. A Hadron. Nucl.},
	month = {10},
	number = {5},
	pages = {434--449},
	title = {{Doppelresonanzexperimente zur Untersuchung der Hyperfeinstruktur des $6s6p$ $^3$P$_1$-Zustandes von $^{171}$Yb und $^{173}$Yb}},
	url = {http://link.springer.com/10.1007/BF01642534},
	volume = {231},
	year = {1970}
}

@article{Wang_AME_2021,
	author = {Wang, Meng and Huang, W.J. and Kondev, F.G. and Audi, G. and Naimi, S.},
	doi = {10.1088/1674-1137/abddaf},
	journal = {Chin. Phys. C},
	month = {3},
	number = {3},
	pages = {030003},
	title = {{The AME 2020 atomic mass evaluation (II). Tables, graphs and references}},
	url = {https://iopscience.iop.org/article/10.1088/1674-1137/abddaf},
	volume = {45},
	year = {2021}
}

@article{Wang_absolute_2015,
	author = {Wang, Jun and Ren, Tongxiang and Lu, Hai and Zhou, Tao and Zhou, Yuanjing},
	doi = {10.1039/C5JA00054H},
	journal = {J. Anal. Spectrom.},
	number = {6},
	pages = {1377--1385},
	title = {{The absolute isotopic composition and atomic weight of ytterbium using multi-collector inductively coupled plasma mass spectrometry and development of an SI-traceable ytterbium isotopic certified reference material}},
	url = {https://xlink.rsc.org/?DOI=C5JA00054H},
	volume = {30},
	year = {2015}
}

@article{Wang_novel_2010,
	author = {Wang, Wen-Li and Xu, Xin-Ye},
	doi = {10.1088/1674-1056/19/12/123202},
	journal = {Chin. Phys. B},
	month = {12},
	number = {12},
	pages = {123202},
	title = {{A novel method to measure the isotope shifts and hyperfine splittings of all ytterbium isotopes for a 399-nm transition}},
	url = {https://iopscience.iop.org/article/10.1088/1674-1056/19/12/123202},
	volume = {19},
	year = {2010}
}

@article{Xu_Measurement_2014,
	author = {Xu, C.-Y. and Singh, J. and Zappala, J. C. and Bailey, K. G. and Dietrich, M. R. and Greene, J. P. and Jiang, W. and Lemke, N. D. and Lu, Z.-T. and Mueller, P. and O'Connor, T. P.},
	doi = {10.1103/PhysRevLett.113.033003},
	journal = {Phys. Rev. Lett.},
	month = {7},
	number = {3},
	pages = {033003},
	title = {{Measurement of the Hyperfine Quenching Rate of the Clock Transition in $^{171}$Yb}},
	url = {https://link.aps.org/doi/10.1103/PhysRevLett.113.033003},
	volume = {113},
	year = {2014}
}

@article{Yamaguchi_High-resolution_2010,
	author = {Yamaguchi, A. and Uetake, S. and Kato, S. and Ito, H. and Takahashi, Y.},
	doi = {10.1088/1367-2630/12/10/103001},
	journal = {New J. Phys.},
	month = {10},
	number = {10},
	pages = {103001},
	title = {{High-resolution laser spectroscopy of a Bose-Einstein condensate using the ultranarrow magnetic quadrupole transition}},
	url = {https://iopscience.iop.org/article/10.1088/1367-2630/12/10/103001},
	volume = {12},
	year = {2010}
}

@article{Yamaguchi_Inelastic_2008,
	author = {Yamaguchi, A. and Uetake, S. and Hashimoto, D. and Doyle, J. M. and Takahashi, Y.},
	doi = {10.1103/PhysRevLett.101.233002},
	journal = {Phys. Rev. Lett.},
	month = {12},
	number = {23},
	pages = {233002},
	title = {{Inelastic Collisions in Optically Trapped Ultracold Metastable Ytterbium}},
	url = {https://link.aps.org/doi/10.1103/PhysRevLett.101.233002},
	volume = {101},
	year = {2008}
}

@phdthesis{Yamaguchi_Metastable_2008,
	author = {Yamaguchi, Atsushi},
	school = {Kyoto University},
	title = {{Metastable State of Ultracold and Quantum Degenerate Ytterbium Atoms: High-Resolution Spectroscopy and Cold Collisions}},
	year = {2008},
	url = {http://hdl.handle.net/2433/124352}
}

@article{Yamamoto_ytterbium_2016,
	author = {Yamamoto, Ryuta and Kobayashi, Jun and Kuno, Takuma and Kato, Kohei and Takahashi, Yoshiro},
	doi = {10.1088/1367-2630/18/2/023016},
	journal = {New J. Phys.},
	month = {2},
	number = {2},
	pages = {023016},
	title = {{An ytterbium quantum gas microscope with narrow-line laser cooling}},
	url = {https://iopscience.iop.org/article/10.1088/1367-2630/18/2/023016},
	volume = {18},
	year = {2016}
}

@article{Yasuda_Improved_2012,
	author = {Yasuda, Masami and Inaba, Hajime and Kohno, Takuya and Tanabe, Takehiko and Nakajima, Yoshiaki and Hosaka, Kazumoto and Akamatsu, Daisuke and Onae, Atsushi and Suzuyama, Tomonari and Amemiya, Masaki and Hong, Feng-Lei},
	doi = {10.1143/APEX.5.102401},
	journal = {Appl. Phys. Express},
	month = {9},
	number = {10},
	pages = {102401},
	title = {{Improved Absolute Frequency Measurement of the $^{171}$Yb Optical Lattice Clock towards a Candidate for the Redefinition of the Second}},
	url = {https://iopscience.iop.org/article/10.1143/APEX.5.102401},
	volume = {5},
	year = {2012}
}

@article{Zhang_Controlling_2020,
	author = {Zhang, Ren and Cheng, Yanting and Zhang, Peng and Zhai, Hui},
	doi = {10.1038/s42254-020-0157-9},
	journal = {Nat. Rev. Phys.},
	month = {3},
	number = {4},
	pages = {213--220},
	title = {{Controlling the interaction of ultracold alkaline-earth atoms}},
	url = {https://www.nature.com/articles/s42254-020-0157-9},
	volume = {2},
	year = {2020}
}

@article{Zhang_Precise_2023,
	author = {Zhang, Ang and Tian, Congcong and Zhu, Qiang and Wang, Bing and Xiong, Dezhi and Xiong, Zhuanxian and He, Lingxiang and Lyu, Baolong},
	doi = {10.1088/1674-1056/aca14e},
	journal = {Chin. Phys. B},
	month = {2},
	number = {2},
	pages = {020601},
	title = {{Precise measurement of $^{171}$Yb magnetic constants for $^1$S$_0$-$^3$P$_0$ clock transition}},
	url = {https://iopscience.iop.org/article/10.1088/1674-1056/aca14e},
	volume = {32},
	year = {2023}
}

@article{Zhao_Absolute_2024,
	author = {Zhao, Chengcheng and Qi, Qichao and Sun, Changyue and Peng, Chengquan and Jin, Taoyun and Zhang, Tao and Lei, Shuai and Xia, Yan and Feng, Suzhen and Xu, Xinye},
	doi = {10.1103/PhysRevA.110.042816},
	journal = {Phys. Rev. A},
	month = {10},
	number = {4},
	pages = {042816},
	title = {{Absolute frequency measurements of the $6s6p$ $^3$P$_0$-$5d6s$ $^3$D$_1$ transitions in $^{170,171,172,173,174,176}$Yb by modulation transfer spectroscopy}},
	url = {https://link.aps.org/doi/10.1103/PhysRevA.110.042816},
	volume = {110},
	year = {2024}
}

@article{Zheng_Magic_2020,
	author = {Zheng, T. A. and Yang, Y. A. and Safronova, M. S. and Safronova, U. I. and Xiong, Zhuan Xian and Xia, T. and Lu, Z. T.},
	doi = {10.1103/PhysRevA.102.062805},
	journal = {Phys. Rev. A},
	number = {6},
	pages = {4--7},
	title = {{Magic wavelengths of the Yb ($6s^2$ $^1$S$_0$-$6s6p$ $^3$P$_1$) intercombination transition}},
	volume = {102},
	year = {2020}
}

@article{Zhou_Characterization_2020,
	author = {Zhou, Min and Zhang, Sheng and Luo, Limeng and Xu, Xinye},
	doi = {10.1103/PhysRevA.101.062506},
	journal = {Phys. Rev. A},
	month = {6},
	number = {6},
	pages = {062506},
	title = {{Characterization of ytterbium resonance lines at 649 nm with modulation-transfer spectroscopy}},
	url = {https://doi.org/10.1103/PhysRevA.101.062506 https://link.aps.org/doi/10.1103/PhysRevA.101.062506},
	volume = {101},
	year = {2020}
}

@article{Zinkstok_Hyperfine_2002,
	author = {Zinkstok, R. and van Duijn, E J and Witte, S. and Hogervorst, W.},
	doi = {10.1088/0953-4075/35/12/305},
	journal = {J. Phys. B Mol. Opt. Phys.},
	month = {6},
	number = {12},
	pages = {305},
	title = {{Hyperfine structure and isotope shift of transitions in Yb I using UV and deep-UV cw laser light and the angular distribution of fluorescence radiation}},
	url = {https://iopscience.iop.org/article/10.1088/0953-4075/35/12/305},
	volume = {35},
	year = {2002}
}

@article{Zmbov_First_1966,
	author = {Zmbov, K. F. and Margrave, J. L.},
	doi = {10.1021/j100881a508},
	journal = {J. Phys. Chem.},
	month = {9},
	number = {9},
	pages = {3014--3017},
	title = {{The First Ionization Potentials of Samarium, Europium, Gadolinium, Dysprosium, Holmium, Erbium, Thulium, and Ytterbium by the Electron-Impact Method}},
	url = {https://pubs.acs.org/doi/abs/10.1021/j100881a508},
	volume = {70},
	year = {1966}
}

@article{vanWijngaarden_Measurement_1994,
	author = {van Wijngaarden, W. A. and Li, J.},
	doi = {10.1364/JOSAB.11.002163},
	journal = {J. Opt. Soc. Am. B},
	month = {11},
	number = {11},
	pages = {2163},
	title = {{Measurement of isotope shifts and hyperfine splittings of ytterbium by means of acousto-optic modulation}},
	url = {https://opg.optica.org/abstract.cfm?URI=josab-11-11-2163},
	volume = {11},
	year = {1994}
}


%% file: miscellaneous.bib
@article{Chaiko_Isotope_1966,
    author = {Chaiko, Y.},
    journal = {Opt. Spectrosc.},
    volume = {20},
    pages = {424},
    year = {1966},
    title = {{Isotope shifts of resonance lines in the spectrum of ytterbium}},
    doi = {}
}

@unpublished{Topper_unpublished,
    author = {T\"opper, O. and Guth\"ohrlein, G. H. and Hillerman, P.},
    title = {},
    note = {unpublished}
}

@article{Komarovskii_Oscillator_1969,
    author = {Komarovskii, V. A. and Penkin, N. P.},
    journal = {Opt. Spectrosc.},
    volume = {26},
    pages = {483},
    year = {1969},
    title = {{Oscillator Strengths of the Spectral Lines of Tm I and Yb I}},
    doi = {},
}

@inproceedings{Lange_1970,
    author = {Lange, W. and Luther, J. and Stendel, A.},
    title = {},
    booktitle = {{Proceedings, 2nd Conference of European Group for Atomic Spectroscopy}},
    year = {1970},
    pages = {31},
    address = {Hanover},
}

@article{Blagoev_1978,
    author = {Blagoev, K. B. and Komarovskii, V. A. and Penkin, N. P.},
    journal = {Opt. Spectrosc.},
    volume = {45},
    pages = {832},
    year = {1978},
    title = {},
    doi = {},
}

@article{Burshtein_Lifetimes_1974,
    author = {Burshtein, M. L. and Verolainen, Ya. F. and Komarovskii, V. A. and Osherovich, A. L. and Penkin, N. P.},
    journal = {Opt. Spectrosc.},
    volume = {37},
    pages = {351},
    year = {1974},
    title = {{Lifetimes of the $^3$P$_1$ level of Yb I and the $^2$P$_{3/2,1/2}$ level of Yb II}},
    doi = {},
}

@misc{NIST_ASD,
    author = {A.~Kramida and {Yu.~Ralchenko} and
    J.~Reader and {and NIST ASD Team}},
    HOWPUBLISHED = {{NIST Atomic Spectra Database
    (ver. 5.12), [Online]. Available:
    {\tt{https://physics.nist.gov/asd}} [2025, August 19].
    National Institute of Standards and Technology,
    Gaithersburg, MD.}},
    year = {2024},
}

@misc{Steck2024CesiumDLine,
  author       = {Steck, Daniel A.},
  title        = {{Cesium D Line Data}},
  howpublished = {Available online at \url{http://steck.us/alkalidata}},
  note         = {Revision 2.3.3, original revision posted 23 January 1998},
  year         = {2024}
}

@misc{Steck2024Rb85DLine,
  author       = {Steck, Daniel A.},
  title        = {{Rubidium 85 D Line Data}},
  howpublished = {Available online at \url{http://steck.us/alkalidata}},
  note         = {Revision 2.3.3, original revision posted 30 April 2008},
  year         = {2024}
}

@misc{Steck2024Rb87DLine,
  author       = {Steck, Daniel A.},
  title        = {{Rubidium 87 D Line Data}},
  howpublished = {Available online at \url{http://steck.us/alkalidata}},
  note         = {Revision 2.3.3, original revision posted 25 September 2001},
  year         = {2024}
}

@misc{Steck2024SodiumDLine,
  author       = {Steck, Daniel A.},
  title        = {{Sodium D Line Data}},
  howpublished = {Available online at \url{http://steck.us/alkalidata}},
  note         = {Revision 2.3.3, original revision posted 27 May 2000},
  year         = {2024}
}

@manual{BuckResearch2012,
  title        = {{CR-1A Hygrometer with Autofill: Operating Manual}},
  author       = {{Buck Research Instruments, LLC}},
  organization = {Buck Research Instruments, LLC},
  address      = {Boulder, CO, USA},
  year         = {2012},
  url          = {https://www.hygrometers.com/wp-content/uploads/CR-1A-users-manual-2009-12.pdf},
}

@misc{CCTF2021PSFS2,
  author       = {{Consultative Committee for Time and Frequency (CCTF)}},
  title        = {{Recommendation PSFS-2}},
  year         = {2021},
  note         = {22nd meeting (session II – online)},
  url          = {https://www.bipm.org/en/committees/cc/cctf/22-_2-2021},
  urldate      = {2025-05-21}
}

@misc{UDportal,
  author       = {Barakhshan, Parinaz and Marrs, Adam and Bhosale, Akshay and Arora, Bindiya and Eigenmann, Rudolf and Safronova, Marianna S.},
  title        = {{Portal for High-Precision Atomic Data and Computation} (version 2.0)},
  howpublished = {University of Delaware, Newark, DE, USA. \url{https://www.udel.edu/atom}},
  note         = {Accessed May 25, 2025. Original data source cited where available.},
  year         = {2022},
  urldate      = {2025-05-25}
}

@misc{SteckBook,
  author       = {Steck, Daniel A.},
  title        = {{Quantum and Atom Optics}},
  howpublished = {Available online at \url{http://steck.us/teaching}},
  note         = {revision 0.16.4, 7 May 2025},
  year         = {2007}
}

@misc{Trassinelli_minimalistic_2024,
	archiveprefix = {arXiv},
	arxivid = {2406.08293},
	author = {Trassinelli, Martino and Maxton, Marleen},
	eprint = {2406.08293},
	month = {12},
	pages = {1--11},
	title = {{A minimalistic and general weighted averaging method for inconsistent data}},
	url = {http://arxiv.org/abs/2406.08293},
	year = {2024}
}

@misc{Pucher_Sr_2025,
	archiveprefix = {arXiv},
	arxivid = {2507.10487},
	author = {Pucher, Sebastian and Kristensen, Sofus Laguna and Kroeze, Ronen M.},
	eprint = {2507.10487},
	month = {7},
	pages = {1--31},
	title = {{$^{88}$Sr Reference Data}},
	url = {http://arxiv.org/abs/2507.10487},
	year = {2025}
}

@misc{Li_Fast_2025,
	archiveprefix = {arXiv},
	arxivid = {2506.15633},
	author = {Li, Yiyi and Bao, Yicheng and Peper, Michael and Li, Chenyuan and Thompson, Jeff D.},
	eprint = {2506.15633},
	pages = {1--16},
	title = {{Fast, continuous and coherent atom replacement in a neutral atom qubit array}},
	url = {http://arxiv.org/abs/2506.15633},
	year = {2025}
}

@misc{Johanning_Resonance_2010,
	archiveprefix = {arXiv},
	arxivid = {0712.0969},
	author = {Johanning, M. and Braun, A. and Eiteneuer, D. and Paape, Chr. and Balzer, Chr. and Neuhauser, W. and Wunderlich, Chr.},
	eprint = {0712.0969},
	month = {6},
	title = {{Resonance enhanced isotope-selective photoionization of YbI for ion trap loading}},
	url = {http://arxiv.org/abs/0712.0969},
	year = {2010}
}

@misc{Aeppli_Atomic_2025,
    archivePrefix = {arXiv},
    arxivid = {2512.21428},
    author = {Aeppli, Alexander and Arthur-Dworschack, Willa J. and Beloy, Kyle and Berry, Caitlin M. and Bothwell, Tobias and Folz, Angela and Fortier, Tara M. and Grogan, Tanner and Hassan, Youssef S. and Hu, Zoey Z. and Hume, David B. and Hunt, Benjamin D. and Kim, Kyungtae and Koepke, Amanda and Lee, Dahyeon and Leibrandt, David R. and Lewis, Ben and Ludlow, Andrew D. and Marshall, Mason C. and Nardelli, Nicholas V. and Ranganath, Harikesh and Castillo, Daniel A. Rodriguez and Sherman, Jeffrey A. and Siegel, Jacob L. and Thornton, Suzanne and Warfield, William and Ye, Jun},
    eprint = {2512.21428},
    month = {dec},
    pages = {1--19},
    title = {{Atomic clock frequency ratios with fractional uncertainty $\leq3.2\times10^{-18}$}},
    url = {http://arxiv.org/abs/2512.21428},
    year = {2025}
}

@misc{Ishiyama_Excluding_2026,
    archivePrefix = {arXiv},
    arxivid = {2601.08487},
    author = {Ishiyama, Taiki and Ono, Koki and Asano, Reiji and Kawase, Hokuto and Takano, Tetsushi and Sunaga, Ayaki and Yamamoto, Yasuhiro and Tanaka, Minoru and Takahashi, Yoshiro},
    eprint = {2601.08487},
    month = {jan},
    title = {{Excluding Hypothetical Light Boson Interpretation of Yb King Plot Nonlinearity with the $^1$S$_0$ $\leftrightarrow$ $^3$P$_2$ Isotope Shift Measurement}},
    url = {http://arxiv.org/abs/2601.08487},
    year = {2026}
}
